\documentclass[aps,reprint,superscriptaddress,nofootinbib,longbibliography]{revtex4-2}
\usepackage{graphicx, color, soul}
\usepackage{xcolor, bbold}      
\usepackage{siunitx}
\usepackage{amsmath, amssymb}
\usepackage{braket}
\usepackage{color,soul,url}
\usepackage{float}
\usepackage[hidelinks]{hyperref}

\usepackage{microtype} 
\usepackage[]{newtxtext} 
\usepackage[]{newtxmath}

\usepackage{hyperref}
\hypersetup{colorlinks,
	linkcolor={blue!75!black!80!yellow},
	citecolor={blue!75!black!80!yellow},
	urlcolor={blue!75!black!80!yellow}
}

\usepackage{booktabs}
\usepackage{colortbl}
\newcommand{\separatorrule}{\arrayrulecolor{black!25}\specialrule{.25pt}{.65\jot}{.65\jot}\arrayrulecolor{black}}

\newcommand{\vk}{\Pi}
\newcommand{\vkk}{{\bf \Pi}}
\newcommand{\ii}{\mathrm{i}}
\newcommand{\tr}{\text{Tr}} 
\newcommand{\A}{\mathcal{A}} 
\newcommand{\F}{\mathcal{F}} 

\newcommand{\matrixel}[3]{\left<#1 \vphantom{#2} \vphantom{#3} \left| #2 \vphantom{#1} \vphantom{#3} \right| #3 \vphantom{#1} \vphantom{#2} \right>}

\begin{document}
\title{Topological Phases of Photonic Crystals under Crystalline Symmetries}

\author{Sachin Vaidya}
\email{sachin4594@gmail.com}
\affiliation{Department of Physics, The Pennsylvania State University, University Park, PA 16802, USA}
\author{Ali Ghorashi}
\affiliation{Department of Physics, Massachusetts Institute of Technology, Cambridge, Massachusetts 02139, USA}
\author{Thomas Christensen}
\affiliation{Department of Physics, Massachusetts Institute of Technology, Cambridge, Massachusetts 02139, USA}
\affiliation{Department of Electrical and Photonics Engineering, Technical University of Denmark, 2800, Denmark}
\author{Mikael C. Rechtsman}
\affiliation{Department of Physics, The Pennsylvania State University, University Park, PA 16802, USA}
\author{Wladimir A. Benalcazar}
\email{benalcazar@emory.edu}
\affiliation{Department of Physics, The Pennsylvania State University, University Park, PA 16802, USA}
\affiliation{Department of Physics, Princeton University, Princeton, New Jersey 08542, USA}
\affiliation{Department of Physics, Emory University, Atlanta, GA 30322, USA}

\date{\today}

\begin{abstract}
Photonic crystals (PhCs) have emerged as a popular platform for realizing various topological phases due to their flexibility and potential for device applications. In this article, we present a comprehensive classification of topological bands in one- and two-dimensional photonic crystals, with and without time-reversal symmetry. Our approach exploits the symmetry representations of field eigenmodes at high-symmetry points in momentum space, allowing for the efficient design of a wide range of topological PhCs. In particular, we show that the complete classification provided here is useful for diagnosing photonic crystal analogs of obstructed atomic limits, fragile phases, and stable topological phases that include bands with Dirac points and Chern numbers.
\end{abstract}
\maketitle


\section{Introduction}
Photonic crystals (PhCs) are periodically patterned dielectric media that can be described by a Maxwell eigenvalue problem~\cite{photoniccrystalsbook, photoniccrystalsbook2}. The periodicity of the dielectric medium acts analogously to a potential for electromagnetic waves and the solutions take the form of Bloch functions that are distributed into photonic bands. Similar to electronic states in conventional solids, PhC eigenmodes can be characterized by topological indices that are global properties across momentum space~\cite{topoPhCReview1, topoPhCReview2, topoPhCReview3}. An important physical manifestation of these topological indices is the existence of states that reside on the boundaries of the system - this is known as the bulk-boundary correspondence. 

A wide variety of topological phases have been realized using PhC-based platforms (as distinct from waveguide-arrays~\cite{rechtsman2013photonic, HOTI_waveguides}, coupled-resonator~\cite{hafezi2013imaging, HOTI_ringresonator} or microwave-circuit~\cite{microwave_HOTI} realizations). In one and two dimensions, this includes analogs of the SSH model with quantized polarization~\cite{1DPhC1, 1DPhC2, 1DPhC3}, Chern insulators~\cite{Chern0,  Chern5, Chern1, Chern2, Chern3, Chern4}, quantum spin-Hall-like phases~\cite{QSH1, QSH2, QSH3, QSH4, QSH5}, Dirac semi-metals~\cite{Dirac1, Dirac2, Dirac3, diracmode1, diracexp1, diracexp2, diracexp3}, valley-Hall phases~\cite{ValleyHall1, ValleyHall2, ValleyHall3, stobbe_valleyhall}, bulk-obstructed higher-order topological insulators (HOTIs)~\cite{HOTI1, HOTI2, HOTI3, HOTI4, HOTI5, HOTI6}, including quadrupolar HOTIs~\cite{Quadrupole1, Quadrupole2, FloquetQuadrupole}, and fragile phases~\cite{FragilePhC}. Several of these have also been proposed for photonic device applications such as for lasing~\cite{1DPhC3, HOTI_Laser1, HOTI_Laser2, PhC_topo_laser1}, harmonic generation~\cite{SHG, THG} and light transport~\cite{TopoTransport1}. Moreover, the flexibility of the PhC-based platform has made it possible to explore the effects of non-linearity~\cite{NL_PhC1, NL_PhC2} and non-Hermiticity~\cite{NH_PhC1, NH_PhC2} alongside topology -- novel physics that is difficult to realize in conventional electronic solids.

Topological systems can be  classified in the tenfold way~\cite{10foldway1, 10foldway2, kitaev2009periodic} by the presence or absence of the three fundamental symmetries: time-reversal, chiral, and particle-hole symmetries. PhCs generally do not possess chiral and particle-hole symmetries and therefore belong in either class A (TR-broken) or class AI (TR-symmetric) of the tenfold way. However, crystalline symmetries enrich this classification and can help identify finer topological distinctions within these classes. 

There are three kinds of topological bands: (i) Obstructed ``atomic limit" (OAL) bands~\cite{TQC1}, that admit exponentially-localized Wannier functions~\cite{MLWF} (such bands are referred to as ``Wannierizable") (ii) fragile bands~\cite{Fragile1, Fragile2} that are non-Wannierizable but become Wannierizable when combined with other atomic limit bands and (iii) stable topological bands that are not Wannierizable. In all cases, topology can be generally identified by computing Berry phases (or, more generally, Wilson loops) over the entire Brillouin zone. In the presence of crystalline symmetries, it is possible to identify and distinguish a subset of them by constructing symmetry-indicator invariants~\cite{Rot_HOTI1, symmetry_indicators_po, benalcazar2014}. Compared to Berry phases, this symmetry-based approach can be substantially less intensive for computation since it only requires looking at the eigenmodes at high-symmetry points of the Brillouin zone (BZ).

In this article, we build on previous works in electronic systems~\cite{benalcazar2014, Rot_HOTI1} and comprehensively develop a complete classification for topological bands in one- and two-dimensional PhCs under crystalline symmetries both with and without time-reversal symmetry. For each point-group symmetry, we exhaustively calculate the topological indices, defined using symmetry-indicator invariants, for the basis set of atomic limits that span the space of all possible atomic limit bands via the procedure of induction of band representations~\cite{canoEBRs, bradlyn2017}. This allows us to establish a bulk-boundary correspondence for OAL bands in PhCs where we show that despite the absence of a Fermi level, the notion of a filling anomaly remains meaningful and can be used to infer the topological origin of boundary states directly from the frequency spectrum of the PhCs. This approach also allows us to diagnose topological bands that are not OALs, namely fragile phases and bands with Dirac points and Chern numbers, which is made possible by exploiting the linear structure of the classification. Based on our classification, we propose a strategy to diagnose and design topological PhCs. Finally, for completeness, we discuss the PhC-based implementations of a few other topological systems that lie outside of this framework but where symmetry plays an important role.

The rest of the paper is organized as follows: In section II, we review the concepts of Berry phases, symmetry-indicator invariants, and filling anomaly for 1D PhCs. In section III, we extend these ideas to 2D PhCs by developing the classification due to rotational symmetries, both with and without TR-symmetry. In section IV, we discuss design and characterization strategies for various topological PhCs using our classification, along with appropriate examples. In section V, we discuss PhC-based implementations of the quantum spin-Hall phases, valley-Hall phases and analogs of insulators with quantized multipole moments, all of which lie outside of this framework.

\section{1D Photonic crystals}
Maxwell's equations with no sources and for a medium that is linear, isotropic, and lossless are~\cite{photoniccrystalsbook, photoniccrystalsbook2}
\begin{align}
\nabla \cdot {\bf H}({\bf r},t) = 0, \nonumber\\
\nabla \times {\bf E}({\bf r},t)+\mu_0 \partial_t {\bf H}({\bf r},t) = 0, \nonumber\\
\nabla  \cdot [\epsilon({\bf r}) {\bf E}({\bf r},t)] = 0, \nonumber\\
\nabla \times {\bf H}({\bf r},t) - \epsilon_0 \epsilon({\bf r}) \partial_t {\bf E}({\bf r},t) = 0,
\label{eq:Maxwell}
\end{align}
where ${\bf E}$ and ${\bf H}$ are the electric and magnetic fields respectively, $\epsilon({\bf r})$ is the dielectric function, and $\epsilon_0$ and $\mu_0$ are the vacuum permitivity and permeability respectively. Expanding the temporal component of the electric and magnetic fields into harmonics as ${\bf H}({\bf r},t) = {\bf H}({\bf r}) \mathrm{e}^{-\mathrm{i}\omega t}, \quad {\bf E}({\bf r},t) = {\bf E}({\bf r}) \mathrm{e}^{-\mathrm{i}\omega t}$, these equations reduce to
\begin{align}
\nabla \times \left(\frac{1}{\epsilon({\bf r})} \nabla \times {\bf H}({\bf r}) \right) = \left(\frac{\omega}{c}\right)^2 {\bf H}({\bf r}), \nonumber\\
\nabla \times \nabla \times {\bf E}({\bf r})  = \left(\frac{\omega}{c}\right)^2 \epsilon({\bf r}) {\bf E}({\bf r}).
\label{eq:MasterEq}
\end{align}

Due to the absence of magneto-electric coupling, we can choose to solve only the equation for $\mathbf{H}(\mathbf{r})$ in Eq.~\eqref{eq:MasterEq} since $\mathbf{E}(\mathbf{r})$ can be found from $\mathbf{H}(\mathbf{r})$ using the last equation in Eq.~\eqref{eq:Maxwell}. 

A 1D PhC, shown schematically in Fig.~\ref{fig:1DPhC_unitcell}(a), is a 3D material characterized by a refractive index that is periodic along one direction ($x$) and is uniform along the other two directions ($y$ and $z$). The magnetic field eigenmode can therefore be written as a plane wave solution in the $y,z$ plane multiplied by an $x$-dependent vector field, $\mathbf{H} = e^{i\mathbf{k_{\parallel}}\cdot\boldsymbol{\rho}}\mathbf{h}(x)$, where $\mathbf{k_{\parallel}}$ is the momentum along the uniform directions and $\boldsymbol{\rho} = y\mathbf{\hat{y}}+z\mathbf{\hat{z}}$. However, we are only concerned with propagation along the periodic direction, which implies that $\mathbf{k_{\parallel}} = 0$. Moreover, since the fields must be perpendicular to the propagation direction, we can define two orthogonal polarizations where the vector fields lie in the $y, z$ plane. Assuming isotropy, we can take these polarized fields to be $\mathbf{h}_z(x) = h_z(x)\mathbf{\hat{z}}$ and $\mathbf{h}_y(x) = h_y(x)\mathbf{\hat{y}}$.
This leads to the following eigenvalue problem for the scalar fields $h_{\xi}(x)$ for $\xi\in\{y,z\}$,
\begin{align}
\hat{\Theta}_1h_\xi(x) =& \left(\frac{\omega}{c}\right)^2 h_\xi(x), \;\; \hat{\Theta}_1 \equiv -\partial_x\left( \frac{1}{\epsilon(x)} \partial_x\right),
\label{eq:eigval_problem}
\end{align}
where $\hat{\Theta}_1$ is the 1D Maxwell operator that plays a role analogous to the Hamiltonian in quantum mechanics. By exploiting the periodicity of the dielectric function, the above equation can be solved using Bloch's theorem. Specifically, the ansatz $h_{\xi,n,k_x}(x) = e^{ik_xx}u_{\xi,n,k_x}(x)$, can be used to solve Eq.~\eqref{eq:eigval_problem}, where $u_{\xi,n,k_x}(x)$ is the periodic part of the field defined over a unit cell. With this, Eq.~\eqref{eq:eigval_problem} can be written as
\begin{align}
\hat{\Theta}_{1,k_x}u_{\xi,n, k_x}(x) = \left(\frac{\omega_n}{c}\right)^2 u_{\xi,n, k_x}(x),
\label{eq:eigval_problem_kspace}
\end{align}
where
\begin{align}
    \hat{\Theta}_{1,k_x} \equiv -(\partial_x+ik_x)\left( \frac{1}{\epsilon(x)} (\partial_x+ik_x)\right).
\end{align}

This yields field solutions distributed across discrete frequency bands labeled by the index $n$ and with their momentum, $k_x$, restricted to lie within the first BZ, as shown in Fig.~\ref{fig:1DPhC_unitcell}(b). It is also useful to define the inner product between two fields over a unit cell (UC) as
\begin{align}
    \braket{u_{\xi,n_1,k_1} | u_{\xi,n_2,k_2}} = \int_\mathrm{UC} u_{\xi,n_1,k_1}^*(x)u_{\xi,n_2,k_2}(x)\ {\rm d}x.
\end{align}

 Like electronic energy bands in conventional solids, the introduction of frequency gaps allows for a topological characterization of isolated individual photonic bands or a group of bands, as discussed in the following section. 

\subsection{Classification due to inversion symmetry}
\label{classification_due_to_inversion_sym_1D}

\begin{figure}[]
\centering
\includegraphics[width=\columnwidth]{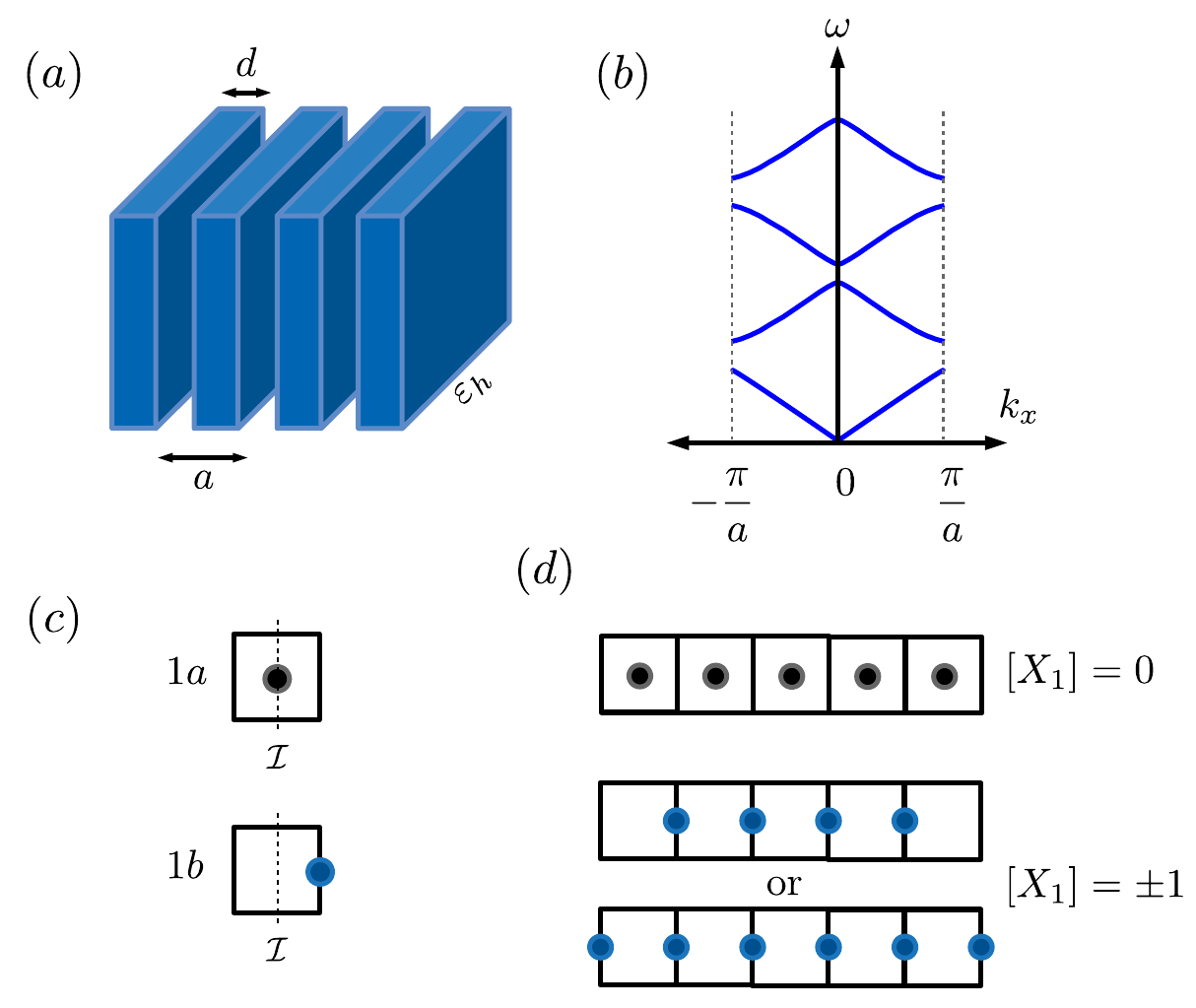}
\caption{(a) Schematic of a 1D PhC made out of alternating layers of dielectric materials of dielectric constants $\varepsilon_h$ and $\varepsilon_l$ with lattice constant $a$. (b) Schematic of the dispersion of light in a 1D PhC. (c) Wannier centers (solid circles) are located at the two possible maximal Wyckoff positions in the inversion-symmetric unit cells (squares). (d) Filling anomaly due to inversion symmetry. The finite trivial system (with $[X_1] = 0$) has a number of states equal to the number of unit cells and is inversion symmetric. The topological system requires at least one more or one fewer state to maintain inversion symmetry.}
\label{fig:1DPhC_unitcell}
\end{figure}

1D PhCs fall into class A or AI of the tenfold way, depending on whether they break or preserve time-reversal symmetry (TRS), respectively. In either case, 1D PhCs are topologically trivial in the absence of other symmetries. However, the presence of inversion symmetry protects two topological phases in both A and AI. 
The invariant for a single band in these phases is the Berry phase
\begin{align}
\theta=\int_{\mathrm{BZ}} \A_{n,k_x}\ {\rm d}k_x
\label{eq:BerryPhase}
\end{align}
where $\A_{n,k}=-\ii\matrixel{u_{\xi,n,k_x}}{\partial_{k_x}}{u_{\xi,n,k_x}}$ is the Berry connection. Under an inversion-symmetric choice of unit cell, the Berry phase is quantized to 0 or $\pi$. This quantization has an intuitive interpretation: in 1D, all photonic bands admit maximally localized Wannier functions whose centers are gauge invariant quantities~\cite{kohn_WF, Wannierm1, Wannier0, PhCWannier1, PhCWannier2, PhCWannier3, PhCWannier4}. Due to inversion symmetry, a single Wannier center (per unit cell) can only be located in two distinct positions in the unit cell, as shown in Fig.~\ref{fig:1DPhC_unitcell}(c). These positions are called maximal Wyckoff positions and are labeled by $1a$ and $1b$. The Berry phase in Eq.~\eqref{eq:BerryPhase} of a single non-degenerate band indicates the location of the (one) Wannier center within each unit cell, where $\theta=0$ and $\pi$ correspond to the Wannier center being located at the position $1a$ (middle of the unit cell) and $1b$ (edge of the unit cell), respectively.

The calculation of Eq.~\eqref{eq:BerryPhase} involves an integral over the entire BZ, but it can be greatly simplified by looking at the representations of the $\mathbf{H}$ or $\mathbf{E}$ fields at the high-symmetry points (HSPs) of the BZ~\cite{Wilson_loop_inversionsym}, which are $\mathbf{\Gamma}$ $(k_x = 0)$ and $\mathbf{X}$ $(k_x = \pi/a)$. Under inversion symmetry $\mathcal{I}: \mathbf{r} \rightarrow -\mathbf{r}$, the 1D Maxwell operator obeys
\begin{equation}
    \hat{\mathcal{I}}\hat{\Theta}_{1,k_x}\hat{\mathcal{I}}^{-1} = \hat{\Theta}_{1,-k_x},
    \label{eq:InversionSymmetry1D}
\end{equation}
where $\hat{\mathcal{I}}$ is the inversion operator. Equation~\eqref{eq:InversionSymmetry1D} implies that $\hat{\Theta}_{1, k_x}$ commutes with $\hat{\mathcal{I}}$ at $\mathbf{\Gamma}$ and $\mathbf{X}$, i.e.,  $[\hat{\Theta}_{1,\mathbf{\Gamma}}, \hat{\mathcal{I}}] = [\hat{\Theta}_{1,\mathbf{X}}, \hat{\mathcal{I}}] = 0$, since these HSPs map to themselves under a negative sign, modulo a reciprocal lattice vector. Thus, the eigenmodes at these HSPs can be labeled by the eigenvalues of $\hat{\mathcal{I}}$, which are $\pm 1$ since $\hat{\mathcal{I}}^2 = 1$. To aid with generalization to 2D later, we denote these eigenvalues at the HSP $\mathbf{\Pi}$ as $\mathbf{\Pi}_{1}=+1$ and $\mathbf{\Pi}_{2}=-1$. We can now define the symmetry-indicator invariant for a set of bands as
\begin{align}
[X_1] \equiv \# \mathbf{X}_{1} - \# \mathbf{\Gamma}_{1}\quad \in \mathbb{Z},
\label{eq:1d_invariant}
\end{align}
where $\# \mathbf{\Pi}_{1}$ is the number of states at the HSP $\mathbf{\Pi}$ with $\mathcal{I}$ eigenvalue +1. The invariant in Eq.~\eqref{eq:1d_invariant} then encodes the value of the Berry phase as~\cite{Wilson_loop_inversionsym, Inv_sym_topo_insulators}
\begin{align}
\frac{\theta}{2\pi}=\frac{1}{2}[X_1]\quad \text{mod 1},
\label{eq:QuantizedPolarization1D}
\end{align}
which provides a $\mathbb{Z}_2$ classification of dipole moments in inversion-symmetric 1D crystals. We note that the Berry phase and the invariant in Eq.~\eqref{eq:1d_invariant} depend on the choice of unit cell.

The bands that originate from localized and symmetric Wannier functions form a representation of the crystal's symmetry group, called a band representation~\cite{canoEBRs}. The values of $[X_1]$ for a single isolated band can be enumerated exhaustively by working out the inverse problem, i.e., given a set of Wannier functions, we can calculate the band representation that such a set leads to. This inverse problem of band topology has been used to classify topological phases in insulators~\cite{bradlyn2017,canoEBRs}. We review this procedure for 1D bands in Appendix~C.

In the next section, we explore the physical consequence of a non-trivial invariant: the presence of boundary states. However, as we shall describe shortly, due to the lack of additional symmetries that impose constraints on the frequency spectrum (such as chiral or particle-hole symmetries), these boundary states need not lie within bandgaps, and the issue of bulk-boundary correspondence is somewhat more subtle in PhCs.

\subsection{Filling anomaly, counting mismatch, and boundary states}

\begin{figure*}[]%
\centering
\includegraphics[width=2\columnwidth]{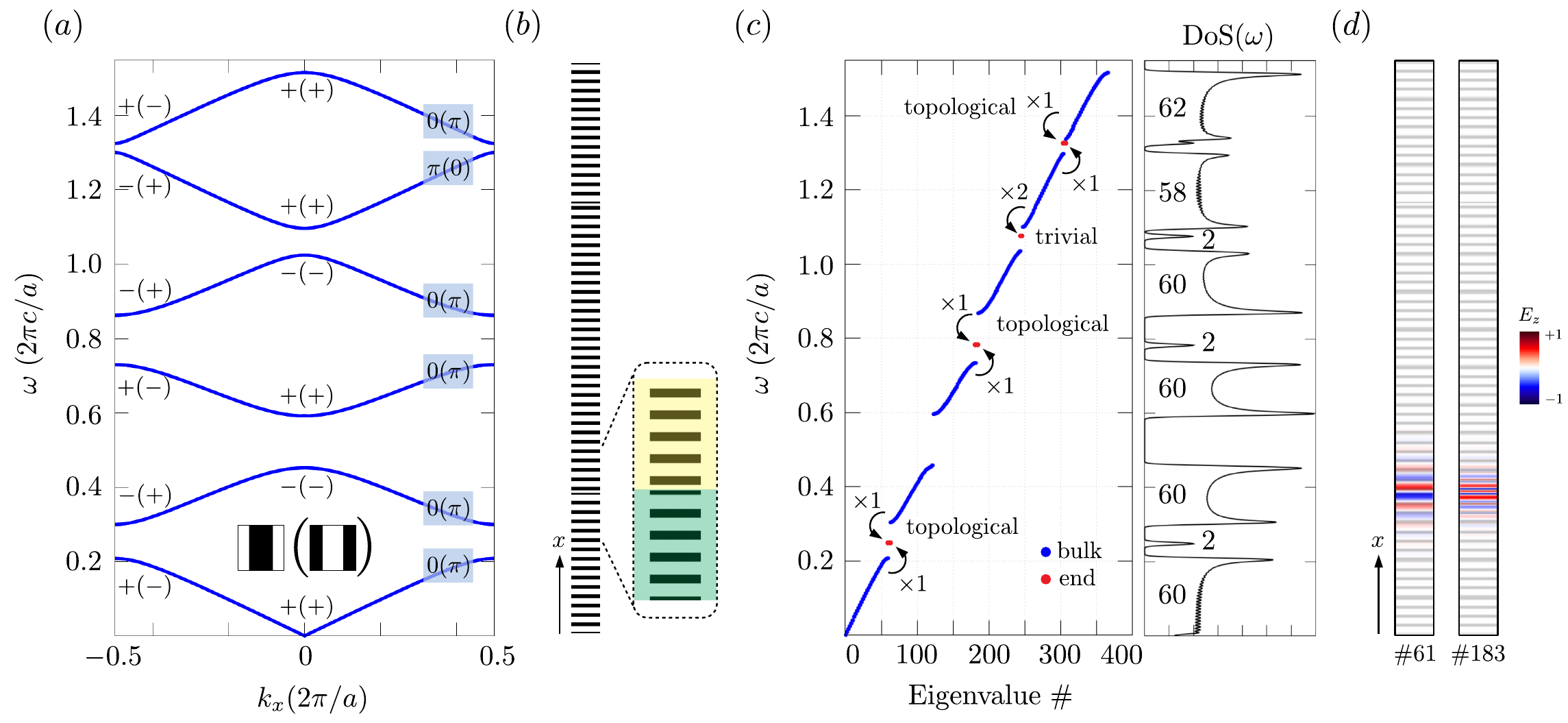}
\caption{(a) The photonic band structure of a 1D PhC with $\varepsilon_h = 6.25$, $\varepsilon_l = 1$ and $d=0.6a$. The two possible types of inversion-symmetric unit cells are shown in the inset. Eigenvalues of $\mathcal{I}$ at the HSPs for both types of unit cells are labeled with $+/-$ signs. The Berry phases for both types of unit cells are shown in blue boxes.  (b) The dielectric profile of a finite system of size 61 unit cells with interfaces between the two types of unit cells. The inset highlights the switch between the unit cell types across the boundary (c) The frequency spectrum for the finite system shown in (b). An odd-integer counting mismatch per band leads to the presence of topological boundary states in the first and third bandgaps. The photonic DoS is also shown in the same figure, labeled with the number of states. (d) The $E_z$ mode profiles of one of the two topological boundary states in the first and third gaps.}
\label{fig:1D_example}
\end{figure*}

The existence of boundary states can be heuristically understood by considering the effect of a boundary between two distinct topological phases. Since the invariants are quantized and can only change at gap closings, a gap-closing point at the boundary is required, resulting in boundary states. For 1D systems with inversion symmetry, such topological boundary states are associated with a \emph{filling anomaly}~\cite{Rot_HOTI1, Rot_HOTI2, filling_anomaly}, which we describe now.

Consider a finite tiling of $N$ inversion-symmetric 1D unit cells which creates two halves or ``sectors" in real space, related by inversion-symmetry, with two boundaries as shown in Fig.~\ref{fig:1DPhC_unitcell}(d). A single isolated band in the bulk gives rise to $N$ states in this finite system. For a trivial bulk band with $[X_1] = 0$, the Wannier centers in the finite tiling must be placed at the $1a$ position of the unit cell, and the number of states that correspond to this bulk band is equal to $N$. However, for a topological bulk band with $[X_1] = \pm 1$, the Wannier centers in the finite tiling must be placed at the $1b$ position of the unit cell, which leads to a difficulty: $N$ states cannot maintain inversion symmetry due to the shifted position of the Wannier centers. Instead, either $N-1$ or $N+1$ (or more generally, $N-\overline{1}_2$ where $\overline{1}_2$ is any integer congruent to $1 \hskip 3 pt \mathrm{mod} \hskip 3 pt 2)$ Wannier centers are necessary to be consistent with inversion symmetry as shown in Fig.~\ref{fig:1DPhC_unitcell}(d). This inability to maintain both the expected number of states and inversion symmetry simultaneously is know as filling anomaly~\cite{Rot_HOTI1}, and leads to the quantization of fractional charge at boundaries in electronic systems and fractional electromagnetic energy density in PhCs. 

Since each Wannier center corresponds to a single state, the filling anomaly also presents a practical way to diagnose non-trivial topology by counting states in the spectrum of a finite system~\cite{Counting_mismatch, Quadrupole1}. The spectral consequence of the filling anomaly is that the states in the finite system within the frequency range of a single topological bulk band must have an odd ($\overline{1}_2$) number of missing or additional states as compared to the number of unit cells. 

The missing states are paired up with missing states from a different topological band in a way that preserves the inversion symmetry of the system and these typically reside inside the bandgap as boundary states. However, due to a lack of additional symmetries that pin these boundary states to the middle of the gap, they could be pushed into a bulk band by inversion-symmetry preserving perturbations to the boundaries. Since such perturbations act identically on both boundaries of the system, the bulk band would gain an odd number of additional states. 

Crucially, regardless of the details of the perturbation, the number of expected states and the actual states within the frequency range of a single topological band will differ by $\overline{1}_2$; we refer to this as a ``counting mismatch". In contrast, trivial boundary states, such as defect states, originate from a single band and would give rise to a counting mismatch of an even (= $\overline{0}_2$) number of states for that band in a finite system with two boundaries related by inversion symmetry. Therefore, the counting mismatch is a $\mathbb{Z}_2$ invariant that can be determined from the frequency spectrum of the PhC and thus can directly reveal the topological nature of bulk bands. We provide a more detailed discussion of the origin of this counting mismatch in Appendix A.

To summarize this argument, in the absence of chiral or particle-hole symmetry, the bulk-boundary correspondence of topological 1D PhCs with inversion symmetry is subtle in that the boundary states \emph{may} or \emph{may not} appear within a bandgap. However, regardless of their location in the frequency spectrum, the states within the frequency range of a topological band in a finite system \emph{must} exhibit an odd-integer counting mismatch. 

We now consider an explicit example of a 1D PhC consisting of alternating layers of $\mathrm{TiO_2}$ $(\varepsilon = 6.25)$ and air $(\varepsilon = 1)$. The $\mathrm{TiO_2}$ layer occupies a filling fraction $d/a = 0.6$ in the unit cell with lattice constant $a$. The first six bands of this 1D PhC are shown in Fig.~\ref{fig:1D_example}(a). Two distinct types of inversion-symmetric unit cells are possible for this PhC, as shown in the inset of Fig.~\ref{fig:1D_example}(a). The two types of unit cells are re-definitions of each other, related by a shift of $a/2$ along the $x$ direction. The eigenvalues of $\mathcal{I}$ at the HSPs $\mathbf{\Gamma}$ and $\mathbf{X}$ for both types of unit cells, as well as the Berry phase calculated using Eq.~\eqref{eq:BerryPhase}, are shown in the same plot. They show that while the band structure is identical for the two types of unit cells, the Berry phases and, correspondingly, the symmetry-indicator invariants are different. This is consistent with the fact that the re-definition of the unit cell shifts the Wannier center from the $1a$ position to the $1b$ position and, therefore, also the Berry phase from $0$ to $\pi$. This implies that if a band in one of the unit cell types is trivial, the corresponding band in the other type is topological.

Next, we simulate a large inversion-symmetric supercell with interfaces between the two types of unit cells in a strip geometry as shown in Fig.~\ref{fig:1D_example}(b). This supercell has two inversion-symmetry-related sectors with two boundaries and consists of a total of 61 unit cells. Therefore we expect to find 61 states per band in the spectrum of this supercell which is shown in Fig.~\ref{fig:1D_example}(c). However, due to the distinct topology of the bands in the two unit cell types, each band in the finite structure exhibits a counting mismatch of $\overline{1}_2$ states. For bands 1 to 4, we find the counting mismatch to be one missing state each and that these mismatched states reside in the bandgaps as boundary states whose field profiles are shown in Fig.~\ref{fig:1D_example}(d). For band 5, we find a counting mismatch of three missing states, two of which reside in the fourth gap and are trivial states since they originate from the same band. The remaining missing state is paired with another state from band 6. However, we can see that this pair of mismatched states does not lie deep inside the fifth bandgap but is instead very close to the band-edge of band 6. Including these states as part of band 6, we find a counting mismatch of one additional state for band 6.

The in-gap topological boundary states discussed above have been directly observed in experiments in 1D PhCs and 1D periodic-dielectric waveguides~\cite{1DPhC1, 1DPhC2, 1DPhC3}. 

Having introduced the notion of topological bands in the presence of crystalline symmetries in 1D, we now extend the topological classification and characterization of photonic bands to 2D.

\section{2D Photonic crystals}
Two-dimensional PhCs consist of a periodic patterning of the dielectric along two directions (e.g., $x$ and $y$) and a uniform dielectric profile along the third direction (e.g., $z$), with wave propagation restricted to lie in the $x,y$ plane. In this setting, the equations in (\ref{eq:MasterEq}) can be simplified by exploiting the mirror symmetry through the $x,y$ plane that sends $z \rightarrow -z$. This separates the states into two orthogonal polarizations: transverse electric (TE) with $\mathbf{E}(\mathbf{r}) = \mathcal{E}_x(x,y)\mathbf{\hat{x}} + \mathcal{E}_y(x,y)\mathbf{\hat{y}}$, $\mathbf{H}(\mathbf{r}) = \mathcal{H}_z(x,y)\mathbf{\hat{z}}$, which is even under the mirror symmetry, and transverse magnetic (TM) with $\mathbf{E}(\mathbf{r}) = \mathcal{E}_z(x,y)\mathbf{\hat{z}}$, $\mathbf{H}(\mathbf{r}) = \mathcal{H}_x(x,y)\mathbf{\hat{x}} + \mathcal{H}_y(x,y)\mathbf{\hat{y}}$, which is odd under the mirror symmetry. For these generally non-degenerate TE and TM polarizations, the eigenvalue problem is most easily solved for the scalar fields $\mathcal{H}_z(x,y)$ and $\mathcal{E}_z(x,y)$ respectively, via~\cite{photoniccrystalsbook2}
\begin{align}
    -\left[ \partial_x\frac{1}{\varepsilon(x,y)}\partial_x + \partial_y\frac{1}{\varepsilon(x,y)}\partial_y \right] \mathcal{H}_z(x, y) &= \frac{\omega^2}{c^2}\mathcal{H}_z(x, y), \nonumber \\
    -\frac{1}{\varepsilon(x,y)}\left( \partial_x^2 + \partial_y^2 \right) \mathcal{E}_z(x, y) &= \frac{\omega^2}{c^2}\mathcal{E}_z(x, y).
\end{align}

As in the 1D case, these eigenvalue problems can be solved using Bloch's theorem, and the solutions are distributed into frequency bands with their momenta restricted to the 2D BZ. Since TE and TM polarizations are orthogonal, we restrict the discussion to a single polarization of choice. We now characterize the topological phases of 2D PhCs by first constructing the topological invariants that classify them under different point group symmetries and then deriving bulk-boundary correspondences and their associated index theorems. 

The classification of PhCs can be divided into whether they obey time-reversal symmetry (TRS) (class AI) or not (class A). In 2D, without additional symmetries, class AI does not host topological phases. In contrast, class A hosts topological phases characterized by the Chern number ($C \in \mathbb{Z}$) that encodes the number of chiral edge states at the boundaries of a finite system. The Chern number also presents an obstruction to the construction of exponentially localized Wannier functions, and hence such bands are referred to as non-Wannierizable~\cite{Vanderbiltbook, Chern_Wannier_breakdown}. 

\begin{figure}[]%
\centering
\includegraphics[width=0.7\columnwidth]{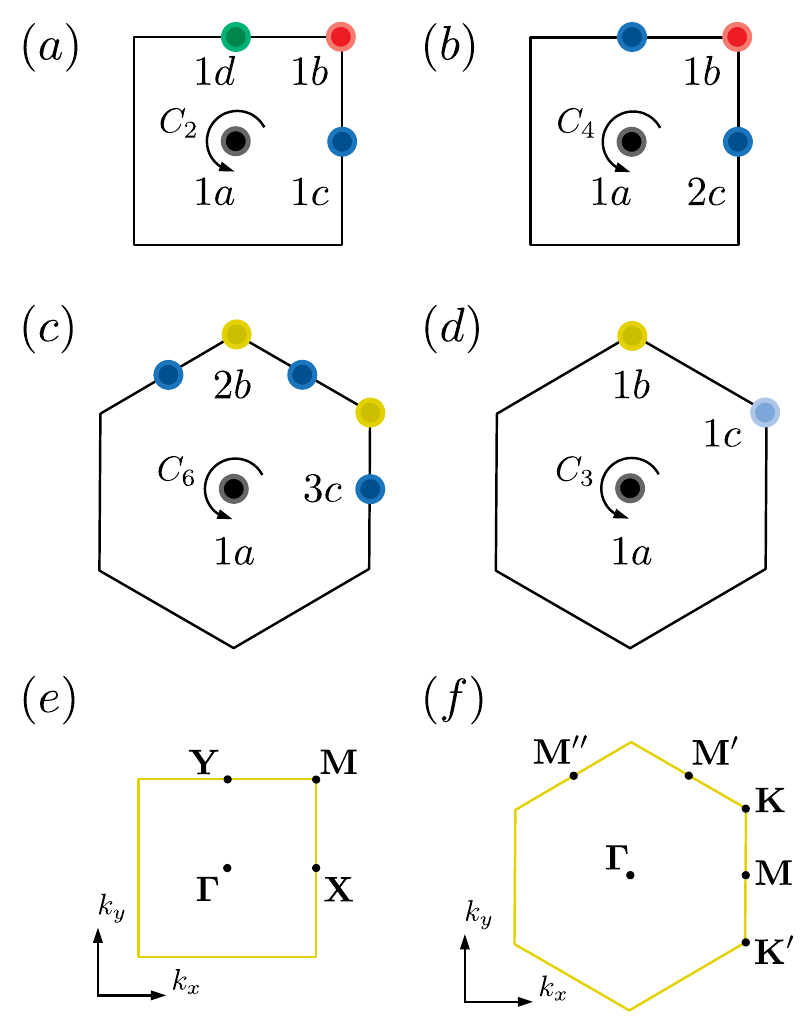}
\caption{Maximal Wyckoff positions for (a) $C_2$ (b) $C_4$ (c) $C_6$ and (d) $C_3$ symmetric unit cells. (e) BZ of a square lattice with possible HSPs. (f) BZ of a triangular lattice with possible HSPs.}
\label{fig:2DWP}
\end{figure}

When the Chern number vanishes, and in the presence of crystalline symmetries, photonic bands may be associated with Wannier centers fixed at maximal Wyckoff positions of the 2D unit cells (Fig.~\ref{fig:2DWP}). As mentioned previously, such bands are collectively called atomic limits; in particular, we use the term `obstructed atomic limits (OAL)' to refer to cases where the Wannier centers are displaced away from the center of the unit cell. Under some circumstances, a symmetry-preserving Wannier representation of bands may not be possible despite their vanishing Chern number. Such bands are termed \emph{fragile} and have the property of admitting a Wannier representation when considered as a set that includes additional specific atomic limit bands~\cite{Fragile1, Fragile2, FragilePhC}.

Similar to 1D, the topology of bands in 2D PhCs can be characterized using Berry phases. However, when bands are degenerate, they must be treated collectively, which requires the use of Wilson loops~\cite{Tutorial_WL1, Tutorial_WL2}. The Wilson loop is defined as
\begin{align}
\mathcal{W}(\mathcal{C})=\mathcal{P}\exp\left[\left(\ii \int_{\mathcal{C}} \boldsymbol{\A}(\mathbf{k})\cdot {\rm d}\mathbf{k} \right)\right],
\label{eq:Wilsonloop}
\end{align}
where $\mathcal{C}$ is a closed contour in $\mathbf{k}$-space, $\mathcal{P}$ denotes a path ordering of the exponential and $\boldsymbol{\A}(\mathbf{k})$ is the multi-band Berry connection given by
\begin{align}
  \boldsymbol{\A}(\mathbf{k}) \equiv \boldsymbol{\A}_{m,n}(\mathbf{k}) = -\ii\matrixel{u_{\mathbf{k},m}}{\nabla_\mathbf{k}}{u_{\mathbf{k},n}}.
 \label{eq:berry_connection}
\end{align}
Here, $m, n$ label the bands in a group of connected bands. When $\mathcal{C}$ is taken to be a non-contractible loop in the Brillouin zone, the Wilson loop eigenvalues are proportional to the expectation values of the position operator of the hybrid Wannier functions in the same direction. Therefore, these eigenvalues can indicate the Wannierizable nature of bands in atomic limit phases or indicate the non-Wannierizable nature of fragile bands or Chern bands by their non-trivial winding numbers~\cite{Vanderbiltbook, MLWF}. Similar to the Berry phase in 1D, the calculations of these Wilson loops can also be simplified by looking at the representations of the eigenmodes at the HSPs of the BZ.

\subsection{Classification due to rotational symmetries}

Consider a projector into the bands of interest given by $P_k = \sum_j \ket{u_{j,k}} \bra{u_{j,k}}$. The eigenvalues of the rotation operator, $\hat{r}_n$, projected into the bands of interest at the HSP $\vkk$, $P_\vkk\hat{r}_n P_\vkk$, are 
\begin{align}
\vkk^{(n)}_p = e^{2\pi \ii (p-1)/n}, \quad \text{for } p=1,2,\ldots n.
\end{align}
Following previous studies on the characterization of the topology of energy bands in condensed matter systems~\cite{Rot_HOTI1}, we define the integer invariants
\begin{align}
[\vk^{(n)}_p] \equiv \# \vkk^{(n)}_p-\#\mathbf{\Gamma}^{(n)}_p \quad \in \mathbb{Z},
\label{eq:RotationInvariants}
\end{align}
where $\# \vkk^{(n)}_p$ is the number of states in the frequency band(s) in question with rotation operator eigenvalue $\vkk^{(n)}_p$.

These invariants can be constructed for 2D lattices with $C_n$ symmetry at all high symmetry points shown in Fig.~\ref{fig:2DWP}(e) and (f). However, some of the invariants in Eq.~\eqref{eq:RotationInvariants} are redundant for three reasons: (i) Rotation symmetry  forces representations at certain HSPs to be the same. Particularly, $C_4$ symmetry forces equal representations at ${\bf X}$ and ${\bf Y}$, while $C_6$ symmetry forces equal representations at ${\bf M}$, ${\bf M'}$, and ${\bf M''}$, as well as at ${\bf K}$ and ${\bf K'}$; (ii) the fact that the number of bands in consideration is constant across the BZ, from which it follows that  $\sum_p \# \mathbf{\Pi}^{(n)}_p = \sum_p \# \mathbf{\Gamma}^{(n)}_p$, or $\sum_p [\Pi^{(n)}_p]=0$; and (iii) the existence of TRS, which implies that the Chern number vanishes and that rotation eigenvalues at ${\bf \Pi}^{(n)}$ and $-{\bf \Pi}^{(n)}$ are related by complex conjugation. This leads to $[M^{(4)}_2] = [M^{(4)}_4]$ $(\text{for } C_4)$, $[K^{(3)}_2]=[K'^{(3)}_3]$ $(\text{for } C_3)$, $[K^{(3)}_3]=[K'^{(3)}_2]$ $(\text{for } C_3)$, $[K^{(3)}_1]=[K'^{(3)}_1]$ $(\text{for } C_3)$ and $[K^{(3)}_2]=[K^{(3)}_3]$ $(\text{for } C_6)$.

Therefore, in the presence of TRS (class AI), the classification for $N$ bands is given by the indices~\cite{Rot_HOTI1}
\begin{align}
\chi_\mathcal{T}^{(2)}&= \left( [X^{(2)}_1],[Y^{(2)}_1],[M^{(2)}_1]; N \right), \nonumber\\
\chi_\mathcal{T}^{(3)}&= \left( [K^{(3)}_1], [K^{(3)}_2]; N \right), \nonumber\\
\chi_\mathcal{T}^{(4)}&= \left( [X^{(2)}_1],[M^{(4)}_1],[M^{(4)}_2]; N \right), \nonumber\\
\chi_\mathcal{T}^{(6)}&= \left( [M^{(2)}_1],[K^{(3)}_1]; N \right).
\label{eq:ClassificationIndicesTRS}
\end{align}
On breaking TRS, the classification of 2D $C_n$-symmetric PhCs must include the Chern number since it can now admit non-zero values. Furthermore, breaking TRS reduces the number of constraints on the invariants (i.e., condition (iii) above is relaxed) and therefore increases the number of invariants required to identify distinct topological phases uniquely. Taking into account these considerations, the most general classification (class A) of 2D $C_n$-symmetric PhCs is given by the indices
\begin{align}
\chi^{(2)}&=\left( C\, \Big| \, \rho^{(2)} \right) = \left( C\, \Big| \,[X^{(2)}_1],[Y^{(2)}_1],[M^{(2)}_1]; N\right), \nonumber\\
\chi^{(3)}&=\left( C\, \Big| \, \rho^{(3)} \right)=\left( C\, \Big| \,[K^{(3)}_1], [K^{(3)}_2],[K'^{(3)}_1], [K'^{(3)}_2]; N \right), \nonumber\\
\chi^{(4)}&=\left( C\, \Big| \, \rho^{(4)} \right)=\left( C\, \Big| \, [X^{(2)}_1],[M^{(4)}_1],[M^{(4)}_2],[M^{(4)}_4]; N \right), \nonumber\\
\chi^{(6)}&=\left( C\, \Big| \, \rho^{(6)} \right)=\left( C\, \Big| \,[M^{(2)}_1],[K^{(3)}_1],[K^{(3)}_2]; N \right),
\label{eq:ClassificationIndicesGeneral}
\end{align}
where $C$ is the Chern number given by
\begin{align}
C = \frac{1}{2\pi} \int_{\mathrm{BZ}}\tr[ \nabla_{\mathbf{k}}\times \boldsymbol{\mathcal{A}}(\mathbf{k})]\ {\rm d}^2{\bf k}.
 \label{eq:Chern}
\end{align}
Similar to the 1D case, we can exhaustively calculate the values of $\chi^{(n)}$ (in the case when $C=0$) or $\chi_\mathcal{T}^{(n)}$ by induction of band representations. To perform this, we require knowledge about the Wannier functions' internal symmetry representation, known as ``site symmetry representation", $\rho(C_n)$, as well as the location of their gauge-invariant centers, the Wannier centers. We provide a detailed derivation of the symmetry-indicator invariants at HSPs and the corresponding indices for all 2D atomic limits, with and without TRS, in Appendix D and show the final results in Tables \ref{tab:C2_Invariants_table}-\ref{tab:C6_Invariants}. Each case in these tables uniquely identifies an atomic limit protected by the corresponding rotational symmetry.

\subsection{Relation between symmetry-indicator invariants and Chern number}

The Chern number mod $n$ can be inferred from the rotation eigenvalues at HSPs of systems with $C_n$ symmetry~\cite{fang_Chern_from_sym}. Using this,
relations between the Chern numbers, Eq.~\eqref{eq:Chern}, and the symmetry indicator invariants can be derived, as done in Appendix B.
These relations take the form of equivalence relations modulo the rotation order of the considered group:
\begin{align}
    C^{(2)} &= -[X_1^{(2)}]-[Y_1^{(2)}]-[M_1^{(2)}] \pmod 2, \nonumber\\
    C^{(3)} &= -[K_1^{(3)}]-2[K_2^{(3)}]+2[K_1^{\prime (3)}]+[K_2^{ \prime (3)}] \pmod 3, \nonumber \\
    C^{(4)} &= 2[M_1^{(4)}]+[M_2^{(4)}]-[M_4^{(4)}]-2[X_1^{(2)}] \pmod 4, \nonumber\\    
    C^{(6)} &= -8[K_1^{(3)}]-4[K_2^{(3)}]+3[M_1^{(2)}] \pmod 6, \nonumber\\
    \label{eq:ChernSymmInd}
\end{align}
Compared to the direct evaluation of Eq.~\eqref{eq:Chern}, these relations provide a fast and simple way to calculate the Chern number mod~$n$ for $C_n$-symmetric PhCs with broken TRS.

\subsection{Index theorems}

\begin{table}[!htb]
	\centering
		\centering
	\begin{tabular}{cccc}
		\toprule 
		WP & Site symm. & $\chi_\mathcal{T}^{(2)}$ & $\chi^{(2)}$\\
		\midrule
		$1a$ & $\rho(C_2)=\text{any}$ &  $(0,0,0; 1)$ & $(0\, | \,0,0,0; 1)$\\
		\separatorrule	
		$1c$ & $\rho(C_2)=+1$ & $(-1,0,-1; 1)$ & $(0\, | \,-1,0,-1; 1)$\\
		$1c$ & $\rho(C_2)=-1$  & $(1,0,1; 1)$ & $(0\, | \,1,0,1; 1)$\\
		\separatorrule
		$1d$ & $\rho(C_2)=+1$ & $(0,-1,-1; 1)$ & $(0\, | \,0,-1,-1; 1)$\\
		$1d$ & $\rho(C_2)=-1$ & $(0,1,1; 1)$ & $(0\, | \,0,1,1; 1)$\\
		\separatorrule
		$1b$ & $\rho(C_2)=+1$ & $(-1,-1,0; 1)$ & $(0\, | \,-1,-1,0; 1)$\\
		$1b$ & $\rho(C_2)=-1$ & $(1,1,0; 1)$ & $(0\, | \,1,1,0; 1)$\\
		\bottomrule
	\end{tabular} 
	\caption{$C_2$ symmetry:  Indices induced from every maximal Wyckoff position (WP).}
	\label{tab:C2_Invariants_table}
 \end{table}
 \begin{table}[!htb]
	\begin{tabular}{cccc}
		\toprule 
		WP & Site symm. & $\chi_\mathcal{T}^{(3)}$ & $\chi^{(3)}$\\
		\midrule
		$1a$ & $\rho(C_3)=\text{any}$ &  $(0,0; 1)$ & $(0\, | \,0,0,0,0; 1)$\\
		\separatorrule
		$1b$ & $\rho(C_3)=+1$ &  $(-1,1; 1)$ & $(0\, | \,-1,1,-1,0; 1)$ \\
		$1b$ & $\rho(C_3)=e^{\ii\frac{2\pi}{3}\sigma_z}$ &  $(1,-1; 2)$ & $(0\, | \,1,-1,1,0; 2)$\\
		$1b$ & $\rho(C_3)=e^{\ii\frac{2\pi}{3}}$ &  $-$ & $(0\, | \,0,-1,1,-1; 1)$\\
		$1b$ & $\rho(C_3)=e^{\ii\frac{4\pi}{3}}$ &  $-$ & $(0\, | \,1,0,0,1; 1)$\\	
		\separatorrule
		$1c$ & $\rho(C_3)=+1$ &  $(-1,0; 1)$ & $(0\, | \,-1,0,-1,1; 1)$\\
		$1c$ & $\rho(C_3)=e^{\ii\frac{2\pi}{3}\sigma_z}$ &  $(1,0; 2)$ & $(0\, | \,1,0,1,-1; 2)$\\
		$1c$ & $\rho(C_3)=e^{\ii\frac{2\pi}{3}}$ &  $-$ & $(0\, | \,1,-1,0,-1; 1)$\\
		$1c$ & $\rho(C_3)=e^{\ii\frac{4\pi}{3}}$ &  $-$ & $(0\, | \,0,1,1,0; 1)$\\
		\bottomrule
	\end{tabular} 
	\caption{$C_3$ symmetry:  Indices induced from every maximal Wyckoff position.}
	\label{tab:OtherInvariants2D}
  \end{table}
  \begin{table}[!htb]
	\begin{tabular}{cccc}
		\toprule 
		WP & Site symm. & $\chi_\mathcal{T}^{(4)}$ & $\chi^{(4)}$\\
		\midrule
		$1a$ & $\rho(C_4)=\text{any}$ &  $(0,0,0; 1)$ & $(0 \, | \, 0,0,0,0; 1)$\\
		\separatorrule
		$2c$ & $\rho(C_2)=+1$ &  $(-1,-1,1; 2)$ & $(0\, | \,-1,-1,1,1; 2)$\\
		$2c$ & $\rho(C_2)=-1$ &  $(1,1,-1; 2)$ & $(0\, | \,1,1,-1,-; 2)$\\
		\separatorrule
		$1b$ & $\rho(C_4)=+1$ &  $(-1,-1,0; 1)$ & $(0\, | \,-1,-1,0,0; 1)$\\
		$1b$ & $\rho(C_4)=-1$ &  $(-1,1,0; 1)$ & $(0\, | \,-1,1,0,0; 1)$\\
		$1b$ & $\rho(C_4)=\ii\sigma_z$ &  $(2,0,0; 2)$ & $(0\, | \,2,0,0,0; 2)$\\
		$1b$ & $\rho(C_4) = +\ii$ & $-$ & $(0\, | \,1,0,-1,1; 1)$\\
		$1b$ & $\rho(C_4) = -\ii$ & $-$ & $(0\, | \,1,0,1,-1; 1)$\\
		\bottomrule
	\end{tabular} 
	\caption{$C_4$ symmetry: Indices induced from every maximal Wyckoff position.}
	\label{tab:C4_Invariants}
 \end{table}
 \begin{table}[!htb]
	\begin{tabular}{cccc}
		\toprule 
		WP & Site symm. & $\chi_\mathcal{T}^{(6)}$ & $\chi^{(6)}$\\
		\midrule
		$1a$ & $\rho(C_6)=\text{any}$ &  $(0,0; 1)$ & $(0\, | \,0,0,0; 1)$\\
		\separatorrule
		$2b$ & $\rho(C_3)=+1$ &  $(0,-2; 2)$ &  $(0\, | \,0,-2,1; 2)$\\
		$2b$ & $\rho(C_3)=e^{\ii\frac{2\pi}{3}\sigma_z}$ &  $(0,2; 4)$ & $(0\, | \,0,2,-1; 4)$ \\
		$2b$ & $\rho(C_3)=e^{\ii\frac{2\pi}{3}}$ &  $-$ & $(0\, | \,0,1,-2; 2)$ \\
		$2b$ & $\rho(C_3)=e^{\ii\frac{4\pi}{3}}$ &  $-$ & $(0\, | \,0,1,1; 2)$ \\
		\separatorrule
		$3c$ & $\rho(C_2)=+1$ &  $(-2,0; 3)$ & $(0\, | \,-2,0,0; 3)$\\
		$3c$ & $\rho(C_2)=-1$ &  $(2,0; 3)$ & $(0\, | \,2,0,0; 3)$\\
		\bottomrule
	\end{tabular} 
	\caption{$C_6$ symmetry:  Indices induced from every maximal Wyckoff position.}
	\label{tab:C6_Invariants}
\end{table}

$C_n$-symmetric PhCs with different $\chi^{(n)}$ or $\chi_\mathcal{T}^{(n)}$ belong to different topological phases, as they cannot be deformed into one another without closing the bulk energy gap or breaking the symmetry\footnote{These are \emph{weak} invariants that depend on the particular choice of a $C_n$-symmetric unit cell. However, once such a choice is made, these invariants can only change discretely at gap-closing points when the system undergoes a symmetry-preserving adiabatic deformation.}~\cite{teo2010,teo2013,benalcazar2014}. Furthermore, for Wannierizable bands, the Wannier center configuration directly determines the existence of a filling anomaly and consequently the possible existence of in-gap edge and corner states. Therefore, finding the symmetry-indicator invariants is useful in establishing a bulk-boundary correspondence for such bands. The presence of edge states is directly related to the dipole moment of the Wannier centers. In 1D, this takes the form of Eq.~\eqref{eq:QuantizedPolarization1D} whereas in 2D, Ref.~\cite{Rot_HOTI1} showed that the bands have dipole moments indicated by
\begin{align}
{\bf P}^{(2)}&=\frac{1}{2}\left( [Y^{(2)}_1]+[M^{(2)}_1] \right) {\bf a}_1+\frac{1}{2}\left( [X^{(2)}_1]+[M^{(2)}_1] \right) {\bf a}_2, \nonumber\\
{\bf P}^{(4)}&= \frac{1}{2}[X_1^{(2)}] ({\bf a}_1+{\bf a}_2), \nonumber\\
{\bf P}^{(6)}&={\bf 0},
\label{eq:WeakInvariantsFromRotationInvariants}
\end{align}
where the superscript $n$ in ${\bf P}^{(n)}$ labels the $C_n$ symmetry. The dipole moments in Eq.~\eqref{eq:WeakInvariantsFromRotationInvariants} are defined modulo 1 and are valid for both TR-symmetric and TR-broken PhCs, as long as the Chern number vanishes in the latter case. ${\bf P}^{(2)}$ is a $\mathbb{Z}_2 \times \mathbb{Z}_2$ index and ${\bf P}^{(4)}$ is a $\mathbb{Z}_2$ index. In the case of $C_3$ symmetry, the dipole moment is given by
\begin{align}
 {\bf P}^{(3)}&=\frac{2}{3}\left( [K^{(3)}_1]+2[K^{(3)}_2] \right)({\bf a}_1+{\bf a}_2) \quad (\text{under TRS}), \nonumber\\
{\bf P}^{(3)}&=\left([K^{(3)}_1]+[K^{(3)}_2]-\frac{2}{3}[K'^{(3)}_1]-\frac{1}{3}[K'^{(3)}_2]\right)({\bf a}_1+{\bf a}_2)\nonumber\\& (\text{under broken TRS}),
\end{align}
where ${\bf P}^{(3)}$ is a $\mathbb{Z}_3$ index for $C_3$ symmetry. 

In all cases, non-trivial ${\bf P}$ is associated with an edge-induced filling anomaly. For 2D spinless systems, such as the PhCs considered here, $\mathcal{I}$ and $C_2$ have identical transformation properties and are isomorphic operations that send $x,y \rightarrow -x, -y$. Therefore, for $C_2$, $C_4$, and $C_6$ symmetries, a non-trivial ${\bf P}$ is associated with a counting mismatch of $\overline{1}_2$ in the edge spectrum since inversion symmetry ($\mathcal{I}$) is a subgroup of these rotations, and an edge supercell (with one periodic direction) can always be chosen such that $\mathcal{I}$ is maintained. In the case of $C_2$ symmetry, the counting mismatch is a $\mathbb{Z}_2 \times \mathbb{Z}_2$ invariant as edge supercells in both directions must be independently considered (i.e., finite-in-$x$, periodic-in-$y$ or finite-in-$y$, periodic-in-$x$). In the case of $C_4$ symmetry, the edge spectrum is identical in both directions, and therefore the counting mismatch is a $\mathbb{Z}_2$ invariant. In the case of $C_6$ symmetry, both ${\bf P}^{(6)}$ and the counting mismatch in the edge spectrum are always trivial. Since $\mathcal{I}$ is not a subgroup of $C_3$ symmetry, an edge supercell can never be chosen such that $\mathcal{I}$ is maintained. Therefore, the counting mismatch cannot distinguish between different values of ${\bf P}^{(3)}$. Instead, in this case, the fractionalization of energy density at the edges must be directly calculated using the eigenmodes of a $C_3$-symmetric finite system.

Additionally, some Wannier center configurations can lead to higher-order topological states. In class AI, these phases are determined by the corner ``charges'' \footnote{In PhCs, the electromagnetic energy density is analogous to electronic charge density fractionally quantized at corners.}
\begin{align}
Q^{(2)}_{\text{corner}, \mathcal{T}}&=\frac{1}{4}\left( -[X^{(2)}_1]-[Y^{(2)}_1]+[M^{(2)}_1] \right), \nonumber\\
Q^{(3)}_{\text{corner}, \mathcal{T}}&=\frac{1}{3}[K^{(3)}_2], \nonumber\\
Q^{(4)}_{\text{corner}, \mathcal{T}}&=\frac{1}{4}\left( [X_1^{(2)}]+2[M_1^{(4)}]+3[M_2^{(4)}] \right), \nonumber\\
Q^{(6)}_{\text{corner}, \mathcal{T}}&=\frac{1}{4}[M^{(2)}_1]+\frac{1}{6}[K^{(3)}_1],
\label{eq:CornerIndices}
\end{align}
as shown initially in Ref.~\cite{Rot_HOTI1}. We extend this to class A, where they are
\begin{align}
Q^{(2)}_\text{corner}&=\frac{1}{4}\left( -[X^{(2)}_1]-[Y^{(2)}_1]+[M^{(2)}_1] \right), \nonumber\\
Q^{(3)}_\text{corner}&=\frac{1}{3}\left( [K^{(3)}_1] + [K^{(3)}_2] - [K'^{(3)}_1] \right), \nonumber\\
Q^{(4)}_\text{corner}&=\frac{1}{4}\left( [X_1^{(2)}]+2[M_1^{(4)}]+\frac{3}{2}[M_2^{(4)}]+\frac{3}{2}[M_4^{(4)}] \right), \nonumber\\
Q^{(6)}_\text{corner}&=\frac{1}{4}[M^{(2)}_1]+\frac{2}{3}[K^{(3)}_1],
\label{eq:CornerIndices_TRS_broken}
\end{align}
$Q^{(n)}_{\text{corner}, \mathcal{T}}$ (for TR-symmetric) or $Q^{(n)}_\text{corner}$ (for TR-broken), are $\mathbb{Z}_n$ topological quantities and are associated with a corner-induced filling anomaly, a counting mismatch of states $\in \{ {\overline{0}_n}, \dots {\overline{n-1}_n} \}$ in a finite system with $n$ symmetry-related sectors and possibly the presence of in-gap corner-localized states. The derivation of these formulae and other details concerning the finite systems where these formulae are valid are given in Appendix E. 

For the formulae in Eq.~\eqref{eq:CornerIndices_TRS_broken}, we have assumed that the Chern number vanishes in the TR-broken case and that the bands are OALs with well-defined Wannier centers. However, it is also possible for fractional charges to localize at disclinations in $C_n$-symmetric systems with non-Wannierizable Chern bands. In such cases, the formulae for disclination charges contain a Chern number contribution and contributions from the symmetry-indicator invariants~\cite{Disclination_charges_Cn}. We note that the formulae in Eq.~\eqref{eq:CornerIndices_TRS_broken} are consistent with the disclination charges given in Ref.~\cite{Disclination_charges_Cn} with a vanishing Chern number contribution as is expected.

Finally, we note that in fermionic systems, where insulating states rely on completely filled bands, a quantization of corner charge requires ${\bf P}^{(n)}={\bf 0}$. In photonic systems, however, we are only concerned with the \emph{existence} of localized states, and the ${\bf P}^{(n)}={\bf 0}$ constraint can be relaxed. Therefore, we also consider cases where ${\bf P}^{(n)}$ and $Q$ can simultaneously admit non-trivial values, leading to both edge and corner states that may be degenerate with each other and/or with the bulk bands. However, their associated counting mismatch remains robust.

\section{Design and Characterization of 2D Topological Photonic Crystals}

In the previous sections, we exhaustively built the topological classifications in class A and AI. We also identified the indices that correspond to OAL phases via the induction of the band representations from the symmetry representation of the Wannier functions and the Wyckoff positions of their Wannier centers. This classification forms a linear algebraic structure, such that when two sets of bands of a $C_n$-symmetric system, in phases $\chi_1$ and $\chi_2$ respectively, are combined, they are in phase $\chi_1+\chi_2$. This observation forms the basis of a strategy we now propose to diagnose and design topological PhCs.

Given a PhC, our starting point is the calculation of the $C_n$ symmetry representations at HSPs for $N$ bands to determine $\rho^{(n)}$ (here, $\rho^{(n)} = \chi^{(n)}_{\mathcal{T}}$ for TR-symmetric systems). $\rho^{(n)}$ can always be expressed as the following linear combination
\begin{align}
\rho^{(n)}=\sum_{p} \alpha_p \hskip 2 pt \rho^{(n)}_{p},
\label{eq:chi_linear_combination}
\end{align}
where $\rho^{(n)}_{p}$ correspond to the indices of atomic limits in Tables~\ref{tab:C2_Invariants_table}--\ref{tab:C6_Invariants}. Since the $\rho^{(n)}_{p}$ for different site symmetry representations for the same Wannier center configuration are linearly dependent, the linear combination in Eq.~\eqref{eq:chi_linear_combination} is non-unique, and all possible linear combinations must be examined to obtain the correct topological characterization.

The topology of this set of $N$ bands can then be determined by the following set of rules \cite{Fragile1, song2018diagnosis}:
(i)~If the bands are in an OAL phase, there exists a linear combination such that the coefficients $\{\alpha_p\}$ are all positive integers (the converse is not true).
(ii)~If a linear combination with positive integer $\{\alpha_p\}$ is impossible and at least one negative integer coefficient is required, the bands are in a fragile topological phase.
(iii)~If a linear combination with integer $\{\alpha_p\}$ is not possible, the bands are either gapless under TRS, in which case we have a Dirac semi-metal phase, or are gapped and have a non-vanishing Chern number under broken TRS. 

In the following sections, we provide examples that illustrate these cases.

\subsection{Example 1: OAL phases with four-fold rotation in class AI}

\begin{figure*}[htb]%
\centering
\includegraphics[width=1\textwidth]{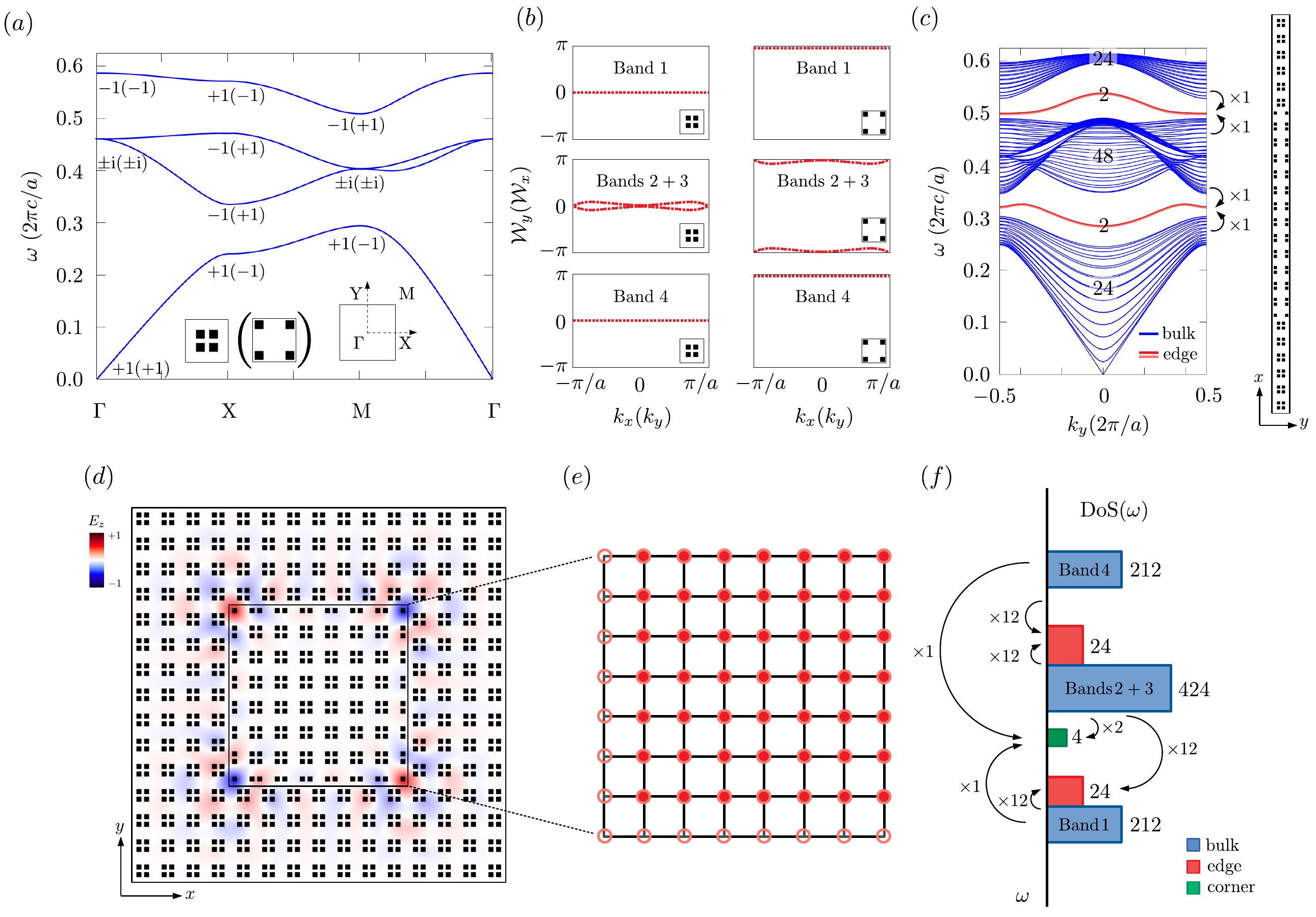}
\caption{(a) TM-polarized band structure of a $C_{4v}$ symmetric PhC with $\varepsilon = 12$. The two possible types of $C_4$-symmetric unit cells are shown in the insets along with the 2D BZ. $C_4$ eigenvalues at $\mathbf{\Gamma}$ and $\mathbf{M}$, $C_2$ eigenvalues at $\mathbf{X}$ are shown for the first four bands. (b) Wilson loop eigenvalues $\mathcal{W}_y (\mathcal{W}_x)$ for bands 1, $2+3$ and 4 along $k_x (k_y)$ for both types of unit cells. (c) Edge spectrum consisting of a total of 25 unit cells of the two types in a strip configuration (shown on the right). An odd-integer counting mismatch per band leads to the presence of edge states in the first and second TM bandgaps. (d) The dielectric and $E_z$ mode profile of one of the four corner modes in a finite system of size $15\times 15$ unit cells consisting of the two types of unit cells in a core-cladding configuration. (e) A tiling of the unit cells with Wannier centers (solid circles) for one band of the inner core of size $7\times 7$. Additional Wannier centers from other bands (hollow circles) are required to maintain $C_4$ symmetry. Counting the Wannier centers along the boundary sites, we see that 24 edge and four corner states are expected in this finite configuration. (f) A schematic of the DoS for the structure in (d). A counting mismatch of states for bands 1 to 4 leads to four degenerate corner states in the first TM bandgap. The counting mismatch for the edge states depends on the system size for such a finite configuration. }
\label{fig:2D_OAL}
\end{figure*}

We now show an example of an OAL phase and its associated boundary signatures in a 2D PhC. Similar OAL phases have been widely implemented in PhCs~\cite{HOTI1, HOTI2, HOTI3, HOTI4, HOTI5, HOTI6, HOTI_Laser1, HOTI_Laser2}. Consider two PhCs with unit cells shown in the inset of Fig.~\ref{fig:2D_OAL}(a), which consist of four dielectric square pillars in a $C_{4v}$-symmetric configuration with $\varepsilon = 12$. These two unit cell choices, referred to as ``expanded" and ``contracted", are related by a half-lattice-constant shift along the $x$ and $y$ directions. We will consider the first four TM bands for the following analysis. 

The symmetry-indicator invariants can be computed using the relevant rotation eigenvalues of the electromagnetic eigenmodes at the HSPs $\mathbf{\Gamma}$, $\mathbf{X}$ and $\mathbf{M}$ for both unit cell types; these are shown in Fig.~\ref{fig:2D_OAL}(a). For the contracted unit cell, bands 1 and 4 have the index $\chi_{\mathcal{T}}^{(4)} = (0,0,0; 1)$, and the pair of degenerate bands $2+3$ have the index $\chi_{\mathcal{T}}^{(4)} = (0,0,0; 2)$. Each of these indices corresponds to Wannier centers located at the $1a$ Wyckoff position in the 2D unit cell shown in Fig.~\ref{fig:2DWP}(b). For the expanded unit cell, bands 1 and 4 have the indices $\chi_{\mathcal{T}}^{(4)} = (-1,-1,0; 1)$ and $\chi_{\mathcal{T}}^{(4)} = (-1,+1,0; 1)$ respectively. Bands $2+3$ have the index $\chi_{\mathcal{T}}^{(4)} = (2,0,0; 2)$. Each of these indices corresponds to the Wannier centers at the $1b$ Wyckoff position. These indices lead to $\mathbf{P}^{(4)} = (1/2, 1/2)$ and $Q^{(4)}_{\text{corner}, \mathcal{T}} = 1/4$ for bands 1 and 4 and $\mathbf{P}^{(4)} = \mathbf{0}$ and $Q^{(4)}_{\text{corner}, \mathcal{T}} = 1/2$ for bands 2+3 .


The Wannierizable nature of these atomic limit bands can also be established by examining the Wilson loops as shown in Fig.~\ref{fig:2D_OAL}(b). Here, the Wilson loop eigenvalues for each band are calculated by integrating the Berry connection along one momentum direction and plotting it as a function of the other momentum. This indicates the locations of the hybrid Wannier centers that are exponentially localized in one spatial direction but delocalized in another. The observed shifts in the Wilson loop eigenvalues between the contracted and expanded unit cells are consistent with the real space shifts that relate the two unit cell types where the Wannier centers reside at the $1a$ and $1b$ positions, respectively.

To illustrate that the dipole moments of the bands lead to edge states, we simulate a finite system consisting of the expanded and contracted unit cells in a strip geometry. The strip geometry is a large supercell along one direction, consisting of an inner domain with the expanded unit cell and an outer domain with the contracted unit cell with periodic boundaries along both directions, as shown in Fig.~\ref{fig:2D_OAL}(c) . We consider a supercell of size $25\times 1$ unit cells, and therefore expect the spectrum to contain 25 states per band. However, due to the non-zero dipole moments, bands 1 and 4 have a counting mismatch of one missing state (= $\overline{1}_2$) each. In contrast, bands 2+3, which have a vanishing dipole moment, exhibit a counting mismatch of two missing states (= $\overline{0}_2$) as shown in the edge spectrum in Fig.~\ref{fig:2D_OAL}(c). 

The non-zero corner charges for bands 1 to 4 will similarly lead to a counting mismatch due to the presence of corner states in a finite system with corners. To illustrate this, we now examine a finite $C_4$-symmetric system in a core-cladding configuration as shown in Fig.~\ref{fig:2D_OAL}(d). This finite system has four symmetry-related sectors with four corners and has a size of $15\times 15=225$ unit cells. Therefore, each band contributes $225$ states to the spectrum of the finite system. However, the non-zero dipole moment of bands of the inner core region leads to edge states on all edges, as we have discussed previously and shown in Fig.~\ref{fig:2D_OAL}(c). In the finite system, these edge states have a size-dependent counting mismatch. If we consider a finite tiling of size $7\times 7$ unit cells that represent the inner core, each with a Wannier center at $1b$, we observe that additional Wannier centers from other bands are required to maintain $C_4$ symmetry, as shown in Fig.~\ref{fig:2D_OAL}(e). Counting the Wannier centers that live on the entire boundary between the core and cladding, we can predict the appearance of $24$ edge states and $4$ corner states. 

In Fig.~\ref{fig:2D_OAL}(f), we show a schematic of the calculated DoS of the full $15\times 15$ finite system, up to the frequency range of the first four TM bands and identify the number of bulk, edge, and corner states from their localization and mode profiles. The state counting in Fig.~\ref{fig:2D_OAL}(f) confirms the predicted $24$ edge states and $4$ corner states. The counting mismatch due to the corners is size-independent and is identified in Fig.~\ref{fig:2D_OAL}(f) as equal to one missing state (= $\overline{1}_4$) each for bands 1 and 4 and two missing states for bands 2+3 (= $\overline{2}_4$), accounting for the expected number of corner states and consistent with the corner charges of the bands. We point out that even if a $C_4$-preserving perturbation to the corners pushes the four corner states into any of the bulk bands, the counting mismatch remains. For example, if the four corner states were pushed into band 1, the counting mismatch for this band would go from one missing state to three additional states, both of which are equal modulo 4 $(\overline{-3}_4 = \overline{1}_4)$.

\begin{figure}[t!]%
\centering
\includegraphics[width=\columnwidth]{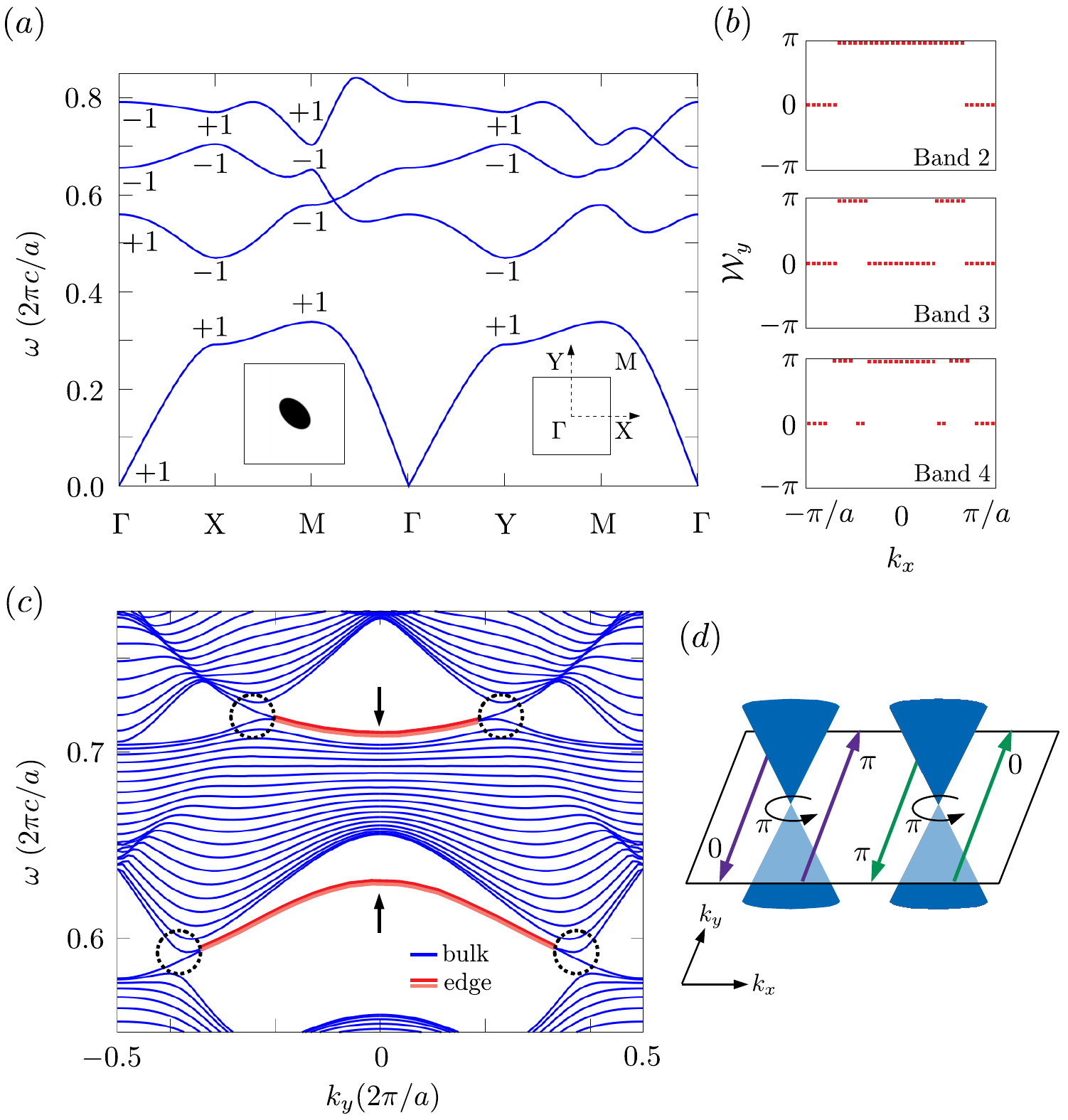}
\caption{(a) TM-polarized band structure of a $C_{2v}$-symmetric PhC whose unit cell is shown in the inset. $C_2$ eigenvalues at $\mathbf{\Gamma}$, $\mathbf{X}$, $\mathbf{Y}$ and $\mathbf{M}$ are shown for the first four bands. (b) Wilson loop eigenvalues $\mathcal{W}_y$ for the bands 2, 3, and 4 plotted as a function of $k_x$. The discontinuities indicate the presence of Dirac points. (c) Edge spectrum of this PhC showing edge states (marked with arrows) whose dispersion terminates at Dirac points (marked with circles) on the left (red) and right (light red) edges. (d) Dirac points are gap-closing points that separate 1D topological phases with different Berry phases. They can also be thought of as sources of $\pi$ Berry phase.}
\label{fig:2D_Dirac}
\end{figure}

\subsection{Example 2: Dirac semi-metal in class AI}

\begin{figure}[htb]%
\centering
\includegraphics[width=\columnwidth]{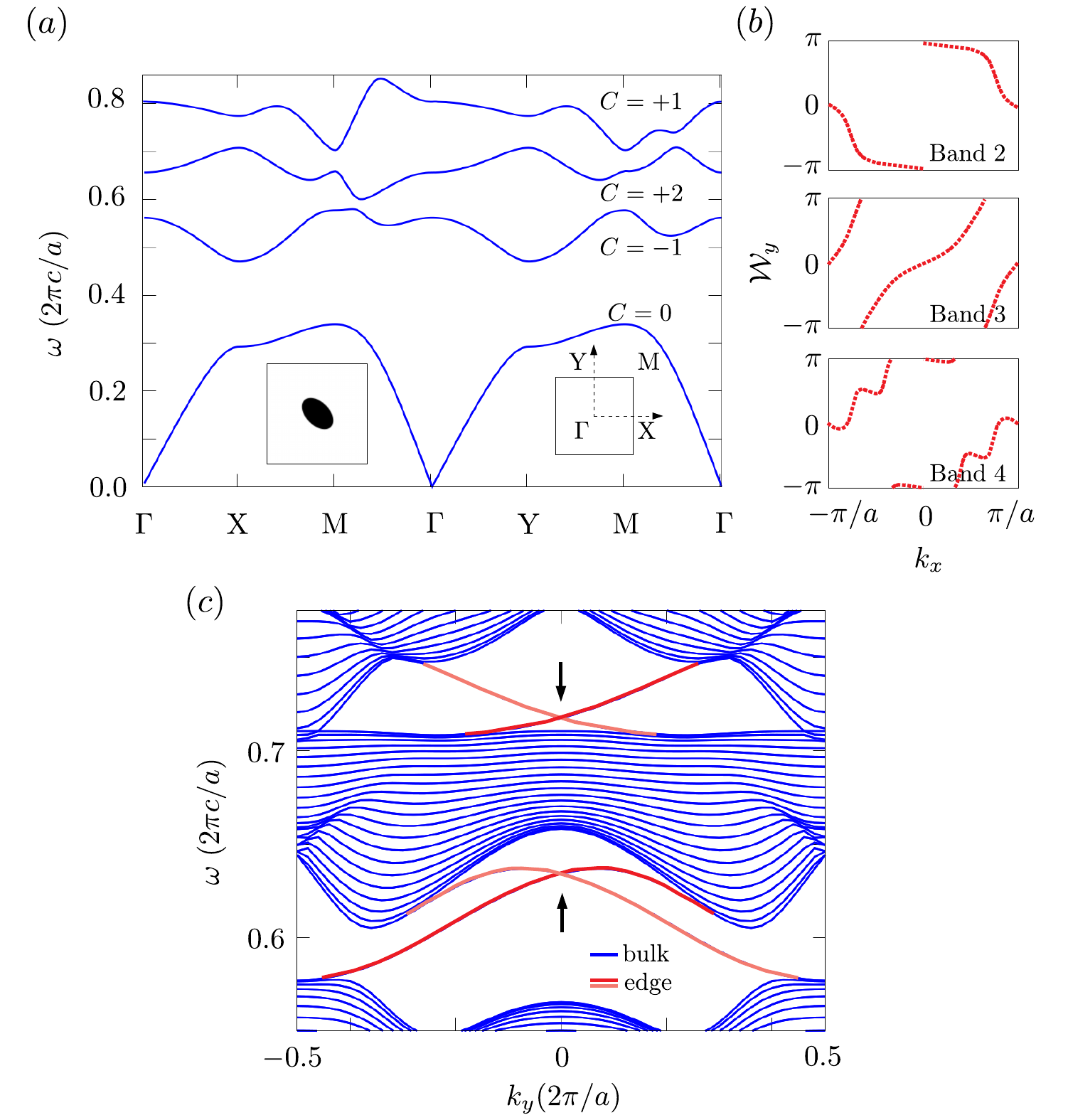}
\caption{(a) TM-polarized band structure of a $C_{2}$-symmetric gyromagnetic PhC whose unit cell is shown in the inset. The Chern numbers for the first four bands are also shown. (b) Wilson loop eigenvalues $\mathcal{W}_y$ for the bands 2, 3, and 4 plotted as a function of $k_x$. The winding of the eigenvalues indicates the non-Wannierizablility of the bands, and the winding number is equal to the Chern number of the band. (c) Edge spectrum showing projected bulk bands (blue) and chiral edge states (red).}
\label{fig:2D_Chern}
\end{figure}

Next, we show the topological characterization of a PhC with Dirac points in class AI. We do this via three distinct perspectives: (1) Examining the symmetry-indicator invariants of 1D subsystems, (2) computing the Wilson loops, and (3) constructing the indices of the 2D bands of the system. 

Consider the $C_2$-symmetric PhC in the inset of Fig.~\ref{fig:2D_Dirac}(a), which consists of an elliptical disc ($\varepsilon = 12$) with its semi-major and semi-minor axes oriented along the diagonals of a square unit cell. This PhC's TM spectrum exhibits two sets of Dirac points along the $\mathbf{\Gamma}-\mathbf{M}$ direction, one between bands 2 and 3, and one between bands 3 and 4, as shown in Fig.~\ref{fig:2D_Dirac}(a). 

We first examine the topology of the gapped phases of 1D subsystems that are obtained by fixing one of the momenta, say $k_y$. In this example, bands 2 and 4 have different $C_2$ eigenvalues (and hence $\mathcal{I}$ eigenvalues in the 1D subsystem) at the $\mathbf{\Gamma}$ and $\mathbf{X}$ points, corresponding to a 1D topological phase at the $k_y = 0$ cut with $[X_1] = 1$ (or equivalently, $\theta=\pi$ from Eq.\eqref{eq:BerryPhase}). On the other hand, these bands have the same $C_2$ eigenvalues at the $\mathbf{Y}$ and $\mathbf{M}$ points, corresponding to a trivial phase at the $k_y = \pi/a$ cut  with $[X_1] = 0$ (or equivalently, $\theta$ $=0$). These Dirac points are thus the required transition points that separate trivial and topological gapped phases of the 1D subsystems.

This change in the topology of the one-dimensional subsystem at the Dirac points can also be seen from the Wilson loop spectrum. The Wilson loop eigenvalues plotted in Fig.~\ref{fig:2D_Dirac}(b) exhibit jump discontinuities from $0$ to $\pi$ at the momenta of the Dirac points, which correspond to a switch in the value of $[X_1]$ from $0$ to $1$. Consequently, edge states only appear in the portion of the 1D edge Brillouin zone that is topologically non-trivial. Fig.~\ref{fig:2D_Dirac}(c) shows the edge spectrum for the PhC with open boundaries along $x$ and periodic boundaries along $y$. 

The Wilson loop can also help diagnose generic Dirac points that may be present in the interior of the Brillouin zone. In the current example, there are two additional pairs of jump discontinuities in the Wilson loop spectrum for band 4 which are due to such generic Dirac points between bands 4 and 5.

Since the bands 2, 3, and 4 are non-degenerate at all HSPs, we can classify them by constructing the 2D indices under TRS from table \ref{tab:C2_Invariants_table}, which are respectively $\chi_{\mathcal{T}}^{(2)} = (-1,-1,-1; 1)$, $\chi_{\mathcal{T}}^{(2)} = (0,0,0; 1)$ and $\chi_{\mathcal{T}}^{(2)} = (1,1,1; 1)$. The indices for bands 2 and 4 are not found in table \ref{tab:C2_Invariants_table}, and expanding these in a linear combination of OALs results in fractional coefficients $\{\alpha_p\}$. Therefore, these are stable topological bands and must contain a gapless point somewhere in the BZ under TRS. In this example, the PhC has Dirac points on high-symmetry lines as seen in Fig.~\ref{fig:2D_Dirac}(a). Band 3 is an example of a situation where stable topological bands could have the same indices as atomic limit bands.

Relevant to PhC design, these invariants can be useful for finding spectrally-isolated Dirac points for applications such as creating cavity states that are algebraically localized to embedded point defects~\cite{diracmode1, diracexp1, diracexp2, diracexp3, vaidya2021point} or enabling large-area single-mode lasing~\cite{dirac_laser1, dirac_laser2}.

\subsection{Example 3: Chern insulator in class A}

Consider the PhC introduced in the previous section. We break TRS for this PhC by introducing non-diagonal terms in the permeability tensor, which correspond to a response of a gyromagnetic material under a magnetic field applied in the $z$-direction. Specifically, we set the permeability tensor to
\begin{align}
\mu = 
\begin{bmatrix}
\mu & \ii \kappa & 0\\
-\ii \kappa & \mu & 0 \\
0 & 0 & \mu_0
\end{bmatrix},
\end{align}
where $\mu = \mu_0$ is the vacuum permeability and $\kappa = 0.25\mu_0$. The Dirac points that were previously protected by a combination of inversion and TRS are now gapped, and bands 2, 3 and 4 are non-degenerate and have the invariants $\chi^{(2)} = (-1\, | \, -1,-1,-1; 1)$, $\chi^{(2)} = (+2\, | \, 0,0,0; 1)$ and $\chi^{(2)} = (+1\, | \, 1,1,1; 1)$, respectively. The first invariant of the listed tuples is the Chern number, obtained as the winding number of the Wilson loop spectrum in Fig.~\ref{fig:2D_Chern}(b). The winding numbers agree with the symmetry-imposed constraints in Eq.~\eqref{eq:ChernSymmInd}. 

The  Chern number leads to chiral edge states at the boundary of a finite system as shown in Fig.~\ref{fig:2D_Chern}(c). These edge states exhibit unidirectional transport and have been observed in gyromagnetic PhCs at microwave frequencies~\cite{Chern1, Chern4}. Proposed applications for these edge states include optical isolators and slow-light devices that could significantly outperform their conventional counterparts~\cite{Slowlight_chern1, Slowlight_chern2, Slowlight_chern3}.

\subsection{Example 4: Fragile phase in class AI}

Fragile phases have bands that exhibit a symmetry-protected winding in their Wilson loop spectrum, indicating that the bands cannot form a symmetry-preserving Wannier representation. However, when considered as a set along with additional atomic limit bands, the full set becomes Wannierizable, and accordingly, the Wilson loop winding is lost. They are characterized by indices that must be written as a linear combination of the invariants in Tables \ref{tab:C2_Invariants_table}-\ref{tab:C6_Invariants} with at least one negative integer coefficient.

\begin{figure}[t]
\centering
\includegraphics[width=\columnwidth]{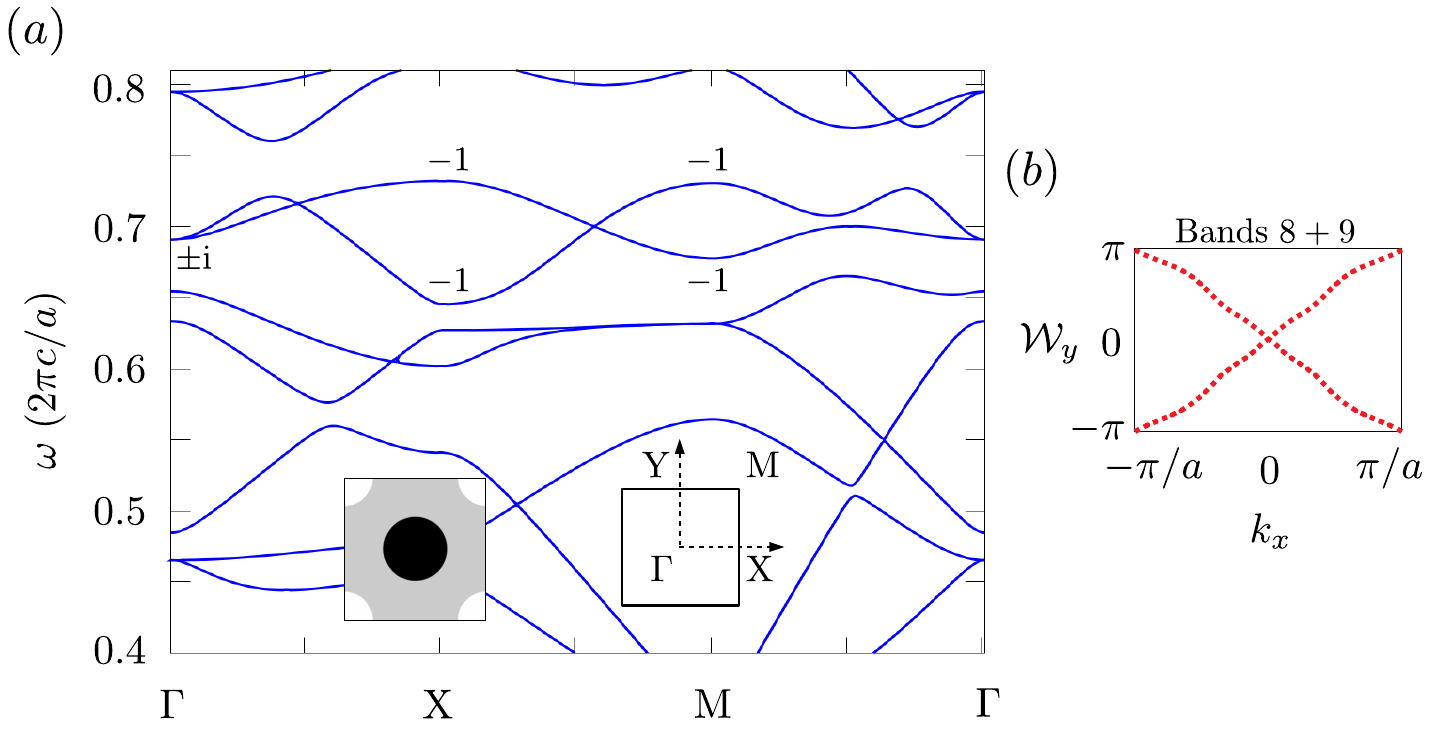}
\caption{(a) TE-polarized band structure of a $C_{4v}$ symmetric PhC with lattice constant, $a$, whose unit cell is shown in the inset. This unit cell consists of dielectric discs of $\varepsilon_1 = 1$ (white) with $r_1 = 0.2a$ and $\varepsilon_2 = 16$ (black) with $r_2 = 0.225a$ in a background material of $\varepsilon_3 = 4$ (gray). $C_4$ eigenvalues at $\mathbf{\Gamma}$ and $\mathbf{M}$, $C_2$ eigenvalues at $\mathbf{X}$ are shown for bands 8 and 9. (b) Wilson loop eigenvalues $\mathcal{W}_y$ for the bands 8+9, plotted as a function of $k_x$. The opposite winding of the eigenvalues indicates the non-Wannierizablility of the bands, particularly that the bands are fragile.}
\label{fig:2D_Fragile}
\end{figure}

We now present a novel PhC design with fragile bands in a $C_{4v}$ symmetry setting whose unit cell is shown in the inset of Fig.~\ref{fig:2D_Fragile}(a). The PhC is composed of three materials, $\varepsilon_1 = 1$ (white), $\varepsilon_2 = 16$ (black) and $\varepsilon_3 = 4$ (gray). We consider the two isolated and degenerate bands, bands $8+9$ in the TE-polarized band structure of this PhC shown in Fig.~\ref{fig:2D_Fragile}(a). Using the relevant rotation eigenvalues of the electromagnetic eigenmodes at the HSPs, we compute the invariant for these bands to be $\chi_{\mathcal{T}}^{(4)} = (0,2,-1; 2)$. Since this invariant is not found in Table \ref{tab:C4_Invariants}, we express it as the following linear combination of OALs from Table \ref{tab:C4_Invariants}: $\chi_{\mathcal{T}}^{(4)} = (0,0,-1; 2) = 1\times(1, 1, -1; 2) + 1\times(-1,-1,0; 1) + (-1)\times(0,0,0;1)$. The requirement of a negative integer coefficient in this expansion indicates that this set of two bands is fragile. The non-Wannierizable nature of these bands is also evident from the Wilson loop spectrum in Fig.~\ref{fig:2D_Fragile}(b) which shows opposite winding of the two eigenvalues.

A different PhC realization of a fragile phase with $C_6$ symmetry was previously reported in~\cite{FragilePhC}. Like OAL phases, fragile PhCs may host corner states resulting from the total corner charge of all Wannierizable components in their decomposition~\cite{Rot_HOTI1, Rot_HOTI3}.

\section{Other topological phases}

Finally, we discuss a selection of other topological phases where crystalline symmetries play a crucial role, but whose realization may not be directly inferred from the topological indices presented so far. 

\subsection{Quantum spin-Hall analogs}
The electronic quantum spin-Hall effect (QSHE) can be thought of as being deformable to two Chern insulators with opposite Chern numbers stacked on top of each other, one for each spin degree of freedom~\cite{QSHE_CM1, QSHE_CM2, QSHE_CM3}. This creates spin-polarized ``helical" edge states on the boundary of a finite sample which are protected against back-scattering due to the Kramers' degeneracy at time-reversal invariant momenta. 

Since the bosonic TR operator squares to $+1$, photons lack the Kramers degeneracy enjoyed by their fermionic counterparts (whose TR operator squares to $-1$). A photonic counterpart to the QSHE consequently necessitates a replacement for Kramers degeneracy. This can be achieved by incorporating spatial symmetries, particularly $C_{6v}$ symmetry, to construct a pseudo-TR operator~\cite{QSH1}. It can be shown that the bulk topology of such a PhC is identical to that of the QSHE by explicit calculation of the pseudo-spin polarized Wilson loop spectrum~\cite{Berry_bands}, where an opposite winding of the two eigenvalues is observed. However, since this winding is enforced by a crystalline symmetry, it is more appropriate to classify these PhCs as fragile phases than as true QSH systems. Nevertheless, such PhCs have states with a well-defined pseudo-spin, analogous to the spin of electrons~\cite{QSH1, QSH2, QSH3, QSH4, QSH5} and exhibit pseudo-spin polarized helical edge states, similar to the QSHE, as shown in Fig.~\ref{fig:Other_topo}(a). The presence of an edge necessarily breaks the $C_{6v}$ symmetry of the bulk and therefore also the pseudo-TR symmetry allowing for the hybridization of the edge states. This opens a gap in the edge spectrum as shown in Fig.~\ref{fig:Other_topo}(a) and allows for the back-scattering of the edge states in the vicinity of the gap.

\subsection{Valley-Hall phases}
As shown previously, Dirac points can be gapped by breaking TRS, thereby creating bands with a non-zero Chern number. Breaking inversion symmetry (i.e., two-fold rotation in a 2D system) can also gap Dirac points and introduce local Berry curvature with boundary manifestations. Reducing $C_6$ symmetry to $C_3$ gaps the Dirac points that generically exist at the $\mathbf{K}$ $(\mathbf{K^{\prime}})$ points of the BZ. This causes the Berry curvature to peak at the ``valleys" formed at the $\mathbf{K}$ $(\mathbf{K^{\prime}})$ points. Due to TRS, the total Berry curvature and the Chern number are identically zero. However, the non-zero local Berry curvature at the $\mathbf{K}$ $(\mathbf{K^{\prime}})$ valleys can be used to define valley Chern numbers such that $C_{\mathbf{K}} = -C_{\mathbf{K^{\prime}}}$. 

\begin{figure}[]%
\centering
\includegraphics[width=\columnwidth]{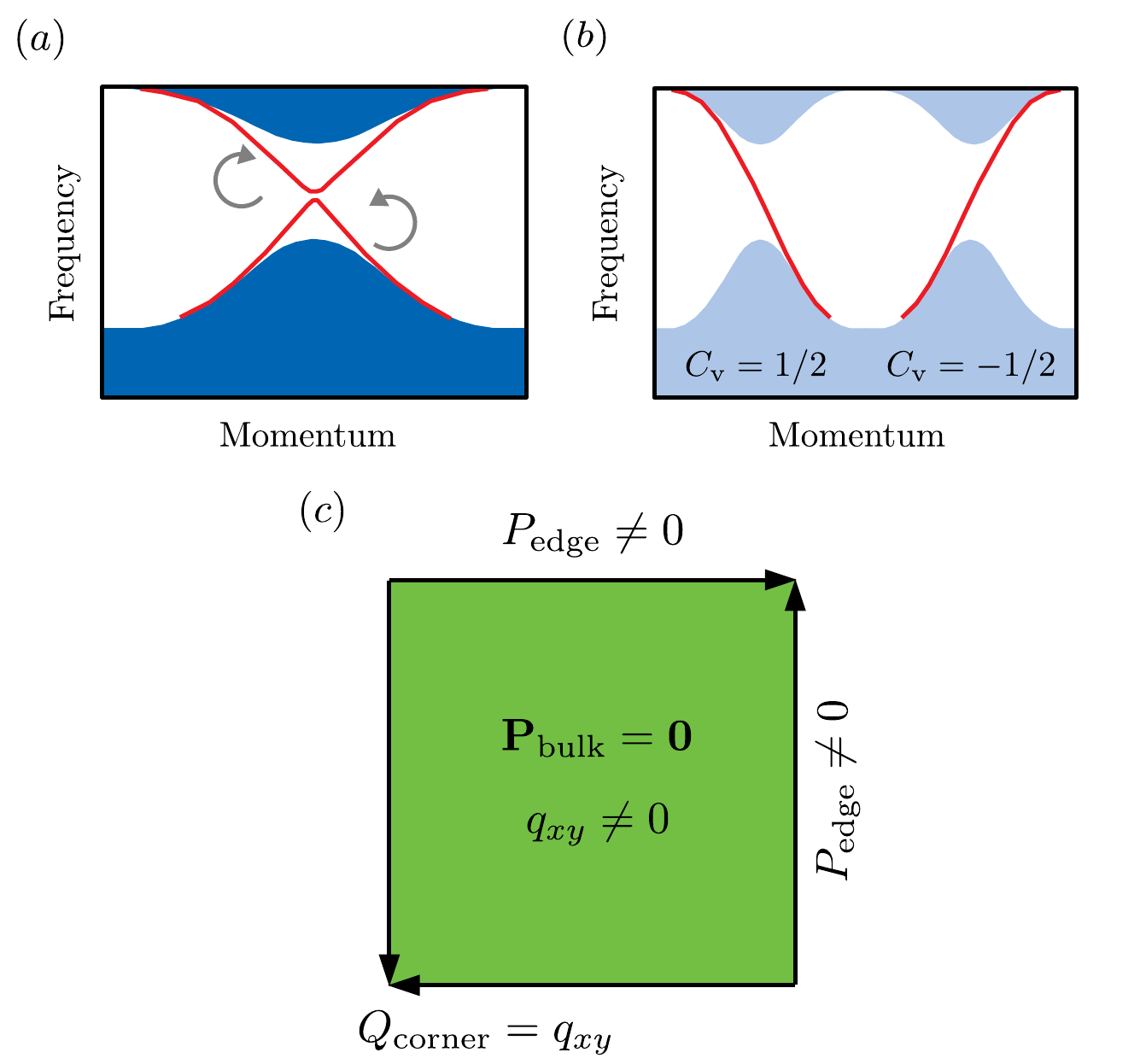}
\caption{(a) Pseudo-spin polarized helical edge states of a quantum spin-Hall analog PhC. (b) Edge states of a valley-Hall PhC. (c) Schematic of a PhC quadrupole insulator with vanishing bulk dipole moment and non-zero bulk quadrupole moment. }
\label{fig:Other_topo}
\end{figure}

In this case, the bulk-boundary correspondence is only well-defined at the boundary between two such systems, one spatially inverted with respect to the other. The edge states that thus emerge have a dispersion as shown in Fig.~\ref{fig:Other_topo}(b) and can generally backscatter, unlike the chiral edge states of a Chern insulator. Certain types of edge geometries and symmetry-preserving perturbations are known to suppress inter-valley scattering, leading to nearly perfect (but incidental) backscatter-free transport in the absence of structural imperfections~\cite{ValleyHall2}. However, in the presence of random disorder, typically introduced by fabrication imperfections, it was recently shown that these valley-Hall edge states may not perform better than conventional edge states for practical light transport~\cite{stobbe_valleyhall}. Valley-Hall edge states have been observed in PhC designs spanning orders of magnitude in frequency~\cite{ValleyHall1, ValleyHall2, ValleyHall3, stobbe_valleyhall}.

\subsection{Quadrupole and octupole topological insulators}
Quadrupole and octupole topological insulators (QTIs and OTIs, respectively) are a final example of crystalline symmetry-protected topological phases which host fractional corner charges, similar to, but ultimately distinct from, OAL insulators~\cite{benalcazar2017quad}. They are $\mathbb{Z}_2$ classified, with fractional corner charges quantized to $\{0, 1/2\}\mod{1}$. The prototypical model is $C_{4v}$ symmetric~\cite{benalcazar2017quadPRB}. Under $C_4$ symmetry, the QTI phase is bulk-obstructed and therefore an atomic limit. However, relaxing $C_{4v}$ down to only reflection symmetries also protects the quantization of corner charge, although their symmetry-indicator invariants due to reflection symmetry vanish. Thus, the protection due to reflection symmetries is more subtle than for OALs; they exhibit a gapped Wilson loop spectrum, not pinned by symmetries, and the change in topology here is accompanied by a gap closing in the Wilson loop spectrum, which implies a gap closing in the edge spectrum, instead of in the bulk spectrum. 

QTI and OTI phases require a set of anti-commuting spatial symmetries that can be achieved by threading a $\pi$-flux in simple tight-binding models. However, PhCs cannot be accurately described by such models. Instead, a quadrupole phase can be achieved by breaking time-reversal symmetry while preserving the product of mirror and time-reversal symmetries~\cite{Quadrupole1}. Alternatively, QTI phases can also be realized in PhCs with anti-commuting glide symmetries~\cite{Quadrupole2}. Topological indices that diagnose the QTI and OTI topologies have been recently demonstrated~\cite{wheeler2019,kang2019}, and follow the natural extension of the index for dipole moments~\cite{resta1998}. Such indices have also recently been used to show that QTIs can also be protected solely by chiral symmetry~\cite{roy2020quad,li2020quad,yang2021quad}, which has allowed the introduction of a $\mathbb{Z}$ classification of higher-order topological insulators in 2D and 3D~\cite{benalcazar2021}.

\section{Discussion}
The past decade has seen the uncovering of a wide range of topological phenomena in PhCs~\cite{topoPhCReview1, topoPhCReview2, topoPhCReview3, Chern0, Chern5, Chern1}, validating the notion that topological band theory is a wave phenomenon, transcending the existence of bound orbital states present in electronic systems. Motivated by these recent developments, we have here extended the use of symmetry-indicator invariants to classify one- and two- dimensional PhCs with crystalline symmetries, with and without time-reversal symmetry. Through various examples, we have also demonstrated that the bulk-boundary correspondence of topological band theory carries over to these systems as well. 

In solids, the atomic ions form potentials that bind electronic orbitals. The electrons in the crystal hop between these orbitals, giving rise to Bloch energy bands that can often be described by simplified tight-binding models, where the hopping terms in the Hamiltonian are given by the overlap integrals between different orbitals. Photonic analogs of solid-state lattices have been achieved in periodic arrays of coupled waveguides, where each waveguide supports a guided mode so that the extended array can be thought of as having inter-orbital hoppings that also lead to tight-binding descriptions~\cite{yariv1991book}. As a result, the electronic theory of non-interacting topological phases carries over directly to this case. In contrast, the PhCs studied in this work are not well-described by simple tight-binding models; instead, they necessarily require a continuum description based on the full-wave solution of Maxwell's equations.

A further crucial difference between PhCs and solids is that PhCs host classical, bosonic waves unlike electrons, which are fermionic and described by a quantum wave function. For topological band theory, this has two consequences. First, there is no Kramers' degeneracy for electromagnetic waves, and thus there is no protection of helical edge states as in the QSHE phase of electronic systems. As discussed in Section V, the edge states in the PhC versions of the QSHE phase are not rigorously protected, in contrast to their electronic counterpart. Second, PhCs lack a notion of band filling and consequently require a subtler interpretation of the filling anomaly, involving instead the fractional quantization of electromagnetic mode density instead of charges~\cite{filling_anomaly}. We have demonstrated that this fractionalization comes from a \emph{counting mismatch} of states and that the boundary-localized states associated with it are the consequence of the conservation of the number of degrees of freedom in the system, and do not require a fixed band-filling (i.e., Fermi level). We have further demonstrated how the topological invariants based on symmetry indicators relate to the presence of counting mismatches and their boundary states. Finally, we have presented a novel $C_{4v}$-symmetric PhC design hosting fragile bands. 

Beyond its appeal as a platform for exploring the new, fundamental physics of topology in a controllable setting, the merger of topology and PhCs hold substantial promise for the development of new technologies and device design strategies. We expect that the algebraic structure of the symmetry-indicator invariants will be useful in this pursuit. 

\section{Acknowledgements}
We acknowledge fruitful discussions with Alexander Cerjan and Marius J{\"u}rgensen. M.C.R.,\ S.V.,\ T.C.,\ and A.G.\ acknowledge the support of the U.S. Office of Naval Research (ONR) Multidisciplinary University Research Initiative (MURI) under Grant No.~N00014-20-1-2325. M.C.R.\ and S.V.\ also acknowledge support from the Charles E. Kaufman Foundation under Grant No.~KA2020-114794. A.G.\ also acknowledges funding from the National Science Foundation's Graduate Research Fellowship. T.C.\ also acknowledges the support of a research grant (project no.~42106) from Villum Fonden. W.A.B.\ acknowledges the support of the Eberly Postdoctoral Fellowship at Penn State University, the Moore Postdoctoral Fellowship at Princeton University, and startup funds from Emory University.

\bibliography{Topo_PhC}

\begin{thebibliography}{110}%
\makeatletter
\providecommand \@ifxundefined [1]{%
 \@ifx{#1\undefined}
}%
\providecommand \@ifnum [1]{%
 \ifnum #1\expandafter \@firstoftwo
 \else \expandafter \@secondoftwo
 \fi
}%
\providecommand \@ifx [1]{%
 \ifx #1\expandafter \@firstoftwo
 \else \expandafter \@secondoftwo
 \fi
}%
\providecommand \natexlab [1]{#1}%
\providecommand \enquote  [1]{``#1''}%
\providecommand \bibnamefont  [1]{#1}%
\providecommand \bibfnamefont [1]{#1}%
\providecommand \citenamefont [1]{#1}%
\providecommand \href@noop [0]{\@secondoftwo}%
\providecommand \href [0]{\begingroup \@sanitize@url \@href}%
\providecommand \@href[1]{\@@startlink{#1}\@@href}%
\providecommand \@@href[1]{\endgroup#1\@@endlink}%
\providecommand \@sanitize@url [0]{\catcode `\\12\catcode `\$12\catcode
  `\&12\catcode `\#12\catcode `\^12\catcode `\_12\catcode `\%12\relax}%
\providecommand \@@startlink[1]{}%
\providecommand \@@endlink[0]{}%
\providecommand \url  [0]{\begingroup\@sanitize@url \@url }%
\providecommand \@url [1]{\endgroup\@href {#1}{\urlprefix }}%
\providecommand \urlprefix  [0]{URL }%
\providecommand \Eprint [0]{\href }%
\providecommand \doibase [0]{https://doi.org/}%
\providecommand \selectlanguage [0]{\@gobble}%
\providecommand \bibinfo  [0]{\@secondoftwo}%
\providecommand \bibfield  [0]{\@secondoftwo}%
\providecommand \translation [1]{[#1]}%
\providecommand \BibitemOpen [0]{}%
\providecommand \bibitemStop [0]{}%
\providecommand \bibitemNoStop [0]{.\EOS\space}%
\providecommand \EOS [0]{\spacefactor3000\relax}%
\providecommand \BibitemShut  [1]{\csname bibitem#1\endcsname}%
\let\auto@bib@innerbib\@empty
\bibitem [{\citenamefont {Joannopoulos}\ \emph {et~al.}(2008)\citenamefont
  {Joannopoulos}, \citenamefont {Johnson}, \citenamefont {Winn},\ and\
  \citenamefont {Meade}}]{photoniccrystalsbook}%
  \BibitemOpen
  \bibfield  {author} {\bibinfo {author} {\bibfnamefont {J.~D.}\ \bibnamefont
  {Joannopoulos}}, \bibinfo {author} {\bibfnamefont {S.~G.}\ \bibnamefont
  {Johnson}}, \bibinfo {author} {\bibfnamefont {J.~N.}\ \bibnamefont {Winn}},\
  and\ \bibinfo {author} {\bibfnamefont {R.~D.}\ \bibnamefont {Meade}},\ }\href
  {http://ab-initio.mit.edu/book} {\emph {\bibinfo {title} {Photonic Crystals:
  Molding the Flow of Light}}},\ \bibinfo {edition} {2nd}\ ed.\ (\bibinfo
  {publisher} {Princeton University Press},\ \bibinfo {year}
  {2008})\BibitemShut {NoStop}%
\bibitem [{\citenamefont {Sakoda}(2005)}]{photoniccrystalsbook2}%
  \BibitemOpen
  \bibfield  {author} {\bibinfo {author} {\bibfnamefont {K.}~\bibnamefont
  {Sakoda}},\ }\href@noop {} {\emph {\bibinfo {title} {Optical Properties of
  Photonic Crystals}}},\ \bibinfo {edition} {2nd}\ ed.\ (\bibinfo  {publisher}
  {Springer Series in Optical Sciences Vol. 80},\ \bibinfo {year}
  {2005})\BibitemShut {NoStop}%
\bibitem [{\citenamefont {Ozawa}\ \emph {et~al.}(2019)\citenamefont {Ozawa},
  \citenamefont {Price}, \citenamefont {Amo}, \citenamefont {Goldman},
  \citenamefont {Hafezi}, \citenamefont {Lu}, \citenamefont {Rechtsman},
  \citenamefont {Schuster}, \citenamefont {Simon}, \citenamefont {Zilberberg}
  \emph {et~al.}}]{topoPhCReview1}%
  \BibitemOpen
  \bibfield  {author} {\bibinfo {author} {\bibfnamefont {T.}~\bibnamefont
  {Ozawa}}, \bibinfo {author} {\bibfnamefont {H.~M.}\ \bibnamefont {Price}},
  \bibinfo {author} {\bibfnamefont {A.}~\bibnamefont {Amo}}, \bibinfo {author}
  {\bibfnamefont {N.}~\bibnamefont {Goldman}}, \bibinfo {author} {\bibfnamefont
  {M.}~\bibnamefont {Hafezi}}, \bibinfo {author} {\bibfnamefont
  {L.}~\bibnamefont {Lu}}, \bibinfo {author} {\bibfnamefont {M.~C.}\
  \bibnamefont {Rechtsman}}, \bibinfo {author} {\bibfnamefont {D.}~\bibnamefont
  {Schuster}}, \bibinfo {author} {\bibfnamefont {J.}~\bibnamefont {Simon}},
  \bibinfo {author} {\bibfnamefont {O.}~\bibnamefont {Zilberberg}}, \emph
  {et~al.},\ }\bibfield  {title} {\bibinfo {title} {Topological photonics},\
  }\href@noop {} {\bibfield  {journal} {\bibinfo  {journal} {Reviews of Modern
  Physics}\ }\textbf {\bibinfo {volume} {91}},\ \bibinfo {pages} {015006}
  (\bibinfo {year} {2019})}\BibitemShut {NoStop}%
\bibitem [{\citenamefont {Lu}\ \emph {et~al.}(2014)\citenamefont {Lu},
  \citenamefont {Joannopoulos},\ and\ \citenamefont
  {Solja{\v{c}}i{\'c}}}]{topoPhCReview2}%
  \BibitemOpen
  \bibfield  {author} {\bibinfo {author} {\bibfnamefont {L.}~\bibnamefont
  {Lu}}, \bibinfo {author} {\bibfnamefont {J.~D.}\ \bibnamefont
  {Joannopoulos}},\ and\ \bibinfo {author} {\bibfnamefont {M.}~\bibnamefont
  {Solja{\v{c}}i{\'c}}},\ }\bibfield  {title} {\bibinfo {title} {Topological
  photonics},\ }\href@noop {} {\bibfield  {journal} {\bibinfo  {journal}
  {Nature Photonics}\ }\textbf {\bibinfo {volume} {8}},\ \bibinfo {pages} {821}
  (\bibinfo {year} {2014})}\BibitemShut {NoStop}%
\bibitem [{\citenamefont {Kim}\ \emph {et~al.}(2020{\natexlab{a}})\citenamefont
  {Kim}, \citenamefont {Jacob},\ and\ \citenamefont {Rho}}]{topoPhCReview3}%
  \BibitemOpen
  \bibfield  {author} {\bibinfo {author} {\bibfnamefont {M.}~\bibnamefont
  {Kim}}, \bibinfo {author} {\bibfnamefont {Z.}~\bibnamefont {Jacob}},\ and\
  \bibinfo {author} {\bibfnamefont {J.}~\bibnamefont {Rho}},\ }\bibfield
  {title} {\bibinfo {title} {Recent advances in {2D}, {3D} and higher-order
  topological photonics},\ }\href@noop {} {\bibfield  {journal} {\bibinfo
  {journal} {Light: Science \& Applications}\ }\textbf {\bibinfo {volume}
  {9}},\ \bibinfo {pages} {1} (\bibinfo {year}
  {2020}{\natexlab{a}})}\BibitemShut {NoStop}%
\bibitem [{\citenamefont {Rechtsman}\ \emph {et~al.}(2013)\citenamefont
  {Rechtsman}, \citenamefont {Zeuner}, \citenamefont {Plotnik}, \citenamefont
  {Lumer}, \citenamefont {Podolsky}, \citenamefont {Dreisow}, \citenamefont
  {Nolte}, \citenamefont {Segev},\ and\ \citenamefont
  {Szameit}}]{rechtsman2013photonic}%
  \BibitemOpen
  \bibfield  {author} {\bibinfo {author} {\bibfnamefont {M.~C.}\ \bibnamefont
  {Rechtsman}}, \bibinfo {author} {\bibfnamefont {J.~M.}\ \bibnamefont
  {Zeuner}}, \bibinfo {author} {\bibfnamefont {Y.}~\bibnamefont {Plotnik}},
  \bibinfo {author} {\bibfnamefont {Y.}~\bibnamefont {Lumer}}, \bibinfo
  {author} {\bibfnamefont {D.}~\bibnamefont {Podolsky}}, \bibinfo {author}
  {\bibfnamefont {F.}~\bibnamefont {Dreisow}}, \bibinfo {author} {\bibfnamefont
  {S.}~\bibnamefont {Nolte}}, \bibinfo {author} {\bibfnamefont
  {M.}~\bibnamefont {Segev}},\ and\ \bibinfo {author} {\bibfnamefont
  {A.}~\bibnamefont {Szameit}},\ }\bibfield  {title} {\bibinfo {title}
  {Photonic floquet topological insulators},\ }\href@noop {} {\bibfield
  {journal} {\bibinfo  {journal} {Nature}\ }\textbf {\bibinfo {volume} {496}},\
  \bibinfo {pages} {196} (\bibinfo {year} {2013})}\BibitemShut {NoStop}%
\bibitem [{\citenamefont {Noh}\ \emph {et~al.}(2018)\citenamefont {Noh},
  \citenamefont {Benalcazar}, \citenamefont {Huang}, \citenamefont {Collins},
  \citenamefont {Chen}, \citenamefont {Hughes},\ and\ \citenamefont
  {Rechtsman}}]{HOTI_waveguides}%
  \BibitemOpen
  \bibfield  {author} {\bibinfo {author} {\bibfnamefont {J.}~\bibnamefont
  {Noh}}, \bibinfo {author} {\bibfnamefont {W.~A.}\ \bibnamefont {Benalcazar}},
  \bibinfo {author} {\bibfnamefont {S.}~\bibnamefont {Huang}}, \bibinfo
  {author} {\bibfnamefont {M.~J.}\ \bibnamefont {Collins}}, \bibinfo {author}
  {\bibfnamefont {K.~P.}\ \bibnamefont {Chen}}, \bibinfo {author}
  {\bibfnamefont {T.~L.}\ \bibnamefont {Hughes}},\ and\ \bibinfo {author}
  {\bibfnamefont {M.~C.}\ \bibnamefont {Rechtsman}},\ }\bibfield  {title}
  {\bibinfo {title} {Topological protection of photonic mid-gap defect modes},\
  }\href@noop {} {\bibfield  {journal} {\bibinfo  {journal} {Nature Photonics}\
  }\textbf {\bibinfo {volume} {12}},\ \bibinfo {pages} {408} (\bibinfo {year}
  {2018})}\BibitemShut {NoStop}%
\bibitem [{\citenamefont {Hafezi}\ \emph {et~al.}(2013)\citenamefont {Hafezi},
  \citenamefont {Mittal}, \citenamefont {Fan}, \citenamefont {Migdall},\ and\
  \citenamefont {Taylor}}]{hafezi2013imaging}%
  \BibitemOpen
  \bibfield  {author} {\bibinfo {author} {\bibfnamefont {M.}~\bibnamefont
  {Hafezi}}, \bibinfo {author} {\bibfnamefont {S.}~\bibnamefont {Mittal}},
  \bibinfo {author} {\bibfnamefont {J.}~\bibnamefont {Fan}}, \bibinfo {author}
  {\bibfnamefont {A.}~\bibnamefont {Migdall}},\ and\ \bibinfo {author}
  {\bibfnamefont {J.}~\bibnamefont {Taylor}},\ }\bibfield  {title} {\bibinfo
  {title} {Imaging topological edge states in silicon photonics},\ }\href@noop
  {} {\bibfield  {journal} {\bibinfo  {journal} {Nature Photonics}\ }\textbf
  {\bibinfo {volume} {7}},\ \bibinfo {pages} {1001} (\bibinfo {year}
  {2013})}\BibitemShut {NoStop}%
\bibitem [{\citenamefont {Mittal}\ \emph {et~al.}(2019)\citenamefont {Mittal},
  \citenamefont {Orre}, \citenamefont {Zhu}, \citenamefont {Gorlach},
  \citenamefont {Poddubny},\ and\ \citenamefont {Hafezi}}]{HOTI_ringresonator}%
  \BibitemOpen
  \bibfield  {author} {\bibinfo {author} {\bibfnamefont {S.}~\bibnamefont
  {Mittal}}, \bibinfo {author} {\bibfnamefont {V.~V.}\ \bibnamefont {Orre}},
  \bibinfo {author} {\bibfnamefont {G.}~\bibnamefont {Zhu}}, \bibinfo {author}
  {\bibfnamefont {M.~A.}\ \bibnamefont {Gorlach}}, \bibinfo {author}
  {\bibfnamefont {A.}~\bibnamefont {Poddubny}},\ and\ \bibinfo {author}
  {\bibfnamefont {M.}~\bibnamefont {Hafezi}},\ }\bibfield  {title} {\bibinfo
  {title} {Photonic quadrupole topological phases},\ }\href@noop {} {\bibfield
  {journal} {\bibinfo  {journal} {Nature Photonics}\ }\textbf {\bibinfo
  {volume} {13}},\ \bibinfo {pages} {692} (\bibinfo {year} {2019})}\BibitemShut
  {NoStop}%
\bibitem [{\citenamefont {Peterson}\ \emph {et~al.}(2018)\citenamefont
  {Peterson}, \citenamefont {Benalcazar}, \citenamefont {Hughes},\ and\
  \citenamefont {Bahl}}]{microwave_HOTI}%
  \BibitemOpen
  \bibfield  {author} {\bibinfo {author} {\bibfnamefont {C.~W.}\ \bibnamefont
  {Peterson}}, \bibinfo {author} {\bibfnamefont {W.~A.}\ \bibnamefont
  {Benalcazar}}, \bibinfo {author} {\bibfnamefont {T.~L.}\ \bibnamefont
  {Hughes}},\ and\ \bibinfo {author} {\bibfnamefont {G.}~\bibnamefont {Bahl}},\
  }\bibfield  {title} {\bibinfo {title} {A quantized microwave quadrupole
  insulator with topologically protected corner states},\ }\href@noop {}
  {\bibfield  {journal} {\bibinfo  {journal} {Nature}\ }\textbf {\bibinfo
  {volume} {555}},\ \bibinfo {pages} {346} (\bibinfo {year}
  {2018})}\BibitemShut {NoStop}%
\bibitem [{\citenamefont {Gao}\ \emph {et~al.}(2018)\citenamefont {Gao},
  \citenamefont {Hu}, \citenamefont {Li}, \citenamefont {Yang}, \citenamefont
  {Chai}, \citenamefont {Xie},\ and\ \citenamefont {Gong}}]{1DPhC1}%
  \BibitemOpen
  \bibfield  {author} {\bibinfo {author} {\bibfnamefont {W.}~\bibnamefont
  {Gao}}, \bibinfo {author} {\bibfnamefont {X.}~\bibnamefont {Hu}}, \bibinfo
  {author} {\bibfnamefont {C.}~\bibnamefont {Li}}, \bibinfo {author}
  {\bibfnamefont {J.}~\bibnamefont {Yang}}, \bibinfo {author} {\bibfnamefont
  {Z.}~\bibnamefont {Chai}}, \bibinfo {author} {\bibfnamefont {J.}~\bibnamefont
  {Xie}},\ and\ \bibinfo {author} {\bibfnamefont {Q.}~\bibnamefont {Gong}},\
  }\bibfield  {title} {\bibinfo {title} {Fano-resonance in one-dimensional
  topological photonic crystal heterostructure},\ }\href@noop {} {\bibfield
  {journal} {\bibinfo  {journal} {Optics Express}\ }\textbf {\bibinfo {volume}
  {26}},\ \bibinfo {pages} {8634} (\bibinfo {year} {2018})}\BibitemShut
  {NoStop}%
\bibitem [{\citenamefont {Li}\ \emph {et~al.}(2018)\citenamefont {Li},
  \citenamefont {Hu}, \citenamefont {Gao}, \citenamefont {Ao}, \citenamefont
  {Chu}, \citenamefont {Yang},\ and\ \citenamefont {Gong}}]{1DPhC2}%
  \BibitemOpen
  \bibfield  {author} {\bibinfo {author} {\bibfnamefont {C.}~\bibnamefont
  {Li}}, \bibinfo {author} {\bibfnamefont {X.}~\bibnamefont {Hu}}, \bibinfo
  {author} {\bibfnamefont {W.}~\bibnamefont {Gao}}, \bibinfo {author}
  {\bibfnamefont {Y.}~\bibnamefont {Ao}}, \bibinfo {author} {\bibfnamefont
  {S.}~\bibnamefont {Chu}}, \bibinfo {author} {\bibfnamefont {H.}~\bibnamefont
  {Yang}},\ and\ \bibinfo {author} {\bibfnamefont {Q.}~\bibnamefont {Gong}},\
  }\bibfield  {title} {\bibinfo {title} {Thermo-optical tunable ultracompact
  chip-integrated 1d photonic topological insulator},\ }\href@noop {}
  {\bibfield  {journal} {\bibinfo  {journal} {Advanced Optical Materials}\
  }\textbf {\bibinfo {volume} {6}},\ \bibinfo {pages} {1701071} (\bibinfo
  {year} {2018})}\BibitemShut {NoStop}%
\bibitem [{\citenamefont {Ota}\ \emph {et~al.}(2018)\citenamefont {Ota},
  \citenamefont {Katsumi}, \citenamefont {Watanabe}, \citenamefont {Iwamoto},\
  and\ \citenamefont {Arakawa}}]{1DPhC3}%
  \BibitemOpen
  \bibfield  {author} {\bibinfo {author} {\bibfnamefont {Y.}~\bibnamefont
  {Ota}}, \bibinfo {author} {\bibfnamefont {R.}~\bibnamefont {Katsumi}},
  \bibinfo {author} {\bibfnamefont {K.}~\bibnamefont {Watanabe}}, \bibinfo
  {author} {\bibfnamefont {S.}~\bibnamefont {Iwamoto}},\ and\ \bibinfo {author}
  {\bibfnamefont {Y.}~\bibnamefont {Arakawa}},\ }\bibfield  {title} {\bibinfo
  {title} {Topological photonic crystal nanocavity laser},\ }\href@noop {}
  {\bibfield  {journal} {\bibinfo  {journal} {Communications Physics}\ }\textbf
  {\bibinfo {volume} {1}},\ \bibinfo {pages} {1} (\bibinfo {year}
  {2018})}\BibitemShut {NoStop}%
\bibitem [{\citenamefont {Raghu}\ and\ \citenamefont {Haldane}(2008)}]{Chern0}%
  \BibitemOpen
  \bibfield  {author} {\bibinfo {author} {\bibfnamefont {S.}~\bibnamefont
  {Raghu}}\ and\ \bibinfo {author} {\bibfnamefont {F.~D.~M.}\ \bibnamefont
  {Haldane}},\ }\bibfield  {title} {\bibinfo {title} {Analogs of
  quantum-hall-effect edge states in photonic crystals},\ }\href@noop {}
  {\bibfield  {journal} {\bibinfo  {journal} {Physical Review A}\ }\textbf
  {\bibinfo {volume} {78}},\ \bibinfo {pages} {033834} (\bibinfo {year}
  {2008})}\BibitemShut {NoStop}%
\bibitem [{\citenamefont {Haldane}\ and\ \citenamefont {Raghu}(2008)}]{Chern5}%
  \BibitemOpen
  \bibfield  {author} {\bibinfo {author} {\bibfnamefont {F.~D.~M.}\
  \bibnamefont {Haldane}}\ and\ \bibinfo {author} {\bibfnamefont
  {S.}~\bibnamefont {Raghu}},\ }\bibfield  {title} {\bibinfo {title} {Possible
  realization of directional optical waveguides in photonic crystals with
  broken time-reversal symmetry},\ }\href@noop {} {\bibfield  {journal}
  {\bibinfo  {journal} {Physical review letters}\ }\textbf {\bibinfo {volume}
  {100}},\ \bibinfo {pages} {013904} (\bibinfo {year} {2008})}\BibitemShut
  {NoStop}%
\bibitem [{\citenamefont {Wang}\ \emph {et~al.}(2009)\citenamefont {Wang},
  \citenamefont {Chong}, \citenamefont {Joannopoulos},\ and\ \citenamefont
  {Solja{\v{c}}i{\'c}}}]{Chern1}%
  \BibitemOpen
  \bibfield  {author} {\bibinfo {author} {\bibfnamefont {Z.}~\bibnamefont
  {Wang}}, \bibinfo {author} {\bibfnamefont {Y.}~\bibnamefont {Chong}},
  \bibinfo {author} {\bibfnamefont {J.~D.}\ \bibnamefont {Joannopoulos}},\ and\
  \bibinfo {author} {\bibfnamefont {M.}~\bibnamefont {Solja{\v{c}}i{\'c}}},\
  }\bibfield  {title} {\bibinfo {title} {Observation of unidirectional
  backscattering-immune topological electromagnetic states},\ }\href@noop {}
  {\bibfield  {journal} {\bibinfo  {journal} {Nature}\ }\textbf {\bibinfo
  {volume} {461}},\ \bibinfo {pages} {772} (\bibinfo {year}
  {2009})}\BibitemShut {NoStop}%
\bibitem [{\citenamefont {Wang}\ \emph {et~al.}(2008)\citenamefont {Wang},
  \citenamefont {Chong}, \citenamefont {Joannopoulos},\ and\ \citenamefont
  {Solja{\v{c}}i{\'c}}}]{Chern2}%
  \BibitemOpen
  \bibfield  {author} {\bibinfo {author} {\bibfnamefont {Z.}~\bibnamefont
  {Wang}}, \bibinfo {author} {\bibfnamefont {Y.}~\bibnamefont {Chong}},
  \bibinfo {author} {\bibfnamefont {J.~D.}\ \bibnamefont {Joannopoulos}},\ and\
  \bibinfo {author} {\bibfnamefont {M.}~\bibnamefont {Solja{\v{c}}i{\'c}}},\
  }\bibfield  {title} {\bibinfo {title} {Reflection-free one-way edge modes in
  a gyromagnetic photonic crystal},\ }\href@noop {} {\bibfield  {journal}
  {\bibinfo  {journal} {Physical Review Letters}\ }\textbf {\bibinfo {volume}
  {100}},\ \bibinfo {pages} {013905} (\bibinfo {year} {2008})}\BibitemShut
  {NoStop}%
\bibitem [{\citenamefont {Liu}\ \emph {et~al.}(2012)\citenamefont {Liu},
  \citenamefont {Shen},\ and\ \citenamefont {He}}]{Chern3}%
  \BibitemOpen
  \bibfield  {author} {\bibinfo {author} {\bibfnamefont {K.}~\bibnamefont
  {Liu}}, \bibinfo {author} {\bibfnamefont {L.}~\bibnamefont {Shen}},\ and\
  \bibinfo {author} {\bibfnamefont {S.}~\bibnamefont {He}},\ }\bibfield
  {title} {\bibinfo {title} {One-way edge mode in a gyromagnetic photonic
  crystal slab},\ }\href@noop {} {\bibfield  {journal} {\bibinfo  {journal}
  {Optics Letters}\ }\textbf {\bibinfo {volume} {37}},\ \bibinfo {pages} {4110}
  (\bibinfo {year} {2012})}\BibitemShut {NoStop}%
\bibitem [{\citenamefont {Skirlo}\ \emph {et~al.}(2015)\citenamefont {Skirlo},
  \citenamefont {Lu}, \citenamefont {Igarashi}, \citenamefont {Yan},
  \citenamefont {Joannopoulos},\ and\ \citenamefont
  {Solja{\v{c}}i{\'c}}}]{Chern4}%
  \BibitemOpen
  \bibfield  {author} {\bibinfo {author} {\bibfnamefont {S.~A.}\ \bibnamefont
  {Skirlo}}, \bibinfo {author} {\bibfnamefont {L.}~\bibnamefont {Lu}}, \bibinfo
  {author} {\bibfnamefont {Y.}~\bibnamefont {Igarashi}}, \bibinfo {author}
  {\bibfnamefont {Q.}~\bibnamefont {Yan}}, \bibinfo {author} {\bibfnamefont
  {J.}~\bibnamefont {Joannopoulos}},\ and\ \bibinfo {author} {\bibfnamefont
  {M.}~\bibnamefont {Solja{\v{c}}i{\'c}}},\ }\bibfield  {title} {\bibinfo
  {title} {Experimental observation of large {Chern} numbers in photonic
  crystals},\ }\href@noop {} {\bibfield  {journal} {\bibinfo  {journal}
  {Physical Review Letters}\ }\textbf {\bibinfo {volume} {115}},\ \bibinfo
  {pages} {253901} (\bibinfo {year} {2015})}\BibitemShut {NoStop}%
\bibitem [{\citenamefont {Wu}\ and\ \citenamefont {Hu}(2015)}]{QSH1}%
  \BibitemOpen
  \bibfield  {author} {\bibinfo {author} {\bibfnamefont {L.-H.}\ \bibnamefont
  {Wu}}\ and\ \bibinfo {author} {\bibfnamefont {X.}~\bibnamefont {Hu}},\
  }\bibfield  {title} {\bibinfo {title} {Scheme for achieving a topological
  photonic crystal by using dielectric material},\ }\href@noop {} {\bibfield
  {journal} {\bibinfo  {journal} {Physical Review Letters}\ }\textbf {\bibinfo
  {volume} {114}},\ \bibinfo {pages} {223901} (\bibinfo {year}
  {2015})}\BibitemShut {NoStop}%
\bibitem [{\citenamefont {Xie}\ \emph {et~al.}(2020)\citenamefont {Xie},
  \citenamefont {Su}, \citenamefont {Wang}, \citenamefont {Liu}, \citenamefont
  {Hu}, \citenamefont {Yu}, \citenamefont {Zhan}, \citenamefont {Lu},
  \citenamefont {Wang},\ and\ \citenamefont {Chen}}]{QSH2}%
  \BibitemOpen
  \bibfield  {author} {\bibinfo {author} {\bibfnamefont {B.}~\bibnamefont
  {Xie}}, \bibinfo {author} {\bibfnamefont {G.}~\bibnamefont {Su}}, \bibinfo
  {author} {\bibfnamefont {H.-F.}\ \bibnamefont {Wang}}, \bibinfo {author}
  {\bibfnamefont {F.}~\bibnamefont {Liu}}, \bibinfo {author} {\bibfnamefont
  {L.}~\bibnamefont {Hu}}, \bibinfo {author} {\bibfnamefont {S.-Y.}\
  \bibnamefont {Yu}}, \bibinfo {author} {\bibfnamefont {P.}~\bibnamefont
  {Zhan}}, \bibinfo {author} {\bibfnamefont {M.-H.}\ \bibnamefont {Lu}},
  \bibinfo {author} {\bibfnamefont {Z.}~\bibnamefont {Wang}},\ and\ \bibinfo
  {author} {\bibfnamefont {Y.-F.}\ \bibnamefont {Chen}},\ }\bibfield  {title}
  {\bibinfo {title} {Higher-order quantum spin {Hall} effect in a photonic
  crystal},\ }\href@noop {} {\bibfield  {journal} {\bibinfo  {journal} {Nature
  Communications}\ }\textbf {\bibinfo {volume} {11}},\ \bibinfo {pages} {1}
  (\bibinfo {year} {2020})}\BibitemShut {NoStop}%
\bibitem [{\citenamefont {Barik}\ \emph {et~al.}(2016)\citenamefont {Barik},
  \citenamefont {Miyake}, \citenamefont {DeGottardi}, \citenamefont {Waks},\
  and\ \citenamefont {Hafezi}}]{QSH3}%
  \BibitemOpen
  \bibfield  {author} {\bibinfo {author} {\bibfnamefont {S.}~\bibnamefont
  {Barik}}, \bibinfo {author} {\bibfnamefont {H.}~\bibnamefont {Miyake}},
  \bibinfo {author} {\bibfnamefont {W.}~\bibnamefont {DeGottardi}}, \bibinfo
  {author} {\bibfnamefont {E.}~\bibnamefont {Waks}},\ and\ \bibinfo {author}
  {\bibfnamefont {M.}~\bibnamefont {Hafezi}},\ }\bibfield  {title} {\bibinfo
  {title} {Two-dimensionally confined topological edge states in photonic
  crystals},\ }\href@noop {} {\bibfield  {journal} {\bibinfo  {journal} {New
  Journal of Physics}\ }\textbf {\bibinfo {volume} {18}},\ \bibinfo {pages}
  {113013} (\bibinfo {year} {2016})}\BibitemShut {NoStop}%
\bibitem [{\citenamefont {Cheng}\ \emph {et~al.}(2016)\citenamefont {Cheng},
  \citenamefont {Jouvaud}, \citenamefont {Ni}, \citenamefont {Mousavi},
  \citenamefont {Genack},\ and\ \citenamefont {Khanikaev}}]{QSH4}%
  \BibitemOpen
  \bibfield  {author} {\bibinfo {author} {\bibfnamefont {X.}~\bibnamefont
  {Cheng}}, \bibinfo {author} {\bibfnamefont {C.}~\bibnamefont {Jouvaud}},
  \bibinfo {author} {\bibfnamefont {X.}~\bibnamefont {Ni}}, \bibinfo {author}
  {\bibfnamefont {S.~H.}\ \bibnamefont {Mousavi}}, \bibinfo {author}
  {\bibfnamefont {A.~Z.}\ \bibnamefont {Genack}},\ and\ \bibinfo {author}
  {\bibfnamefont {A.~B.}\ \bibnamefont {Khanikaev}},\ }\bibfield  {title}
  {\bibinfo {title} {Robust reconfigurable electromagnetic pathways within a
  photonic topological insulator},\ }\href@noop {} {\bibfield  {journal}
  {\bibinfo  {journal} {Nature Materials}\ }\textbf {\bibinfo {volume} {15}},\
  \bibinfo {pages} {542} (\bibinfo {year} {2016})}\BibitemShut {NoStop}%
\bibitem [{\citenamefont {Mattei}\ \emph {et~al.}(2022)\citenamefont {Mattei},
  \citenamefont {Liu}, \citenamefont {Mazzei~Capote}, \citenamefont {Liu},
  \citenamefont {Hacha}, \citenamefont {Osswald}, \citenamefont {Yu},\ and\
  \citenamefont {Goldsmith}}]{QSH5}%
  \BibitemOpen
  \bibfield  {author} {\bibinfo {author} {\bibfnamefont {M.~S.}\ \bibnamefont
  {Mattei}}, \bibinfo {author} {\bibfnamefont {B.}~\bibnamefont {Liu}},
  \bibinfo {author} {\bibfnamefont {G.~A.}\ \bibnamefont {Mazzei~Capote}},
  \bibinfo {author} {\bibfnamefont {Z.}~\bibnamefont {Liu}}, \bibinfo {author}
  {\bibfnamefont {B.~G.}\ \bibnamefont {Hacha}}, \bibinfo {author}
  {\bibfnamefont {T.}~\bibnamefont {Osswald}}, \bibinfo {author} {\bibfnamefont
  {Z.}~\bibnamefont {Yu}},\ and\ \bibinfo {author} {\bibfnamefont {R.~H.}\
  \bibnamefont {Goldsmith}},\ }\bibfield  {title} {\bibinfo {title}
  {Three-dimensional printed planar polymer photonic topological insulator
  waveguides and their robustness to lattice defects},\ }\href@noop {}
  {\bibfield  {journal} {\bibinfo  {journal} {ACS Photonics}\ }\textbf
  {\bibinfo {volume} {9}},\ \bibinfo {pages} {1793} (\bibinfo {year}
  {2022})}\BibitemShut {NoStop}%
\bibitem [{\citenamefont {He}\ and\ \citenamefont {Chan}(2015)}]{Dirac1}%
  \BibitemOpen
  \bibfield  {author} {\bibinfo {author} {\bibfnamefont {W.-Y.}\ \bibnamefont
  {He}}\ and\ \bibinfo {author} {\bibfnamefont {C.~T.}\ \bibnamefont {Chan}},\
  }\bibfield  {title} {\bibinfo {title} {The emergence of {Dirac} points in
  photonic crystals with mirror symmetry},\ }\href@noop {} {\bibfield
  {journal} {\bibinfo  {journal} {Scientific reports}\ }\textbf {\bibinfo
  {volume} {5}},\ \bibinfo {pages} {1} (\bibinfo {year} {2015})}\BibitemShut
  {NoStop}%
\bibitem [{\citenamefont {Lin}\ and\ \citenamefont {Lu}(2020)}]{Dirac2}%
  \BibitemOpen
  \bibfield  {author} {\bibinfo {author} {\bibfnamefont {H.}~\bibnamefont
  {Lin}}\ and\ \bibinfo {author} {\bibfnamefont {L.}~\bibnamefont {Lu}},\
  }\bibfield  {title} {\bibinfo {title} {Dirac-vortex topological photonic
  crystal fibre},\ }\href@noop {} {\bibfield  {journal} {\bibinfo  {journal}
  {Light: Science \& Applications}\ }\textbf {\bibinfo {volume} {9}},\ \bibinfo
  {pages} {1} (\bibinfo {year} {2020})}\BibitemShut {NoStop}%
\bibitem [{\citenamefont {Huang}\ \emph {et~al.}(2011)\citenamefont {Huang},
  \citenamefont {Lai}, \citenamefont {Hang}, \citenamefont {Zheng},\ and\
  \citenamefont {Chan}}]{Dirac3}%
  \BibitemOpen
  \bibfield  {author} {\bibinfo {author} {\bibfnamefont {X.}~\bibnamefont
  {Huang}}, \bibinfo {author} {\bibfnamefont {Y.}~\bibnamefont {Lai}}, \bibinfo
  {author} {\bibfnamefont {Z.~H.}\ \bibnamefont {Hang}}, \bibinfo {author}
  {\bibfnamefont {H.}~\bibnamefont {Zheng}},\ and\ \bibinfo {author}
  {\bibfnamefont {C.}~\bibnamefont {Chan}},\ }\bibfield  {title} {\bibinfo
  {title} {Dirac cones induced by accidental degeneracy in photonic crystals
  and zero-refractive-index materials},\ }\href@noop {} {\bibfield  {journal}
  {\bibinfo  {journal} {Nature Materials}\ }\textbf {\bibinfo {volume} {10}},\
  \bibinfo {pages} {582} (\bibinfo {year} {2011})}\BibitemShut {NoStop}%
\bibitem [{\citenamefont {Xie}\ \emph {et~al.}(2014)\citenamefont {Xie},
  \citenamefont {Jiang}, \citenamefont {Boardman}, \citenamefont {Liu},
  \citenamefont {Wu}, \citenamefont {Xie}, \citenamefont {Jiang}, \citenamefont
  {Xu}, \citenamefont {Yu},\ and\ \citenamefont {Davis}}]{diracmode1}%
  \BibitemOpen
  \bibfield  {author} {\bibinfo {author} {\bibfnamefont {K.}~\bibnamefont
  {Xie}}, \bibinfo {author} {\bibfnamefont {H.}~\bibnamefont {Jiang}}, \bibinfo
  {author} {\bibfnamefont {A.~D.}\ \bibnamefont {Boardman}}, \bibinfo {author}
  {\bibfnamefont {Y.}~\bibnamefont {Liu}}, \bibinfo {author} {\bibfnamefont
  {Z.}~\bibnamefont {Wu}}, \bibinfo {author} {\bibfnamefont {M.}~\bibnamefont
  {Xie}}, \bibinfo {author} {\bibfnamefont {P.}~\bibnamefont {Jiang}}, \bibinfo
  {author} {\bibfnamefont {Q.}~\bibnamefont {Xu}}, \bibinfo {author}
  {\bibfnamefont {M.}~\bibnamefont {Yu}},\ and\ \bibinfo {author}
  {\bibfnamefont {L.~E.}\ \bibnamefont {Davis}},\ }\bibfield  {title} {\bibinfo
  {title} {Trapped photons at a {Dirac} point: a new horizon for photonic
  crystals},\ }\href {https://doi.org/https://doi.org/10.1002/lpor.201300186}
  {\bibfield  {journal} {\bibinfo  {journal} {Laser \& Photonics Reviews}\
  }\textbf {\bibinfo {volume} {8}},\ \bibinfo {pages} {583} (\bibinfo {year}
  {2014})}\BibitemShut {NoStop}%
\bibitem [{\citenamefont {Hu}\ \emph {et~al.}(2018)\citenamefont {Hu},
  \citenamefont {Xie}, \citenamefont {Hu}, \citenamefont {Mao}, \citenamefont
  {Xia}, \citenamefont {Jiang}, \citenamefont {Zhang}, \citenamefont {Wen},\
  and\ \citenamefont {Chen}}]{diracexp1}%
  \BibitemOpen
  \bibfield  {author} {\bibinfo {author} {\bibfnamefont {L.}~\bibnamefont
  {Hu}}, \bibinfo {author} {\bibfnamefont {K.}~\bibnamefont {Xie}}, \bibinfo
  {author} {\bibfnamefont {Z.}~\bibnamefont {Hu}}, \bibinfo {author}
  {\bibfnamefont {Q.}~\bibnamefont {Mao}}, \bibinfo {author} {\bibfnamefont
  {J.}~\bibnamefont {Xia}}, \bibinfo {author} {\bibfnamefont {H.}~\bibnamefont
  {Jiang}}, \bibinfo {author} {\bibfnamefont {J.}~\bibnamefont {Zhang}},
  \bibinfo {author} {\bibfnamefont {J.}~\bibnamefont {Wen}},\ and\ \bibinfo
  {author} {\bibfnamefont {J.}~\bibnamefont {Chen}},\ }\bibfield  {title}
  {\bibinfo {title} {Experimental observation of wave localization at the
  {Dirac} frequency in a two-dimensional photonic crystal microcavity},\ }\href
  {https://doi.org/10.1364/OE.26.008213} {\bibfield  {journal} {\bibinfo
  {journal} {Opt. Express}\ }\textbf {\bibinfo {volume} {26}},\ \bibinfo
  {pages} {8213} (\bibinfo {year} {2018})}\BibitemShut {NoStop}%
\bibitem [{\citenamefont {Yang}\ \emph {et~al.}(2019)\citenamefont {Yang},
  \citenamefont {Qian}, \citenamefont {Tang}, \citenamefont {Yuan},
  \citenamefont {Zhang}, \citenamefont {Huang}, \citenamefont {Shi},
  \citenamefont {Qian},\ and\ \citenamefont {Yang}}]{diracexp2}%
  \BibitemOpen
  \bibfield  {author} {\bibinfo {author} {\bibfnamefont {L.}~\bibnamefont
  {Yang}}, \bibinfo {author} {\bibfnamefont {G.}~\bibnamefont {Qian}}, \bibinfo
  {author} {\bibfnamefont {G.}~\bibnamefont {Tang}}, \bibinfo {author}
  {\bibfnamefont {F.}~\bibnamefont {Yuan}}, \bibinfo {author} {\bibfnamefont
  {Z.}~\bibnamefont {Zhang}}, \bibinfo {author} {\bibfnamefont
  {K.}~\bibnamefont {Huang}}, \bibinfo {author} {\bibfnamefont
  {Z.}~\bibnamefont {Shi}}, \bibinfo {author} {\bibfnamefont {Q.}~\bibnamefont
  {Qian}},\ and\ \bibinfo {author} {\bibfnamefont {Z.}~\bibnamefont {Yang}},\
  }\bibfield  {title} {\bibinfo {title} {Observation of {Dirac} mode in
  modified honeycomb hollow core photonic crystal fiber},\ }\href
  {https://doi.org/https://doi.org/10.1016/j.optmat.2019.01.029} {\bibfield
  {journal} {\bibinfo  {journal} {Optical Materials}\ }\textbf {\bibinfo
  {volume} {89}},\ \bibinfo {pages} {203} (\bibinfo {year} {2019})}\BibitemShut
  {NoStop}%
\bibitem [{\citenamefont {Xie}\ \emph {et~al.}(2015)\citenamefont {Xie},
  \citenamefont {Zhang}, \citenamefont {Boardman}, \citenamefont {Jiang},
  \citenamefont {Hu}, \citenamefont {Liu}, \citenamefont {Xie}, \citenamefont
  {Mao}, \citenamefont {Hu}, \citenamefont {Li}, \citenamefont {Yang},
  \citenamefont {Wen},\ and\ \citenamefont {Wang}}]{diracexp3}%
  \BibitemOpen
  \bibfield  {author} {\bibinfo {author} {\bibfnamefont {K.}~\bibnamefont
  {Xie}}, \bibinfo {author} {\bibfnamefont {W.}~\bibnamefont {Zhang}}, \bibinfo
  {author} {\bibfnamefont {A.~D.}\ \bibnamefont {Boardman}}, \bibinfo {author}
  {\bibfnamefont {H.}~\bibnamefont {Jiang}}, \bibinfo {author} {\bibfnamefont
  {Z.}~\bibnamefont {Hu}}, \bibinfo {author} {\bibfnamefont {Y.}~\bibnamefont
  {Liu}}, \bibinfo {author} {\bibfnamefont {M.}~\bibnamefont {Xie}}, \bibinfo
  {author} {\bibfnamefont {Q.}~\bibnamefont {Mao}}, \bibinfo {author}
  {\bibfnamefont {L.}~\bibnamefont {Hu}}, \bibinfo {author} {\bibfnamefont
  {Q.}~\bibnamefont {Li}}, \bibinfo {author} {\bibfnamefont {T.}~\bibnamefont
  {Yang}}, \bibinfo {author} {\bibfnamefont {F.}~\bibnamefont {Wen}},\ and\
  \bibinfo {author} {\bibfnamefont {E.}~\bibnamefont {Wang}},\ }\bibfield
  {title} {\bibinfo {title} {Fiber guiding at the {Dirac} frequency beyond
  photonic bandgaps},\ }\href {https://doi.org/10.1038/lsa.2015.77} {\bibfield
  {journal} {\bibinfo  {journal} {Light: Science {\&} Applications}\ }\textbf
  {\bibinfo {volume} {4}},\ \bibinfo {pages} {e304} (\bibinfo {year}
  {2015})}\BibitemShut {NoStop}%
\bibitem [{\citenamefont {Shalaev}\ \emph
  {et~al.}(2019{\natexlab{a}})\citenamefont {Shalaev}, \citenamefont
  {Walasik},\ and\ \citenamefont {Litchinitser}}]{ValleyHall1}%
  \BibitemOpen
  \bibfield  {author} {\bibinfo {author} {\bibfnamefont {M.~I.}\ \bibnamefont
  {Shalaev}}, \bibinfo {author} {\bibfnamefont {W.}~\bibnamefont {Walasik}},\
  and\ \bibinfo {author} {\bibfnamefont {N.~M.}\ \bibnamefont {Litchinitser}},\
  }\bibfield  {title} {\bibinfo {title} {Optically tunable topological photonic
  crystal},\ }\href@noop {} {\bibfield  {journal} {\bibinfo  {journal}
  {Optica}\ }\textbf {\bibinfo {volume} {6}},\ \bibinfo {pages} {839} (\bibinfo
  {year} {2019}{\natexlab{a}})}\BibitemShut {NoStop}%
\bibitem [{\citenamefont {Ma}\ and\ \citenamefont
  {Shvets}(2016)}]{ValleyHall2}%
  \BibitemOpen
  \bibfield  {author} {\bibinfo {author} {\bibfnamefont {T.}~\bibnamefont
  {Ma}}\ and\ \bibinfo {author} {\bibfnamefont {G.}~\bibnamefont {Shvets}},\
  }\bibfield  {title} {\bibinfo {title} {All-si valley-hall photonic
  topological insulator},\ }\href@noop {} {\bibfield  {journal} {\bibinfo
  {journal} {New Journal of Physics}\ }\textbf {\bibinfo {volume} {18}},\
  \bibinfo {pages} {025012} (\bibinfo {year} {2016})}\BibitemShut {NoStop}%
\bibitem [{\citenamefont {Xi}\ \emph {et~al.}(2020)\citenamefont {Xi},
  \citenamefont {Ye},\ and\ \citenamefont {Wu}}]{ValleyHall3}%
  \BibitemOpen
  \bibfield  {author} {\bibinfo {author} {\bibfnamefont {X.}~\bibnamefont
  {Xi}}, \bibinfo {author} {\bibfnamefont {K.-P.}\ \bibnamefont {Ye}},\ and\
  \bibinfo {author} {\bibfnamefont {R.-X.}\ \bibnamefont {Wu}},\ }\bibfield
  {title} {\bibinfo {title} {Topological photonic crystal of large valley
  {Chern} numbers},\ }\href@noop {} {\bibfield  {journal} {\bibinfo  {journal}
  {Photonics Research}\ }\textbf {\bibinfo {volume} {8}},\ \bibinfo {pages}
  {B1} (\bibinfo {year} {2020})}\BibitemShut {NoStop}%
\bibitem [{\citenamefont {Rosiek}\ \emph {et~al.}(2022)\citenamefont {Rosiek},
  \citenamefont {Arregui}, \citenamefont {Vladimirova}, \citenamefont
  {Albrechtsen}, \citenamefont {Lahijani}, \citenamefont {Christiansen},\ and\
  \citenamefont {Stobbe}}]{stobbe_valleyhall}%
  \BibitemOpen
  \bibfield  {author} {\bibinfo {author} {\bibfnamefont {C.~A.}\ \bibnamefont
  {Rosiek}}, \bibinfo {author} {\bibfnamefont {G.}~\bibnamefont {Arregui}},
  \bibinfo {author} {\bibfnamefont {A.}~\bibnamefont {Vladimirova}}, \bibinfo
  {author} {\bibfnamefont {M.}~\bibnamefont {Albrechtsen}}, \bibinfo {author}
  {\bibfnamefont {B.~V.}\ \bibnamefont {Lahijani}}, \bibinfo {author}
  {\bibfnamefont {R.~E.}\ \bibnamefont {Christiansen}},\ and\ \bibinfo {author}
  {\bibfnamefont {S.}~\bibnamefont {Stobbe}},\ }\href
  {https://doi.org/10.48550/ARXIV.2206.11741} {\bibinfo {title} {Observation of
  strong backscattering in valley-hall photonic topological interface modes}}
  (\bibinfo {year} {2022})\BibitemShut {NoStop}%
\bibitem [{\citenamefont {Chen}\ \emph {et~al.}(2019)\citenamefont {Chen},
  \citenamefont {Deng}, \citenamefont {Shi}, \citenamefont {Zhao},
  \citenamefont {Chen},\ and\ \citenamefont {Dong}}]{HOTI1}%
  \BibitemOpen
  \bibfield  {author} {\bibinfo {author} {\bibfnamefont {X.-D.}\ \bibnamefont
  {Chen}}, \bibinfo {author} {\bibfnamefont {W.-M.}\ \bibnamefont {Deng}},
  \bibinfo {author} {\bibfnamefont {F.-L.}\ \bibnamefont {Shi}}, \bibinfo
  {author} {\bibfnamefont {F.-L.}\ \bibnamefont {Zhao}}, \bibinfo {author}
  {\bibfnamefont {M.}~\bibnamefont {Chen}},\ and\ \bibinfo {author}
  {\bibfnamefont {J.-W.}\ \bibnamefont {Dong}},\ }\bibfield  {title} {\bibinfo
  {title} {Direct observation of corner states in second-order topological
  photonic crystal slabs},\ }\href@noop {} {\bibfield  {journal} {\bibinfo
  {journal} {Physical Review Letters}\ }\textbf {\bibinfo {volume} {122}},\
  \bibinfo {pages} {233902} (\bibinfo {year} {2019})}\BibitemShut {NoStop}%
\bibitem [{\citenamefont {Xie}\ \emph {et~al.}(2019)\citenamefont {Xie},
  \citenamefont {Su}, \citenamefont {Wang}, \citenamefont {Su}, \citenamefont
  {Shen}, \citenamefont {Zhan}, \citenamefont {Lu}, \citenamefont {Wang},\ and\
  \citenamefont {Chen}}]{HOTI2}%
  \BibitemOpen
  \bibfield  {author} {\bibinfo {author} {\bibfnamefont {B.-Y.}\ \bibnamefont
  {Xie}}, \bibinfo {author} {\bibfnamefont {G.-X.}\ \bibnamefont {Su}},
  \bibinfo {author} {\bibfnamefont {H.-F.}\ \bibnamefont {Wang}}, \bibinfo
  {author} {\bibfnamefont {H.}~\bibnamefont {Su}}, \bibinfo {author}
  {\bibfnamefont {X.-P.}\ \bibnamefont {Shen}}, \bibinfo {author}
  {\bibfnamefont {P.}~\bibnamefont {Zhan}}, \bibinfo {author} {\bibfnamefont
  {M.-H.}\ \bibnamefont {Lu}}, \bibinfo {author} {\bibfnamefont {Z.-L.}\
  \bibnamefont {Wang}},\ and\ \bibinfo {author} {\bibfnamefont {Y.-F.}\
  \bibnamefont {Chen}},\ }\bibfield  {title} {\bibinfo {title} {Visualization
  of higher-order topological insulating phases in two-dimensional dielectric
  photonic crystals},\ }\href@noop {} {\bibfield  {journal} {\bibinfo
  {journal} {Physical Review Letters}\ }\textbf {\bibinfo {volume} {122}},\
  \bibinfo {pages} {233903} (\bibinfo {year} {2019})}\BibitemShut {NoStop}%
\bibitem [{\citenamefont {Xie}\ \emph {et~al.}(2018)\citenamefont {Xie},
  \citenamefont {Wang}, \citenamefont {Wang}, \citenamefont {Zhu},
  \citenamefont {Jiang}, \citenamefont {Lu},\ and\ \citenamefont
  {Chen}}]{HOTI3}%
  \BibitemOpen
  \bibfield  {author} {\bibinfo {author} {\bibfnamefont {B.-Y.}\ \bibnamefont
  {Xie}}, \bibinfo {author} {\bibfnamefont {H.-F.}\ \bibnamefont {Wang}},
  \bibinfo {author} {\bibfnamefont {H.-X.}\ \bibnamefont {Wang}}, \bibinfo
  {author} {\bibfnamefont {X.-Y.}\ \bibnamefont {Zhu}}, \bibinfo {author}
  {\bibfnamefont {J.-H.}\ \bibnamefont {Jiang}}, \bibinfo {author}
  {\bibfnamefont {M.-H.}\ \bibnamefont {Lu}},\ and\ \bibinfo {author}
  {\bibfnamefont {Y.-F.}\ \bibnamefont {Chen}},\ }\bibfield  {title} {\bibinfo
  {title} {Second-order photonic topological insulator with corner states},\
  }\href@noop {} {\bibfield  {journal} {\bibinfo  {journal} {Physical Review
  B}\ }\textbf {\bibinfo {volume} {98}},\ \bibinfo {pages} {205147} (\bibinfo
  {year} {2018})}\BibitemShut {NoStop}%
\bibitem [{\citenamefont {Ota}\ \emph {et~al.}(2019)\citenamefont {Ota},
  \citenamefont {Liu}, \citenamefont {Katsumi}, \citenamefont {Watanabe},
  \citenamefont {Wakabayashi}, \citenamefont {Arakawa},\ and\ \citenamefont
  {Iwamoto}}]{HOTI4}%
  \BibitemOpen
  \bibfield  {author} {\bibinfo {author} {\bibfnamefont {Y.}~\bibnamefont
  {Ota}}, \bibinfo {author} {\bibfnamefont {F.}~\bibnamefont {Liu}}, \bibinfo
  {author} {\bibfnamefont {R.}~\bibnamefont {Katsumi}}, \bibinfo {author}
  {\bibfnamefont {K.}~\bibnamefont {Watanabe}}, \bibinfo {author}
  {\bibfnamefont {K.}~\bibnamefont {Wakabayashi}}, \bibinfo {author}
  {\bibfnamefont {Y.}~\bibnamefont {Arakawa}},\ and\ \bibinfo {author}
  {\bibfnamefont {S.}~\bibnamefont {Iwamoto}},\ }\bibfield  {title} {\bibinfo
  {title} {Photonic crystal nanocavity based on a topological corner state},\
  }\href@noop {} {\bibfield  {journal} {\bibinfo  {journal} {Optica}\ }\textbf
  {\bibinfo {volume} {6}},\ \bibinfo {pages} {786} (\bibinfo {year}
  {2019})}\BibitemShut {NoStop}%
\bibitem [{\citenamefont {Liu}\ \emph {et~al.}(2018)\citenamefont {Liu},
  \citenamefont {Deng},\ and\ \citenamefont {Wakabayashi}}]{HOTI5}%
  \BibitemOpen
  \bibfield  {author} {\bibinfo {author} {\bibfnamefont {F.}~\bibnamefont
  {Liu}}, \bibinfo {author} {\bibfnamefont {H.-Y.}\ \bibnamefont {Deng}},\ and\
  \bibinfo {author} {\bibfnamefont {K.}~\bibnamefont {Wakabayashi}},\
  }\bibfield  {title} {\bibinfo {title} {Topological photonic crystals with
  zero berry curvature},\ }\href@noop {} {\bibfield  {journal} {\bibinfo
  {journal} {Physical Review B}\ }\textbf {\bibinfo {volume} {97}},\ \bibinfo
  {pages} {035442} (\bibinfo {year} {2018})}\BibitemShut {NoStop}%
\bibitem [{\citenamefont {Luo}\ \emph {et~al.}(2021)\citenamefont {Luo},
  \citenamefont {Du}, \citenamefont {Guo}, \citenamefont {Liu}, \citenamefont
  {Zhang},\ and\ \citenamefont {Guo}}]{HOTI6}%
  \BibitemOpen
  \bibfield  {author} {\bibinfo {author} {\bibfnamefont {J.}~\bibnamefont
  {Luo}}, \bibinfo {author} {\bibfnamefont {Z.}~\bibnamefont {Du}}, \bibinfo
  {author} {\bibfnamefont {Y.}~\bibnamefont {Guo}}, \bibinfo {author}
  {\bibfnamefont {C.}~\bibnamefont {Liu}}, \bibinfo {author} {\bibfnamefont
  {W.}~\bibnamefont {Zhang}},\ and\ \bibinfo {author} {\bibfnamefont
  {X.}~\bibnamefont {Guo}},\ }\bibfield  {title} {\bibinfo {title}
  {Multi-class, multi-functional design of photonic topological insulators by
  rational symmetry-indicators engineering},\ }\href@noop {} {\bibfield
  {journal} {\bibinfo  {journal} {Nanophotonics}\ } (\bibinfo {year}
  {2021})}\BibitemShut {NoStop}%
\bibitem [{\citenamefont {He}\ \emph {et~al.}(2020)\citenamefont {He},
  \citenamefont {Addison}, \citenamefont {Mele},\ and\ \citenamefont
  {Zhen}}]{Quadrupole1}%
  \BibitemOpen
  \bibfield  {author} {\bibinfo {author} {\bibfnamefont {L.}~\bibnamefont
  {He}}, \bibinfo {author} {\bibfnamefont {Z.}~\bibnamefont {Addison}},
  \bibinfo {author} {\bibfnamefont {E.~J.}\ \bibnamefont {Mele}},\ and\
  \bibinfo {author} {\bibfnamefont {B.}~\bibnamefont {Zhen}},\ }\bibfield
  {title} {\bibinfo {title} {Quadrupole topological photonic crystals},\
  }\href@noop {} {\bibfield  {journal} {\bibinfo  {journal} {Nature
  Communications}\ }\textbf {\bibinfo {volume} {11}},\ \bibinfo {pages} {1}
  (\bibinfo {year} {2020})}\BibitemShut {NoStop}%
\bibitem [{\citenamefont {Zhou}\ \emph {et~al.}(2020)\citenamefont {Zhou},
  \citenamefont {Lin}, \citenamefont {Lu}, \citenamefont {Lai}, \citenamefont
  {Hou},\ and\ \citenamefont {Jiang}}]{Quadrupole2}%
  \BibitemOpen
  \bibfield  {author} {\bibinfo {author} {\bibfnamefont {X.}~\bibnamefont
  {Zhou}}, \bibinfo {author} {\bibfnamefont {Z.-K.}\ \bibnamefont {Lin}},
  \bibinfo {author} {\bibfnamefont {W.}~\bibnamefont {Lu}}, \bibinfo {author}
  {\bibfnamefont {Y.}~\bibnamefont {Lai}}, \bibinfo {author} {\bibfnamefont
  {B.}~\bibnamefont {Hou}},\ and\ \bibinfo {author} {\bibfnamefont {J.-H.}\
  \bibnamefont {Jiang}},\ }\bibfield  {title} {\bibinfo {title} {Twisted
  quadrupole topological photonic crystals},\ }\href@noop {} {\bibfield
  {journal} {\bibinfo  {journal} {Laser \& Photonics Reviews}\ }\textbf
  {\bibinfo {volume} {14}},\ \bibinfo {pages} {2000010} (\bibinfo {year}
  {2020})}\BibitemShut {NoStop}%
\bibitem [{\citenamefont {Jin}\ \emph {et~al.}(2021)\citenamefont {Jin},
  \citenamefont {He}, \citenamefont {Lu}, \citenamefont {Mele},\ and\
  \citenamefont {Zhen}}]{FloquetQuadrupole}%
  \BibitemOpen
  \bibfield  {author} {\bibinfo {author} {\bibfnamefont {J.}~\bibnamefont
  {Jin}}, \bibinfo {author} {\bibfnamefont {L.}~\bibnamefont {He}}, \bibinfo
  {author} {\bibfnamefont {J.}~\bibnamefont {Lu}}, \bibinfo {author}
  {\bibfnamefont {E.~J.}\ \bibnamefont {Mele}},\ and\ \bibinfo {author}
  {\bibfnamefont {B.}~\bibnamefont {Zhen}},\ }\bibfield  {title} {\bibinfo
  {title} {Floquet quadrupole photonic crystals protected by space-time
  symmetry},\ }\href@noop {} {\bibfield  {journal} {\bibinfo  {journal} {arXiv
  preprint arXiv:2103.01198}\ } (\bibinfo {year} {2021})}\BibitemShut {NoStop}%
\bibitem [{\citenamefont {de~Paz}\ \emph {et~al.}(2019)\citenamefont {de~Paz},
  \citenamefont {Vergniory}, \citenamefont {Bercioux}, \citenamefont
  {Garc{\'\i}a-Etxarri},\ and\ \citenamefont {Bradlyn}}]{FragilePhC}%
  \BibitemOpen
  \bibfield  {author} {\bibinfo {author} {\bibfnamefont {M.~B.}\ \bibnamefont
  {de~Paz}}, \bibinfo {author} {\bibfnamefont {M.~G.}\ \bibnamefont
  {Vergniory}}, \bibinfo {author} {\bibfnamefont {D.}~\bibnamefont {Bercioux}},
  \bibinfo {author} {\bibfnamefont {A.}~\bibnamefont {Garc{\'\i}a-Etxarri}},\
  and\ \bibinfo {author} {\bibfnamefont {B.}~\bibnamefont {Bradlyn}},\
  }\bibfield  {title} {\bibinfo {title} {Engineering fragile topology in
  photonic crystals: Topological quantum chemistry of light},\ }\href@noop {}
  {\bibfield  {journal} {\bibinfo  {journal} {Physical Review Research}\
  }\textbf {\bibinfo {volume} {1}},\ \bibinfo {pages} {032005} (\bibinfo {year}
  {2019})}\BibitemShut {NoStop}%
\bibitem [{\citenamefont {Han}\ \emph {et~al.}(2020)\citenamefont {Han},
  \citenamefont {Kang},\ and\ \citenamefont {Jeon}}]{HOTI_Laser1}%
  \BibitemOpen
  \bibfield  {author} {\bibinfo {author} {\bibfnamefont {C.}~\bibnamefont
  {Han}}, \bibinfo {author} {\bibfnamefont {M.}~\bibnamefont {Kang}},\ and\
  \bibinfo {author} {\bibfnamefont {H.}~\bibnamefont {Jeon}},\ }\bibfield
  {title} {\bibinfo {title} {Lasing at multidimensional topological states in a
  two-dimensional photonic crystal structure},\ }\href@noop {} {\bibfield
  {journal} {\bibinfo  {journal} {ACS Photonics}\ }\textbf {\bibinfo {volume}
  {7}},\ \bibinfo {pages} {2027} (\bibinfo {year} {2020})}\BibitemShut
  {NoStop}%
\bibitem [{\citenamefont {Kim}\ \emph {et~al.}(2020{\natexlab{b}})\citenamefont
  {Kim}, \citenamefont {Hwang}, \citenamefont {Smirnova}, \citenamefont
  {Jeong}, \citenamefont {Kivshar},\ and\ \citenamefont {Park}}]{HOTI_Laser2}%
  \BibitemOpen
  \bibfield  {author} {\bibinfo {author} {\bibfnamefont {H.-R.}\ \bibnamefont
  {Kim}}, \bibinfo {author} {\bibfnamefont {M.-S.}\ \bibnamefont {Hwang}},
  \bibinfo {author} {\bibfnamefont {D.}~\bibnamefont {Smirnova}}, \bibinfo
  {author} {\bibfnamefont {K.-Y.}\ \bibnamefont {Jeong}}, \bibinfo {author}
  {\bibfnamefont {Y.}~\bibnamefont {Kivshar}},\ and\ \bibinfo {author}
  {\bibfnamefont {H.-G.}\ \bibnamefont {Park}},\ }\bibfield  {title} {\bibinfo
  {title} {Multipolar lasing modes from topological corner states},\
  }\href@noop {} {\bibfield  {journal} {\bibinfo  {journal} {Nature
  Communications}\ }\textbf {\bibinfo {volume} {11}},\ \bibinfo {pages} {1}
  (\bibinfo {year} {2020}{\natexlab{b}})}\BibitemShut {NoStop}%
\bibitem [{\citenamefont {Bahari}\ \emph {et~al.}(2017)\citenamefont {Bahari},
  \citenamefont {Ndao}, \citenamefont {Vallini}, \citenamefont {El~Amili},
  \citenamefont {Fainman},\ and\ \citenamefont {Kant{\'e}}}]{PhC_topo_laser1}%
  \BibitemOpen
  \bibfield  {author} {\bibinfo {author} {\bibfnamefont {B.}~\bibnamefont
  {Bahari}}, \bibinfo {author} {\bibfnamefont {A.}~\bibnamefont {Ndao}},
  \bibinfo {author} {\bibfnamefont {F.}~\bibnamefont {Vallini}}, \bibinfo
  {author} {\bibfnamefont {A.}~\bibnamefont {El~Amili}}, \bibinfo {author}
  {\bibfnamefont {Y.}~\bibnamefont {Fainman}},\ and\ \bibinfo {author}
  {\bibfnamefont {B.}~\bibnamefont {Kant{\'e}}},\ }\bibfield  {title} {\bibinfo
  {title} {Nonreciprocal lasing in topological cavities of arbitrary
  geometries},\ }\href@noop {} {\bibfield  {journal} {\bibinfo  {journal}
  {Science}\ }\textbf {\bibinfo {volume} {358}},\ \bibinfo {pages} {636}
  (\bibinfo {year} {2017})}\BibitemShut {NoStop}%
\bibitem [{\citenamefont {Chen}\ \emph
  {et~al.}(2021{\natexlab{a}})\citenamefont {Chen}, \citenamefont {Lan},
  \citenamefont {Li},\ and\ \citenamefont {Zhu}}]{SHG}%
  \BibitemOpen
  \bibfield  {author} {\bibinfo {author} {\bibfnamefont {Y.}~\bibnamefont
  {Chen}}, \bibinfo {author} {\bibfnamefont {Z.}~\bibnamefont {Lan}}, \bibinfo
  {author} {\bibfnamefont {J.}~\bibnamefont {Li}},\ and\ \bibinfo {author}
  {\bibfnamefont {J.}~\bibnamefont {Zhu}},\ }\bibfield  {title} {\bibinfo
  {title} {Topologically protected second harmonic generation via doubly
  resonant high-order photonic modes},\ }\href@noop {} {\bibfield  {journal}
  {\bibinfo  {journal} {Physical Review B}\ }\textbf {\bibinfo {volume}
  {104}},\ \bibinfo {pages} {155421} (\bibinfo {year}
  {2021}{\natexlab{a}})}\BibitemShut {NoStop}%
\bibitem [{\citenamefont {Smirnova}\ \emph {et~al.}(2019)\citenamefont
  {Smirnova}, \citenamefont {Kruk}, \citenamefont {Leykam}, \citenamefont
  {Melik-Gaykazyan}, \citenamefont {Choi},\ and\ \citenamefont
  {Kivshar}}]{THG}%
  \BibitemOpen
  \bibfield  {author} {\bibinfo {author} {\bibfnamefont {D.}~\bibnamefont
  {Smirnova}}, \bibinfo {author} {\bibfnamefont {S.}~\bibnamefont {Kruk}},
  \bibinfo {author} {\bibfnamefont {D.}~\bibnamefont {Leykam}}, \bibinfo
  {author} {\bibfnamefont {E.}~\bibnamefont {Melik-Gaykazyan}}, \bibinfo
  {author} {\bibfnamefont {D.-Y.}\ \bibnamefont {Choi}},\ and\ \bibinfo
  {author} {\bibfnamefont {Y.}~\bibnamefont {Kivshar}},\ }\bibfield  {title}
  {\bibinfo {title} {Third-harmonic generation in photonic topological
  metasurfaces},\ }\href@noop {} {\bibfield  {journal} {\bibinfo  {journal}
  {Physical Review Letters}\ }\textbf {\bibinfo {volume} {123}},\ \bibinfo
  {pages} {103901} (\bibinfo {year} {2019})}\BibitemShut {NoStop}%
\bibitem [{\citenamefont {Iwamoto}\ \emph {et~al.}(2021)\citenamefont
  {Iwamoto}, \citenamefont {Ota},\ and\ \citenamefont
  {Arakawa}}]{TopoTransport1}%
  \BibitemOpen
  \bibfield  {author} {\bibinfo {author} {\bibfnamefont {S.}~\bibnamefont
  {Iwamoto}}, \bibinfo {author} {\bibfnamefont {Y.}~\bibnamefont {Ota}},\ and\
  \bibinfo {author} {\bibfnamefont {Y.}~\bibnamefont {Arakawa}},\ }\bibfield
  {title} {\bibinfo {title} {Recent progress in topological waveguides and
  nanocavities in a semiconductor photonic crystal platform},\ }\href@noop {}
  {\bibfield  {journal} {\bibinfo  {journal} {Optical Materials Express}\
  }\textbf {\bibinfo {volume} {11}},\ \bibinfo {pages} {319} (\bibinfo {year}
  {2021})}\BibitemShut {NoStop}%
\bibitem [{\citenamefont {Lan}\ \emph {et~al.}(2020)\citenamefont {Lan},
  \citenamefont {You},\ and\ \citenamefont {Panoiu}}]{NL_PhC1}%
  \BibitemOpen
  \bibfield  {author} {\bibinfo {author} {\bibfnamefont {Z.}~\bibnamefont
  {Lan}}, \bibinfo {author} {\bibfnamefont {J.~W.}\ \bibnamefont {You}},\ and\
  \bibinfo {author} {\bibfnamefont {N.~C.}\ \bibnamefont {Panoiu}},\ }\bibfield
   {title} {\bibinfo {title} {Nonlinear one-way edge-mode interactions for
  frequency mixing in topological photonic crystals},\ }\href@noop {}
  {\bibfield  {journal} {\bibinfo  {journal} {Physical Review B}\ }\textbf
  {\bibinfo {volume} {101}},\ \bibinfo {pages} {155422} (\bibinfo {year}
  {2020})}\BibitemShut {NoStop}%
\bibitem [{\citenamefont {Shalaev}\ \emph
  {et~al.}(2019{\natexlab{b}})\citenamefont {Shalaev}, \citenamefont
  {Walasik},\ and\ \citenamefont {Litchinitser}}]{NL_PhC2}%
  \BibitemOpen
  \bibfield  {author} {\bibinfo {author} {\bibfnamefont {M.~I.}\ \bibnamefont
  {Shalaev}}, \bibinfo {author} {\bibfnamefont {W.}~\bibnamefont {Walasik}},\
  and\ \bibinfo {author} {\bibfnamefont {N.~M.}\ \bibnamefont {Litchinitser}},\
  }\bibfield  {title} {\bibinfo {title} {Optically tunable topological photonic
  crystal},\ }\href@noop {} {\bibfield  {journal} {\bibinfo  {journal}
  {Optica}\ }\textbf {\bibinfo {volume} {6}},\ \bibinfo {pages} {839} (\bibinfo
  {year} {2019}{\natexlab{b}})}\BibitemShut {NoStop}%
\bibitem [{\citenamefont {Zhong}\ \emph {et~al.}(2021)\citenamefont {Zhong},
  \citenamefont {Wang}, \citenamefont {Park}, \citenamefont {Asadchy},
  \citenamefont {Wojcik}, \citenamefont {Dutt},\ and\ \citenamefont
  {Fan}}]{NH_PhC1}%
  \BibitemOpen
  \bibfield  {author} {\bibinfo {author} {\bibfnamefont {J.}~\bibnamefont
  {Zhong}}, \bibinfo {author} {\bibfnamefont {K.}~\bibnamefont {Wang}},
  \bibinfo {author} {\bibfnamefont {Y.}~\bibnamefont {Park}}, \bibinfo {author}
  {\bibfnamefont {V.}~\bibnamefont {Asadchy}}, \bibinfo {author} {\bibfnamefont
  {C.~C.}\ \bibnamefont {Wojcik}}, \bibinfo {author} {\bibfnamefont
  {A.}~\bibnamefont {Dutt}},\ and\ \bibinfo {author} {\bibfnamefont
  {S.}~\bibnamefont {Fan}},\ }\bibfield  {title} {\bibinfo {title} {Nontrivial
  point-gap topology and non-hermitian skin effect in photonic crystals},\
  }\href@noop {} {\bibfield  {journal} {\bibinfo  {journal} {Physical Review
  B}\ }\textbf {\bibinfo {volume} {104}},\ \bibinfo {pages} {125416} (\bibinfo
  {year} {2021})}\BibitemShut {NoStop}%
\bibitem [{\citenamefont {Chen}\ \emph
  {et~al.}(2021{\natexlab{b}})\citenamefont {Chen}, \citenamefont {Jiang},
  \citenamefont {Zhang}, \citenamefont {Zhao}, \citenamefont {Lan},\ and\
  \citenamefont {Wei}}]{NH_PhC2}%
  \BibitemOpen
  \bibfield  {author} {\bibinfo {author} {\bibfnamefont {M.~L.}\ \bibnamefont
  {Chen}}, \bibinfo {author} {\bibfnamefont {L.~J.}\ \bibnamefont {Jiang}},
  \bibinfo {author} {\bibfnamefont {S.}~\bibnamefont {Zhang}}, \bibinfo
  {author} {\bibfnamefont {R.}~\bibnamefont {Zhao}}, \bibinfo {author}
  {\bibfnamefont {Z.}~\bibnamefont {Lan}},\ and\ \bibinfo {author}
  {\bibfnamefont {E.}~\bibnamefont {Wei}},\ }\bibfield  {title} {\bibinfo
  {title} {Comparative study of hermitian and non-hermitian topological
  dielectric photonic crystals},\ }\href@noop {} {\bibfield  {journal}
  {\bibinfo  {journal} {Physical Review A}\ }\textbf {\bibinfo {volume}
  {104}},\ \bibinfo {pages} {033501} (\bibinfo {year}
  {2021}{\natexlab{b}})}\BibitemShut {NoStop}%
\bibitem [{\citenamefont {Ryu}\ \emph {et~al.}(2010)\citenamefont {Ryu},
  \citenamefont {Schnyder}, \citenamefont {Furusaki},\ and\ \citenamefont
  {Ludwig}}]{10foldway1}%
  \BibitemOpen
  \bibfield  {author} {\bibinfo {author} {\bibfnamefont {S.}~\bibnamefont
  {Ryu}}, \bibinfo {author} {\bibfnamefont {A.~P.}\ \bibnamefont {Schnyder}},
  \bibinfo {author} {\bibfnamefont {A.}~\bibnamefont {Furusaki}},\ and\
  \bibinfo {author} {\bibfnamefont {A.~W.}\ \bibnamefont {Ludwig}},\ }\bibfield
   {title} {\bibinfo {title} {Topological insulators and superconductors:
  tenfold way and dimensional hierarchy},\ }\href@noop {} {\bibfield  {journal}
  {\bibinfo  {journal} {New Journal of Physics}\ }\textbf {\bibinfo {volume}
  {12}},\ \bibinfo {pages} {065010} (\bibinfo {year} {2010})}\BibitemShut
  {NoStop}%
\bibitem [{\citenamefont {Chiu}\ \emph {et~al.}(2016)\citenamefont {Chiu},
  \citenamefont {Teo}, \citenamefont {Schnyder},\ and\ \citenamefont
  {Ryu}}]{10foldway2}%
  \BibitemOpen
  \bibfield  {author} {\bibinfo {author} {\bibfnamefont {C.-K.}\ \bibnamefont
  {Chiu}}, \bibinfo {author} {\bibfnamefont {J.~C.}\ \bibnamefont {Teo}},
  \bibinfo {author} {\bibfnamefont {A.~P.}\ \bibnamefont {Schnyder}},\ and\
  \bibinfo {author} {\bibfnamefont {S.}~\bibnamefont {Ryu}},\ }\bibfield
  {title} {\bibinfo {title} {Classification of topological quantum matter with
  symmetries},\ }\href@noop {} {\bibfield  {journal} {\bibinfo  {journal}
  {Reviews of Modern Physics}\ }\textbf {\bibinfo {volume} {88}},\ \bibinfo
  {pages} {035005} (\bibinfo {year} {2016})}\BibitemShut {NoStop}%
\bibitem [{\citenamefont {Kitaev}(2009)}]{kitaev2009periodic}%
  \BibitemOpen
  \bibfield  {author} {\bibinfo {author} {\bibfnamefont {A.}~\bibnamefont
  {Kitaev}},\ }\bibfield  {title} {\bibinfo {title} {Periodic table for
  topological insulators and superconductors},\ }in\ \href@noop {} {\emph
  {\bibinfo {booktitle} {AIP conference proceedings}}},\ Vol.\ \bibinfo
  {volume} {1134}\ (\bibinfo {organization} {American Institute of Physics},\
  \bibinfo {year} {2009})\ pp.\ \bibinfo {pages} {22--30}\BibitemShut {NoStop}%
\bibitem [{\citenamefont {Bradlyn}\ \emph
  {et~al.}(2017{\natexlab{a}})\citenamefont {Bradlyn}, \citenamefont {Elcoro},
  \citenamefont {Cano}, \citenamefont {Vergniory}, \citenamefont {Wang},
  \citenamefont {Felser}, \citenamefont {Aroyo},\ and\ \citenamefont
  {Bernevig}}]{TQC1}%
  \BibitemOpen
  \bibfield  {author} {\bibinfo {author} {\bibfnamefont {B.}~\bibnamefont
  {Bradlyn}}, \bibinfo {author} {\bibfnamefont {L.}~\bibnamefont {Elcoro}},
  \bibinfo {author} {\bibfnamefont {J.}~\bibnamefont {Cano}}, \bibinfo {author}
  {\bibfnamefont {M.}~\bibnamefont {Vergniory}}, \bibinfo {author}
  {\bibfnamefont {Z.}~\bibnamefont {Wang}}, \bibinfo {author} {\bibfnamefont
  {C.}~\bibnamefont {Felser}}, \bibinfo {author} {\bibfnamefont {M.~I.}\
  \bibnamefont {Aroyo}},\ and\ \bibinfo {author} {\bibfnamefont {B.~A.}\
  \bibnamefont {Bernevig}},\ }\bibfield  {title} {\bibinfo {title} {Topological
  quantum chemistry},\ }\href@noop {} {\bibfield  {journal} {\bibinfo
  {journal} {Nature}\ }\textbf {\bibinfo {volume} {547}},\ \bibinfo {pages}
  {298} (\bibinfo {year} {2017}{\natexlab{a}})}\BibitemShut {NoStop}%
\bibitem [{\citenamefont {Marzari}\ \emph {et~al.}(2012)\citenamefont
  {Marzari}, \citenamefont {Mostofi}, \citenamefont {Yates}, \citenamefont
  {Souza},\ and\ \citenamefont {Vanderbilt}}]{MLWF}%
  \BibitemOpen
  \bibfield  {author} {\bibinfo {author} {\bibfnamefont {N.}~\bibnamefont
  {Marzari}}, \bibinfo {author} {\bibfnamefont {A.~A.}\ \bibnamefont
  {Mostofi}}, \bibinfo {author} {\bibfnamefont {J.~R.}\ \bibnamefont {Yates}},
  \bibinfo {author} {\bibfnamefont {I.}~\bibnamefont {Souza}},\ and\ \bibinfo
  {author} {\bibfnamefont {D.}~\bibnamefont {Vanderbilt}},\ }\bibfield  {title}
  {\bibinfo {title} {Maximally localized wannier functions: Theory and
  applications},\ }\href@noop {} {\bibfield  {journal} {\bibinfo  {journal}
  {Reviews of Modern Physics}\ }\textbf {\bibinfo {volume} {84}},\ \bibinfo
  {pages} {1419} (\bibinfo {year} {2012})}\BibitemShut {NoStop}%
\bibitem [{\citenamefont {Po}\ \emph {et~al.}(2018)\citenamefont {Po},
  \citenamefont {Watanabe},\ and\ \citenamefont {Vishwanath}}]{Fragile1}%
  \BibitemOpen
  \bibfield  {author} {\bibinfo {author} {\bibfnamefont {H.~C.}\ \bibnamefont
  {Po}}, \bibinfo {author} {\bibfnamefont {H.}~\bibnamefont {Watanabe}},\ and\
  \bibinfo {author} {\bibfnamefont {A.}~\bibnamefont {Vishwanath}},\ }\bibfield
   {title} {\bibinfo {title} {Fragile topology and wannier obstructions},\
  }\href@noop {} {\bibfield  {journal} {\bibinfo  {journal} {Physical Review
  Letters}\ }\textbf {\bibinfo {volume} {121}},\ \bibinfo {pages} {126402}
  (\bibinfo {year} {2018})}\BibitemShut {NoStop}%
\bibitem [{\citenamefont {Song}\ \emph {et~al.}(2020)\citenamefont {Song},
  \citenamefont {Elcoro},\ and\ \citenamefont {Bernevig}}]{Fragile2}%
  \BibitemOpen
  \bibfield  {author} {\bibinfo {author} {\bibfnamefont {Z.-D.}\ \bibnamefont
  {Song}}, \bibinfo {author} {\bibfnamefont {L.}~\bibnamefont {Elcoro}},\ and\
  \bibinfo {author} {\bibfnamefont {B.~A.}\ \bibnamefont {Bernevig}},\
  }\bibfield  {title} {\bibinfo {title} {Twisted bulk-boundary correspondence
  of fragile topology},\ }\href@noop {} {\bibfield  {journal} {\bibinfo
  {journal} {Science}\ }\textbf {\bibinfo {volume} {367}},\ \bibinfo {pages}
  {794} (\bibinfo {year} {2020})}\BibitemShut {NoStop}%
\bibitem [{\citenamefont {Benalcazar}\ \emph {et~al.}(2019)\citenamefont
  {Benalcazar}, \citenamefont {Li},\ and\ \citenamefont {Hughes}}]{Rot_HOTI1}%
  \BibitemOpen
  \bibfield  {author} {\bibinfo {author} {\bibfnamefont {W.~A.}\ \bibnamefont
  {Benalcazar}}, \bibinfo {author} {\bibfnamefont {T.}~\bibnamefont {Li}},\
  and\ \bibinfo {author} {\bibfnamefont {T.~L.}\ \bibnamefont {Hughes}},\
  }\bibfield  {title} {\bibinfo {title} {Quantization of fractional corner
  charge in $c_n$-symmetric higher-order topological crystalline insulators},\
  }\href@noop {} {\bibfield  {journal} {\bibinfo  {journal} {Physical Review
  B}\ }\textbf {\bibinfo {volume} {99}},\ \bibinfo {pages} {245151} (\bibinfo
  {year} {2019})}\BibitemShut {NoStop}%
\bibitem [{\citenamefont {Po}\ \emph {et~al.}(2017)\citenamefont {Po},
  \citenamefont {Vishwanath},\ and\ \citenamefont
  {Watanabe}}]{symmetry_indicators_po}%
  \BibitemOpen
  \bibfield  {author} {\bibinfo {author} {\bibfnamefont {H.~C.}\ \bibnamefont
  {Po}}, \bibinfo {author} {\bibfnamefont {A.}~\bibnamefont {Vishwanath}},\
  and\ \bibinfo {author} {\bibfnamefont {H.}~\bibnamefont {Watanabe}},\
  }\bibfield  {title} {\bibinfo {title} {Symmetry-based indicators of band
  topology in the 230 space groups},\ }\href@noop {} {\bibfield  {journal}
  {\bibinfo  {journal} {Nature Communications}\ }\textbf {\bibinfo {volume}
  {8}},\ \bibinfo {pages} {1} (\bibinfo {year} {2017})}\BibitemShut {NoStop}%
\bibitem [{\citenamefont {Benalcazar}\ \emph {et~al.}(2014)\citenamefont
  {Benalcazar}, \citenamefont {Teo},\ and\ \citenamefont
  {Hughes}}]{benalcazar2014}%
  \BibitemOpen
  \bibfield  {author} {\bibinfo {author} {\bibfnamefont {W.~A.}\ \bibnamefont
  {Benalcazar}}, \bibinfo {author} {\bibfnamefont {J.~C.~Y.}\ \bibnamefont
  {Teo}},\ and\ \bibinfo {author} {\bibfnamefont {T.~L.}\ \bibnamefont
  {Hughes}},\ }\bibfield  {title} {\bibinfo {title} {Classification of
  two-dimensional topological crystalline superconductors and majorana bound
  states at disclinations},\ }\href
  {https://doi.org/10.1103/PhysRevB.89.224503} {\bibfield  {journal} {\bibinfo
  {journal} {Phys. Rev. B}\ }\textbf {\bibinfo {volume} {89}},\ \bibinfo
  {pages} {224503} (\bibinfo {year} {2014})}\BibitemShut {NoStop}%
\bibitem [{\citenamefont {Cano}\ \emph {et~al.}(2018)\citenamefont {Cano},
  \citenamefont {Bradlyn}, \citenamefont {Wang}, \citenamefont {Elcoro},
  \citenamefont {Vergniory}, \citenamefont {Felser}, \citenamefont {Aroyo},\
  and\ \citenamefont {Bernevig}}]{canoEBRs}%
  \BibitemOpen
  \bibfield  {author} {\bibinfo {author} {\bibfnamefont {J.}~\bibnamefont
  {Cano}}, \bibinfo {author} {\bibfnamefont {B.}~\bibnamefont {Bradlyn}},
  \bibinfo {author} {\bibfnamefont {Z.}~\bibnamefont {Wang}}, \bibinfo {author}
  {\bibfnamefont {L.}~\bibnamefont {Elcoro}}, \bibinfo {author} {\bibfnamefont
  {M.}~\bibnamefont {Vergniory}}, \bibinfo {author} {\bibfnamefont
  {C.}~\bibnamefont {Felser}}, \bibinfo {author} {\bibfnamefont {M.~I.}\
  \bibnamefont {Aroyo}},\ and\ \bibinfo {author} {\bibfnamefont {B.~A.}\
  \bibnamefont {Bernevig}},\ }\bibfield  {title} {\bibinfo {title} {Building
  blocks of topological quantum chemistry: Elementary band representations},\
  }\href@noop {} {\bibfield  {journal} {\bibinfo  {journal} {Physical Review
  B}\ }\textbf {\bibinfo {volume} {97}},\ \bibinfo {pages} {035139} (\bibinfo
  {year} {2018})}\BibitemShut {NoStop}%
\bibitem [{\citenamefont {Bradlyn}\ \emph
  {et~al.}(2017{\natexlab{b}})\citenamefont {Bradlyn}, \citenamefont {Elcoro},
  \citenamefont {Cano}, \citenamefont {Vergniory}, \citenamefont {Wang},
  \citenamefont {Felser}, \citenamefont {Aroyo},\ and\ \citenamefont
  {Bernevig}}]{bradlyn2017}%
  \BibitemOpen
  \bibfield  {author} {\bibinfo {author} {\bibfnamefont {B.}~\bibnamefont
  {Bradlyn}}, \bibinfo {author} {\bibfnamefont {L.}~\bibnamefont {Elcoro}},
  \bibinfo {author} {\bibfnamefont {J.}~\bibnamefont {Cano}}, \bibinfo {author}
  {\bibfnamefont {M.~G.}\ \bibnamefont {Vergniory}}, \bibinfo {author}
  {\bibfnamefont {Z.}~\bibnamefont {Wang}}, \bibinfo {author} {\bibfnamefont
  {C.}~\bibnamefont {Felser}}, \bibinfo {author} {\bibfnamefont {M.~I.}\
  \bibnamefont {Aroyo}},\ and\ \bibinfo {author} {\bibfnamefont {B.~A.}\
  \bibnamefont {Bernevig}},\ }\bibfield  {title} {\bibinfo {title} {Topological
  quantum chemistry},\ }\href {http://dx.doi.org/10.1038/nature23268}
  {\bibfield  {journal} {\bibinfo  {journal} {Nature}\ }\textbf {\bibinfo
  {volume} {547}},\ \bibinfo {pages} {298 EP } (\bibinfo {year}
  {2017}{\natexlab{b}})}\BibitemShut {NoStop}%
\bibitem [{\citenamefont {Kohn}(1959)}]{kohn_WF}%
  \BibitemOpen
  \bibfield  {author} {\bibinfo {author} {\bibfnamefont {W.}~\bibnamefont
  {Kohn}},\ }\bibfield  {title} {\bibinfo {title} {Analytic properties of bloch
  waves and wannier functions},\ }\href@noop {} {\bibfield  {journal} {\bibinfo
   {journal} {Physical Review}\ }\textbf {\bibinfo {volume} {115}},\ \bibinfo
  {pages} {809} (\bibinfo {year} {1959})}\BibitemShut {NoStop}%
\bibitem [{\citenamefont {Tanaue}\ and\ \citenamefont
  {Bruno-Alfonso}(2020)}]{Wannierm1}%
  \BibitemOpen
  \bibfield  {author} {\bibinfo {author} {\bibfnamefont {H.~B.}\ \bibnamefont
  {Tanaue}}\ and\ \bibinfo {author} {\bibfnamefont {A.}~\bibnamefont
  {Bruno-Alfonso}},\ }\bibfield  {title} {\bibinfo {title} {Wannier-function
  expansion of localized modes in 1d photonic crystals without inversion
  symmetry},\ }\href@noop {} {\bibfield  {journal} {\bibinfo  {journal} {JOSA
  B}\ }\textbf {\bibinfo {volume} {37}},\ \bibinfo {pages} {3698} (\bibinfo
  {year} {2020})}\BibitemShut {NoStop}%
\bibitem [{\citenamefont {Romano}\ \emph {et~al.}(2010)\citenamefont {Romano},
  \citenamefont {Nacbar},\ and\ \citenamefont {Bruno-Alfonso}}]{Wannier0}%
  \BibitemOpen
  \bibfield  {author} {\bibinfo {author} {\bibfnamefont {M.~C.}\ \bibnamefont
  {Romano}}, \bibinfo {author} {\bibfnamefont {D.~R.}\ \bibnamefont {Nacbar}},\
  and\ \bibinfo {author} {\bibfnamefont {A.}~\bibnamefont {Bruno-Alfonso}},\
  }\bibfield  {title} {\bibinfo {title} {Wannier functions of a one-dimensional
  photonic crystal with inversion symmetry},\ }\href@noop {} {\bibfield
  {journal} {\bibinfo  {journal} {Journal of Physics B: Atomic, Molecular and
  Optical Physics}\ }\textbf {\bibinfo {volume} {43}},\ \bibinfo {pages}
  {215403} (\bibinfo {year} {2010})}\BibitemShut {NoStop}%
\bibitem [{\citenamefont {Busch}\ \emph {et~al.}(2003)\citenamefont {Busch},
  \citenamefont {Mingaleev}, \citenamefont {Garcia-Martin}, \citenamefont
  {Schillinger},\ and\ \citenamefont {Hermann}}]{PhCWannier1}%
  \BibitemOpen
  \bibfield  {author} {\bibinfo {author} {\bibfnamefont {K.}~\bibnamefont
  {Busch}}, \bibinfo {author} {\bibfnamefont {S.~F.}\ \bibnamefont
  {Mingaleev}}, \bibinfo {author} {\bibfnamefont {A.}~\bibnamefont
  {Garcia-Martin}}, \bibinfo {author} {\bibfnamefont {M.}~\bibnamefont
  {Schillinger}},\ and\ \bibinfo {author} {\bibfnamefont {D.}~\bibnamefont
  {Hermann}},\ }\bibfield  {title} {\bibinfo {title} {The wannier function
  approach to photonic crystal circuits},\ }\href@noop {} {\bibfield  {journal}
  {\bibinfo  {journal} {Journal of Physics: Condensed Matter}\ }\textbf
  {\bibinfo {volume} {15}},\ \bibinfo {pages} {R1233} (\bibinfo {year}
  {2003})}\BibitemShut {NoStop}%
\bibitem [{\citenamefont {Romano}\ \emph {et~al.}(2018)\citenamefont {Romano},
  \citenamefont {Vellasco-Gomes},\ and\ \citenamefont
  {Bruno-Alfonso}}]{PhCWannier2}%
  \BibitemOpen
  \bibfield  {author} {\bibinfo {author} {\bibfnamefont {M.~C.}\ \bibnamefont
  {Romano}}, \bibinfo {author} {\bibfnamefont {A.}~\bibnamefont
  {Vellasco-Gomes}},\ and\ \bibinfo {author} {\bibfnamefont {A.}~\bibnamefont
  {Bruno-Alfonso}},\ }\bibfield  {title} {\bibinfo {title} {Wannier functions
  and the calculation of localized modes in one-dimensional photonic
  crystals},\ }\href@noop {} {\bibfield  {journal} {\bibinfo  {journal} {JOSA
  B}\ }\textbf {\bibinfo {volume} {35}},\ \bibinfo {pages} {826} (\bibinfo
  {year} {2018})}\BibitemShut {NoStop}%
\bibitem [{\citenamefont {Stollenwerk}\ \emph {et~al.}(2011)\citenamefont
  {Stollenwerk}, \citenamefont {Chigrin},\ and\ \citenamefont
  {Kroha}}]{PhCWannier3}%
  \BibitemOpen
  \bibfield  {author} {\bibinfo {author} {\bibfnamefont {T.}~\bibnamefont
  {Stollenwerk}}, \bibinfo {author} {\bibfnamefont {D.~N.}\ \bibnamefont
  {Chigrin}},\ and\ \bibinfo {author} {\bibfnamefont {J.}~\bibnamefont
  {Kroha}},\ }\bibfield  {title} {\bibinfo {title} {Efficient construction of
  maximally localized photonic wannier functions: locality criterion and
  initial conditions},\ }\href@noop {} {\bibfield  {journal} {\bibinfo
  {journal} {JOSA B}\ }\textbf {\bibinfo {volume} {28}},\ \bibinfo {pages}
  {1951} (\bibinfo {year} {2011})}\BibitemShut {NoStop}%
\bibitem [{\citenamefont {Gupta}\ and\ \citenamefont
  {Bradlyn}(2022)}]{PhCWannier4}%
  \BibitemOpen
  \bibfield  {author} {\bibinfo {author} {\bibfnamefont {V.}~\bibnamefont
  {Gupta}}\ and\ \bibinfo {author} {\bibfnamefont {B.}~\bibnamefont
  {Bradlyn}},\ }\bibfield  {title} {\bibinfo {title} {Wannier-function methods
  for topological modes in one-dimensional photonic crystals},\ }\href@noop {}
  {\bibfield  {journal} {\bibinfo  {journal} {Physical Review A}\ }\textbf
  {\bibinfo {volume} {105}},\ \bibinfo {pages} {053521} (\bibinfo {year}
  {2022})}\BibitemShut {NoStop}%
\bibitem [{\citenamefont {Alexandradinata}\ \emph {et~al.}(2014)\citenamefont
  {Alexandradinata}, \citenamefont {Dai},\ and\ \citenamefont
  {Bernevig}}]{Wilson_loop_inversionsym}%
  \BibitemOpen
  \bibfield  {author} {\bibinfo {author} {\bibfnamefont {A.}~\bibnamefont
  {Alexandradinata}}, \bibinfo {author} {\bibfnamefont {X.}~\bibnamefont
  {Dai}},\ and\ \bibinfo {author} {\bibfnamefont {B.~A.}\ \bibnamefont
  {Bernevig}},\ }\bibfield  {title} {\bibinfo {title} {Wilson-loop
  characterization of inversion-symmetric topological insulators},\ }\href@noop
  {} {\bibfield  {journal} {\bibinfo  {journal} {Physical Review B}\ }\textbf
  {\bibinfo {volume} {89}},\ \bibinfo {pages} {155114} (\bibinfo {year}
  {2014})}\BibitemShut {NoStop}%
\bibitem [{\citenamefont {Hughes}\ \emph {et~al.}(2011)\citenamefont {Hughes},
  \citenamefont {Prodan},\ and\ \citenamefont
  {Bernevig}}]{Inv_sym_topo_insulators}%
  \BibitemOpen
  \bibfield  {author} {\bibinfo {author} {\bibfnamefont {T.~L.}\ \bibnamefont
  {Hughes}}, \bibinfo {author} {\bibfnamefont {E.}~\bibnamefont {Prodan}},\
  and\ \bibinfo {author} {\bibfnamefont {B.~A.}\ \bibnamefont {Bernevig}},\
  }\bibfield  {title} {\bibinfo {title} {Inversion-symmetric topological
  insulators},\ }\href {https://doi.org/10.1103/PhysRevB.83.245132} {\bibfield
  {journal} {\bibinfo  {journal} {Phys. Rev. B}\ }\textbf {\bibinfo {volume}
  {83}},\ \bibinfo {pages} {245132} (\bibinfo {year} {2011})}\BibitemShut
  {NoStop}%
\bibitem [{\citenamefont {Fang}\ and\ \citenamefont {Cano}(2021)}]{Rot_HOTI2}%
  \BibitemOpen
  \bibfield  {author} {\bibinfo {author} {\bibfnamefont {Y.}~\bibnamefont
  {Fang}}\ and\ \bibinfo {author} {\bibfnamefont {J.}~\bibnamefont {Cano}},\
  }\bibfield  {title} {\bibinfo {title} {Filling anomaly for general two-and
  three-dimensional $c_4$ symmetric lattices},\ }\href@noop {} {\bibfield
  {journal} {\bibinfo  {journal} {Physical Review B}\ }\textbf {\bibinfo
  {volume} {103}},\ \bibinfo {pages} {165109} (\bibinfo {year}
  {2021})}\BibitemShut {NoStop}%
\bibitem [{\citenamefont {Peterson}\ \emph {et~al.}(2020)\citenamefont
  {Peterson}, \citenamefont {Li}, \citenamefont {Benalcazar}, \citenamefont
  {Hughes},\ and\ \citenamefont {Bahl}}]{filling_anomaly}%
  \BibitemOpen
  \bibfield  {author} {\bibinfo {author} {\bibfnamefont {C.~W.}\ \bibnamefont
  {Peterson}}, \bibinfo {author} {\bibfnamefont {T.}~\bibnamefont {Li}},
  \bibinfo {author} {\bibfnamefont {W.~A.}\ \bibnamefont {Benalcazar}},
  \bibinfo {author} {\bibfnamefont {T.~L.}\ \bibnamefont {Hughes}},\ and\
  \bibinfo {author} {\bibfnamefont {G.}~\bibnamefont {Bahl}},\ }\bibfield
  {title} {\bibinfo {title} {A fractional corner anomaly reveals higher-order
  topology},\ }\href@noop {} {\bibfield  {journal} {\bibinfo  {journal}
  {Science}\ }\textbf {\bibinfo {volume} {368}},\ \bibinfo {pages} {1114}
  (\bibinfo {year} {2020})}\BibitemShut {NoStop}%
\bibitem [{\citenamefont {Song}\ \emph {et~al.}(2017)\citenamefont {Song},
  \citenamefont {Fang},\ and\ \citenamefont {Fang}}]{Counting_mismatch}%
  \BibitemOpen
  \bibfield  {author} {\bibinfo {author} {\bibfnamefont {Z.}~\bibnamefont
  {Song}}, \bibinfo {author} {\bibfnamefont {Z.}~\bibnamefont {Fang}},\ and\
  \bibinfo {author} {\bibfnamefont {C.}~\bibnamefont {Fang}},\ }\bibfield
  {title} {\bibinfo {title} {(d- 2)-dimensional edge states of rotation
  symmetry protected topological states},\ }\href@noop {} {\bibfield  {journal}
  {\bibinfo  {journal} {Physical Review Letters}\ }\textbf {\bibinfo {volume}
  {119}},\ \bibinfo {pages} {246402} (\bibinfo {year} {2017})}\BibitemShut
  {NoStop}%
\bibitem [{\citenamefont {Vanderbilt}(2018)}]{Vanderbiltbook}%
  \BibitemOpen
  \bibfield  {author} {\bibinfo {author} {\bibfnamefont {D.}~\bibnamefont
  {Vanderbilt}},\ }\href@noop {} {\emph {\bibinfo {title} {Berry Phases in
  Electronic Structure Theory: Electric Polarization, Orbital Magnetization and
  Topological Insulators}}}\ (\bibinfo  {publisher} {Cambridge University
  Press},\ \bibinfo {year} {2018})\BibitemShut {NoStop}%
\bibitem [{\citenamefont {Thonhauser}\ and\ \citenamefont
  {Vanderbilt}(2006)}]{Chern_Wannier_breakdown}%
  \BibitemOpen
  \bibfield  {author} {\bibinfo {author} {\bibfnamefont {T.}~\bibnamefont
  {Thonhauser}}\ and\ \bibinfo {author} {\bibfnamefont {D.}~\bibnamefont
  {Vanderbilt}},\ }\bibfield  {title} {\bibinfo {title}
  {Insulator/chern-insulator transition in the haldane model},\ }\href@noop {}
  {\bibfield  {journal} {\bibinfo  {journal} {Physical Review B}\ }\textbf
  {\bibinfo {volume} {74}},\ \bibinfo {pages} {235111} (\bibinfo {year}
  {2006})}\BibitemShut {NoStop}%
\bibitem [{\citenamefont {Wang}\ \emph {et~al.}(2019)\citenamefont {Wang},
  \citenamefont {Guo},\ and\ \citenamefont {Jiang}}]{Tutorial_WL1}%
  \BibitemOpen
  \bibfield  {author} {\bibinfo {author} {\bibfnamefont {H.-X.}\ \bibnamefont
  {Wang}}, \bibinfo {author} {\bibfnamefont {G.-Y.}\ \bibnamefont {Guo}},\ and\
  \bibinfo {author} {\bibfnamefont {J.-H.}\ \bibnamefont {Jiang}},\ }\bibfield
  {title} {\bibinfo {title} {Band topology in classical waves: Wilson-loop
  approach to topological numbers and fragile topology},\ }\href@noop {}
  {\bibfield  {journal} {\bibinfo  {journal} {New Journal of Physics}\ }\textbf
  {\bibinfo {volume} {21}},\ \bibinfo {pages} {093029} (\bibinfo {year}
  {2019})}\BibitemShut {NoStop}%
\bibitem [{\citenamefont {Blanco~de Paz}\ \emph {et~al.}(2020)\citenamefont
  {Blanco~de Paz}, \citenamefont {Devescovi}, \citenamefont {Giedke},
  \citenamefont {Saenz}, \citenamefont {Vergniory}, \citenamefont {Bradlyn},
  \citenamefont {Bercioux},\ and\ \citenamefont
  {Garc{\'\i}a-Etxarri}}]{Tutorial_WL2}%
  \BibitemOpen
  \bibfield  {author} {\bibinfo {author} {\bibfnamefont {M.}~\bibnamefont
  {Blanco~de Paz}}, \bibinfo {author} {\bibfnamefont {C.}~\bibnamefont
  {Devescovi}}, \bibinfo {author} {\bibfnamefont {G.}~\bibnamefont {Giedke}},
  \bibinfo {author} {\bibfnamefont {J.~J.}\ \bibnamefont {Saenz}}, \bibinfo
  {author} {\bibfnamefont {M.~G.}\ \bibnamefont {Vergniory}}, \bibinfo {author}
  {\bibfnamefont {B.}~\bibnamefont {Bradlyn}}, \bibinfo {author} {\bibfnamefont
  {D.}~\bibnamefont {Bercioux}},\ and\ \bibinfo {author} {\bibfnamefont
  {A.}~\bibnamefont {Garc{\'\i}a-Etxarri}},\ }\bibfield  {title} {\bibinfo
  {title} {Tutorial: Computing topological invariants in 2d photonic
  crystals},\ }\href@noop {} {\bibfield  {journal} {\bibinfo  {journal}
  {Advanced Quantum Technologies}\ }\textbf {\bibinfo {volume} {3}},\ \bibinfo
  {pages} {1900117} (\bibinfo {year} {2020})}\BibitemShut {NoStop}%
\bibitem [{\citenamefont {Fang}\ \emph {et~al.}(2012)\citenamefont {Fang},
  \citenamefont {Gilbert},\ and\ \citenamefont
  {Bernevig}}]{fang_Chern_from_sym}%
  \BibitemOpen
  \bibfield  {author} {\bibinfo {author} {\bibfnamefont {C.}~\bibnamefont
  {Fang}}, \bibinfo {author} {\bibfnamefont {M.~J.}\ \bibnamefont {Gilbert}},\
  and\ \bibinfo {author} {\bibfnamefont {B.~A.}\ \bibnamefont {Bernevig}},\
  }\bibfield  {title} {\bibinfo {title} {Bulk topological invariants in
  noninteracting point group symmetric insulators},\ }\href@noop {} {\bibfield
  {journal} {\bibinfo  {journal} {Physical Review B}\ }\textbf {\bibinfo
  {volume} {86}},\ \bibinfo {pages} {115112} (\bibinfo {year}
  {2012})}\BibitemShut {NoStop}%
\bibitem [{\citenamefont {Teo}\ and\ \citenamefont {Kane}(2010)}]{teo2010}%
  \BibitemOpen
  \bibfield  {author} {\bibinfo {author} {\bibfnamefont {J.~C.~Y.}\
  \bibnamefont {Teo}}\ and\ \bibinfo {author} {\bibfnamefont {C.~L.}\
  \bibnamefont {Kane}},\ }\bibfield  {title} {\bibinfo {title} {Topological
  defects and gapless modes in insulators and superconductors},\ }\href
  {https://doi.org/10.1103/PhysRevB.82.115120} {\bibfield  {journal} {\bibinfo
  {journal} {Phys. Rev. B}\ }\textbf {\bibinfo {volume} {82}},\ \bibinfo
  {pages} {115120} (\bibinfo {year} {2010})}\BibitemShut {NoStop}%
\bibitem [{\citenamefont {Teo}\ and\ \citenamefont {Hughes}(2013)}]{teo2013}%
  \BibitemOpen
  \bibfield  {author} {\bibinfo {author} {\bibfnamefont {J.~C.~Y.}\
  \bibnamefont {Teo}}\ and\ \bibinfo {author} {\bibfnamefont {T.~L.}\
  \bibnamefont {Hughes}},\ }\bibfield  {title} {\bibinfo {title} {Existence of
  majorana-fermion bound states on disclinations and the classification of
  topological crystalline superconductors in two dimensions},\ }\href
  {https://doi.org/10.1103/PhysRevLett.111.047006} {\bibfield  {journal}
  {\bibinfo  {journal} {Phys. Rev. Lett.}\ }\textbf {\bibinfo {volume} {111}},\
  \bibinfo {pages} {047006} (\bibinfo {year} {2013})}\BibitemShut {NoStop}%
\bibitem [{\citenamefont {Li}\ \emph {et~al.}(2020{\natexlab{a}})\citenamefont
  {Li}, \citenamefont {Zhu}, \citenamefont {Benalcazar},\ and\ \citenamefont
  {Hughes}}]{Disclination_charges_Cn}%
  \BibitemOpen
  \bibfield  {author} {\bibinfo {author} {\bibfnamefont {T.}~\bibnamefont
  {Li}}, \bibinfo {author} {\bibfnamefont {P.}~\bibnamefont {Zhu}}, \bibinfo
  {author} {\bibfnamefont {W.~A.}\ \bibnamefont {Benalcazar}},\ and\ \bibinfo
  {author} {\bibfnamefont {T.~L.}\ \bibnamefont {Hughes}},\ }\bibfield  {title}
  {\bibinfo {title} {Fractional disclination charge in two-dimensional c
  n-symmetric topological crystalline insulators},\ }\href@noop {} {\bibfield
  {journal} {\bibinfo  {journal} {Physical Review B}\ }\textbf {\bibinfo
  {volume} {101}},\ \bibinfo {pages} {115115} (\bibinfo {year}
  {2020}{\natexlab{a}})}\BibitemShut {NoStop}%
\bibitem [{\citenamefont {Song}\ \emph {et~al.}(2018)\citenamefont {Song},
  \citenamefont {Zhang},\ and\ \citenamefont {Fang}}]{song2018diagnosis}%
  \BibitemOpen
  \bibfield  {author} {\bibinfo {author} {\bibfnamefont {Z.}~\bibnamefont
  {Song}}, \bibinfo {author} {\bibfnamefont {T.}~\bibnamefont {Zhang}},\ and\
  \bibinfo {author} {\bibfnamefont {C.}~\bibnamefont {Fang}},\ }\bibfield
  {title} {\bibinfo {title} {Diagnosis for nonmagnetic topological semimetals
  in the absence of spin-orbital coupling},\ }\href@noop {} {\bibfield
  {journal} {\bibinfo  {journal} {Physical Review X}\ }\textbf {\bibinfo
  {volume} {8}},\ \bibinfo {pages} {031069} (\bibinfo {year}
  {2018})}\BibitemShut {NoStop}%
\bibitem [{\citenamefont {Vaidya}\ \emph {et~al.}(2021)\citenamefont {Vaidya},
  \citenamefont {Benalcazar}, \citenamefont {Cerjan},\ and\ \citenamefont
  {Rechtsman}}]{vaidya2021point}%
  \BibitemOpen
  \bibfield  {author} {\bibinfo {author} {\bibfnamefont {S.}~\bibnamefont
  {Vaidya}}, \bibinfo {author} {\bibfnamefont {W.~A.}\ \bibnamefont
  {Benalcazar}}, \bibinfo {author} {\bibfnamefont {A.}~\bibnamefont {Cerjan}},\
  and\ \bibinfo {author} {\bibfnamefont {M.~C.}\ \bibnamefont {Rechtsman}},\
  }\bibfield  {title} {\bibinfo {title} {Point-defect-localized bound states in
  the continuum in photonic crystals and structured fibers},\ }\href@noop {}
  {\bibfield  {journal} {\bibinfo  {journal} {Physical Review Letters}\
  }\textbf {\bibinfo {volume} {127}},\ \bibinfo {pages} {023605} (\bibinfo
  {year} {2021})}\BibitemShut {NoStop}%
\bibitem [{\citenamefont {Chua}\ \emph {et~al.}(2014)\citenamefont {Chua},
  \citenamefont {Lu}, \citenamefont {Bravo-Abad}, \citenamefont
  {Joannopoulos},\ and\ \citenamefont {Solja{\v{c}}i{\'c}}}]{dirac_laser1}%
  \BibitemOpen
  \bibfield  {author} {\bibinfo {author} {\bibfnamefont {S.-L.}\ \bibnamefont
  {Chua}}, \bibinfo {author} {\bibfnamefont {L.}~\bibnamefont {Lu}}, \bibinfo
  {author} {\bibfnamefont {J.}~\bibnamefont {Bravo-Abad}}, \bibinfo {author}
  {\bibfnamefont {J.~D.}\ \bibnamefont {Joannopoulos}},\ and\ \bibinfo {author}
  {\bibfnamefont {M.}~\bibnamefont {Solja{\v{c}}i{\'c}}},\ }\bibfield  {title}
  {\bibinfo {title} {Larger-area single-mode photonic crystal surface-emitting
  lasers enabled by an accidental {Dirac} point},\ }\href@noop {} {\bibfield
  {journal} {\bibinfo  {journal} {Optics Letters}\ }\textbf {\bibinfo {volume}
  {39}},\ \bibinfo {pages} {2072} (\bibinfo {year} {2014})}\BibitemShut
  {NoStop}%
\bibitem [{\citenamefont {Bravo-Abad}\ \emph {et~al.}(2012)\citenamefont
  {Bravo-Abad}, \citenamefont {Joannopoulos},\ and\ \citenamefont
  {Solja{\v{c}}i{\'c}}}]{dirac_laser2}%
  \BibitemOpen
  \bibfield  {author} {\bibinfo {author} {\bibfnamefont {J.}~\bibnamefont
  {Bravo-Abad}}, \bibinfo {author} {\bibfnamefont {J.~D.}\ \bibnamefont
  {Joannopoulos}},\ and\ \bibinfo {author} {\bibfnamefont {M.}~\bibnamefont
  {Solja{\v{c}}i{\'c}}},\ }\bibfield  {title} {\bibinfo {title} {Enabling
  single-mode behavior over large areas with photonic {Dirac} cones},\
  }\href@noop {} {\bibfield  {journal} {\bibinfo  {journal} {Proceedings of the
  National Academy of Sciences}\ }\textbf {\bibinfo {volume} {109}},\ \bibinfo
  {pages} {9761} (\bibinfo {year} {2012})}\BibitemShut {NoStop}%
\bibitem [{\citenamefont {Guglielmon}\ and\ \citenamefont
  {Rechtsman}(2019)}]{Slowlight_chern1}%
  \BibitemOpen
  \bibfield  {author} {\bibinfo {author} {\bibfnamefont {J.}~\bibnamefont
  {Guglielmon}}\ and\ \bibinfo {author} {\bibfnamefont {M.~C.}\ \bibnamefont
  {Rechtsman}},\ }\bibfield  {title} {\bibinfo {title} {Broadband topological
  slow light through higher momentum-space winding},\ }\href@noop {} {\bibfield
   {journal} {\bibinfo  {journal} {Physical Review Letters}\ }\textbf {\bibinfo
  {volume} {122}},\ \bibinfo {pages} {153904} (\bibinfo {year}
  {2019})}\BibitemShut {NoStop}%
\bibitem [{\citenamefont {Mann}\ and\ \citenamefont
  {Al{\`u}}(2021)}]{Slowlight_chern2}%
  \BibitemOpen
  \bibfield  {author} {\bibinfo {author} {\bibfnamefont {S.~A.}\ \bibnamefont
  {Mann}}\ and\ \bibinfo {author} {\bibfnamefont {A.}~\bibnamefont {Al{\`u}}},\
  }\bibfield  {title} {\bibinfo {title} {Broadband topological slow light
  through brillouin zone winding},\ }\href@noop {} {\bibfield  {journal}
  {\bibinfo  {journal} {Physical Review Letters}\ }\textbf {\bibinfo {volume}
  {127}},\ \bibinfo {pages} {123601} (\bibinfo {year} {2021})}\BibitemShut
  {NoStop}%
\bibitem [{\citenamefont {Yu}\ \emph {et~al.}(2021)\citenamefont {Yu},
  \citenamefont {Xue},\ and\ \citenamefont {Zhang}}]{Slowlight_chern3}%
  \BibitemOpen
  \bibfield  {author} {\bibinfo {author} {\bibfnamefont {L.}~\bibnamefont
  {Yu}}, \bibinfo {author} {\bibfnamefont {H.}~\bibnamefont {Xue}},\ and\
  \bibinfo {author} {\bibfnamefont {B.}~\bibnamefont {Zhang}},\ }\bibfield
  {title} {\bibinfo {title} {Topological slow light via coupling chiral edge
  modes with flatbands},\ }\href@noop {} {\bibfield  {journal} {\bibinfo
  {journal} {Applied Physics Letters}\ }\textbf {\bibinfo {volume} {118}},\
  \bibinfo {pages} {071102} (\bibinfo {year} {2021})}\BibitemShut {NoStop}%
\bibitem [{\citenamefont {Liu}\ \emph {et~al.}(2019)\citenamefont {Liu},
  \citenamefont {Vishwanath},\ and\ \citenamefont {Khalaf}}]{Rot_HOTI3}%
  \BibitemOpen
  \bibfield  {author} {\bibinfo {author} {\bibfnamefont {S.}~\bibnamefont
  {Liu}}, \bibinfo {author} {\bibfnamefont {A.}~\bibnamefont {Vishwanath}},\
  and\ \bibinfo {author} {\bibfnamefont {E.}~\bibnamefont {Khalaf}},\
  }\bibfield  {title} {\bibinfo {title} {Shift insulators: Rotation-protected
  two-dimensional topological crystalline insulators},\ }\href@noop {}
  {\bibfield  {journal} {\bibinfo  {journal} {Physical Review X}\ }\textbf
  {\bibinfo {volume} {9}},\ \bibinfo {pages} {031003} (\bibinfo {year}
  {2019})}\BibitemShut {NoStop}%
\bibitem [{\citenamefont {Bernevig}\ and\ \citenamefont
  {Zhang}(2006)}]{QSHE_CM1}%
  \BibitemOpen
  \bibfield  {author} {\bibinfo {author} {\bibfnamefont {B.~A.}\ \bibnamefont
  {Bernevig}}\ and\ \bibinfo {author} {\bibfnamefont {S.-C.}\ \bibnamefont
  {Zhang}},\ }\bibfield  {title} {\bibinfo {title} {Quantum spin {Hall}
  effect},\ }\href@noop {} {\bibfield  {journal} {\bibinfo  {journal} {Physical
  Review Letters}\ }\textbf {\bibinfo {volume} {96}},\ \bibinfo {pages}
  {106802} (\bibinfo {year} {2006})}\BibitemShut {NoStop}%
\bibitem [{\citenamefont {Kane}\ and\ \citenamefont {Mele}(2005)}]{QSHE_CM2}%
  \BibitemOpen
  \bibfield  {author} {\bibinfo {author} {\bibfnamefont {C.~L.}\ \bibnamefont
  {Kane}}\ and\ \bibinfo {author} {\bibfnamefont {E.~J.}\ \bibnamefont
  {Mele}},\ }\bibfield  {title} {\bibinfo {title} {Quantum spin {Hall} effect
  in graphene},\ }\href@noop {} {\bibfield  {journal} {\bibinfo  {journal}
  {Physical Review Letters}\ }\textbf {\bibinfo {volume} {95}},\ \bibinfo
  {pages} {226801} (\bibinfo {year} {2005})}\BibitemShut {NoStop}%
\bibitem [{\citenamefont {Bernevig}\ \emph {et~al.}(2006)\citenamefont
  {Bernevig}, \citenamefont {Hughes},\ and\ \citenamefont {Zhang}}]{QSHE_CM3}%
  \BibitemOpen
  \bibfield  {author} {\bibinfo {author} {\bibfnamefont {B.~A.}\ \bibnamefont
  {Bernevig}}, \bibinfo {author} {\bibfnamefont {T.~L.}\ \bibnamefont
  {Hughes}},\ and\ \bibinfo {author} {\bibfnamefont {S.-C.}\ \bibnamefont
  {Zhang}},\ }\bibfield  {title} {\bibinfo {title} {Quantum spin {Hall} effect
  and topological phase transition in {HgTe} quantum wells},\ }\href@noop {}
  {\bibfield  {journal} {\bibinfo  {journal} {science}\ }\textbf {\bibinfo
  {volume} {314}},\ \bibinfo {pages} {1757} (\bibinfo {year}
  {2006})}\BibitemShut {NoStop}%
\bibitem [{\citenamefont {Palmer}\ and\ \citenamefont
  {Giannini}(2021)}]{Berry_bands}%
  \BibitemOpen
  \bibfield  {author} {\bibinfo {author} {\bibfnamefont {S.~J.}\ \bibnamefont
  {Palmer}}\ and\ \bibinfo {author} {\bibfnamefont {V.}~\bibnamefont
  {Giannini}},\ }\bibfield  {title} {\bibinfo {title} {Berry bands and
  pseudo-spin of topological photonic phases},\ }\href@noop {} {\bibfield
  {journal} {\bibinfo  {journal} {Physical Review Research}\ }\textbf {\bibinfo
  {volume} {3}},\ \bibinfo {pages} {L022013} (\bibinfo {year}
  {2021})}\BibitemShut {NoStop}%
\bibitem [{\citenamefont {Benalcazar}\ \emph
  {et~al.}(2017{\natexlab{a}})\citenamefont {Benalcazar}, \citenamefont
  {Bernevig},\ and\ \citenamefont {Hughes}}]{benalcazar2017quad}%
  \BibitemOpen
  \bibfield  {author} {\bibinfo {author} {\bibfnamefont {W.~A.}\ \bibnamefont
  {Benalcazar}}, \bibinfo {author} {\bibfnamefont {B.~A.}\ \bibnamefont
  {Bernevig}},\ and\ \bibinfo {author} {\bibfnamefont {T.~L.}\ \bibnamefont
  {Hughes}},\ }\bibfield  {title} {\bibinfo {title} {Quantized electric
  multipole insulators},\ }\href {https://doi.org/10.1126/science.aah6442}
  {\bibfield  {journal} {\bibinfo  {journal} {Science}\ }\textbf {\bibinfo
  {volume} {357}},\ \bibinfo {pages} {61} (\bibinfo {year}
  {2017}{\natexlab{a}})}\BibitemShut {NoStop}%
\bibitem [{\citenamefont {Benalcazar}\ \emph
  {et~al.}(2017{\natexlab{b}})\citenamefont {Benalcazar}, \citenamefont
  {Bernevig},\ and\ \citenamefont {Hughes}}]{benalcazar2017quadPRB}%
  \BibitemOpen
  \bibfield  {author} {\bibinfo {author} {\bibfnamefont {W.~A.}\ \bibnamefont
  {Benalcazar}}, \bibinfo {author} {\bibfnamefont {B.~A.}\ \bibnamefont
  {Bernevig}},\ and\ \bibinfo {author} {\bibfnamefont {T.~L.}\ \bibnamefont
  {Hughes}},\ }\bibfield  {title} {\bibinfo {title} {Electric multipole
  moments, topological multipole moment pumping, and chiral hinge states in
  crystalline insulators},\ }\href {https://doi.org/10.1103/PhysRevB.96.245115}
  {\bibfield  {journal} {\bibinfo  {journal} {Phys. Rev. B}\ }\textbf {\bibinfo
  {volume} {96}},\ \bibinfo {pages} {245115} (\bibinfo {year}
  {2017}{\natexlab{b}})}\BibitemShut {NoStop}%
\bibitem [{\citenamefont {Wheeler}\ \emph {et~al.}(2019)\citenamefont
  {Wheeler}, \citenamefont {Wagner},\ and\ \citenamefont
  {Hughes}}]{wheeler2019}%
  \BibitemOpen
  \bibfield  {author} {\bibinfo {author} {\bibfnamefont {W.~A.}\ \bibnamefont
  {Wheeler}}, \bibinfo {author} {\bibfnamefont {L.~K.}\ \bibnamefont
  {Wagner}},\ and\ \bibinfo {author} {\bibfnamefont {T.~L.}\ \bibnamefont
  {Hughes}},\ }\bibfield  {title} {\bibinfo {title} {Many-body electric
  multipole operators in extended systems},\ }\href
  {https://doi.org/10.1103/PhysRevB.100.245135} {\bibfield  {journal} {\bibinfo
   {journal} {Phys. Rev. B}\ }\textbf {\bibinfo {volume} {100}},\ \bibinfo
  {pages} {245135} (\bibinfo {year} {2019})}\BibitemShut {NoStop}%
\bibitem [{\citenamefont {Kang}\ \emph {et~al.}(2019)\citenamefont {Kang},
  \citenamefont {Shiozaki},\ and\ \citenamefont {Cho}}]{kang2019}%
  \BibitemOpen
  \bibfield  {author} {\bibinfo {author} {\bibfnamefont {B.}~\bibnamefont
  {Kang}}, \bibinfo {author} {\bibfnamefont {K.}~\bibnamefont {Shiozaki}},\
  and\ \bibinfo {author} {\bibfnamefont {G.~Y.}\ \bibnamefont {Cho}},\
  }\bibfield  {title} {\bibinfo {title} {Many-body order parameters for
  multipoles in solids},\ }\href {https://doi.org/10.1103/PhysRevB.100.245134}
  {\bibfield  {journal} {\bibinfo  {journal} {Phys. Rev. B}\ }\textbf {\bibinfo
  {volume} {100}},\ \bibinfo {pages} {245134} (\bibinfo {year}
  {2019})}\BibitemShut {NoStop}%
\bibitem [{\citenamefont {Resta}(1998)}]{resta1998}%
  \BibitemOpen
  \bibfield  {author} {\bibinfo {author} {\bibfnamefont {R.}~\bibnamefont
  {Resta}},\ }\bibfield  {title} {\bibinfo {title} {Quantum-mechanical position
  operator in extended systems},\ }\href
  {https://doi.org/10.1103/PhysRevLett.80.1800} {\bibfield  {journal} {\bibinfo
   {journal} {Phys. Rev. Lett.}\ }\textbf {\bibinfo {volume} {80}},\ \bibinfo
  {pages} {1800} (\bibinfo {year} {1998})}\BibitemShut {NoStop}%
\bibitem [{\citenamefont {Agarwala}\ \emph {et~al.}(2020)\citenamefont
  {Agarwala}, \citenamefont {Juri\ifmmode \check{c}\else
  \v{c}\fi{}i\ifmmode~\acute{c}\else \'{c}\fi{}},\ and\ \citenamefont
  {Roy}}]{roy2020quad}%
  \BibitemOpen
  \bibfield  {author} {\bibinfo {author} {\bibfnamefont {A.}~\bibnamefont
  {Agarwala}}, \bibinfo {author} {\bibfnamefont {V.}~\bibnamefont {Juri\ifmmode
  \check{c}\else \v{c}\fi{}i\ifmmode~\acute{c}\else \'{c}\fi{}}},\ and\
  \bibinfo {author} {\bibfnamefont {B.}~\bibnamefont {Roy}},\ }\bibfield
  {title} {\bibinfo {title} {Higher-order topological insulators in amorphous
  solids},\ }\href {https://doi.org/10.1103/PhysRevResearch.2.012067}
  {\bibfield  {journal} {\bibinfo  {journal} {Phys. Rev. Research}\ }\textbf
  {\bibinfo {volume} {2}},\ \bibinfo {pages} {012067} (\bibinfo {year}
  {2020})}\BibitemShut {NoStop}%
\bibitem [{\citenamefont {Li}\ \emph {et~al.}(2020{\natexlab{b}})\citenamefont
  {Li}, \citenamefont {Fu}, \citenamefont {Hu}, \citenamefont {Li},\ and\
  \citenamefont {Shen}}]{li2020quad}%
  \BibitemOpen
  \bibfield  {author} {\bibinfo {author} {\bibfnamefont {C.-A.}\ \bibnamefont
  {Li}}, \bibinfo {author} {\bibfnamefont {B.}~\bibnamefont {Fu}}, \bibinfo
  {author} {\bibfnamefont {Z.-A.}\ \bibnamefont {Hu}}, \bibinfo {author}
  {\bibfnamefont {J.}~\bibnamefont {Li}},\ and\ \bibinfo {author}
  {\bibfnamefont {S.-Q.}\ \bibnamefont {Shen}},\ }\bibfield  {title} {\bibinfo
  {title} {Topological phase transitions in disordered electric quadrupole
  insulators},\ }\href {https://doi.org/10.1103/PhysRevLett.125.166801}
  {\bibfield  {journal} {\bibinfo  {journal} {Phys. Rev. Lett.}\ }\textbf
  {\bibinfo {volume} {125}},\ \bibinfo {pages} {166801} (\bibinfo {year}
  {2020}{\natexlab{b}})}\BibitemShut {NoStop}%
\bibitem [{\citenamefont {Yang}\ \emph {et~al.}(2021)\citenamefont {Yang},
  \citenamefont {Li}, \citenamefont {Duan},\ and\ \citenamefont
  {Xu}}]{yang2021quad}%
  \BibitemOpen
  \bibfield  {author} {\bibinfo {author} {\bibfnamefont {Y.-B.}\ \bibnamefont
  {Yang}}, \bibinfo {author} {\bibfnamefont {K.}~\bibnamefont {Li}}, \bibinfo
  {author} {\bibfnamefont {L.-M.}\ \bibnamefont {Duan}},\ and\ \bibinfo
  {author} {\bibfnamefont {Y.}~\bibnamefont {Xu}},\ }\bibfield  {title}
  {\bibinfo {title} {Higher-order topological {Anderson} insulators},\ }\href
  {https://doi.org/10.1103/PhysRevB.103.085408} {\bibfield  {journal} {\bibinfo
   {journal} {Phys. Rev. B}\ }\textbf {\bibinfo {volume} {103}},\ \bibinfo
  {pages} {085408} (\bibinfo {year} {2021})}\BibitemShut {NoStop}%
\bibitem [{\citenamefont {Benalcazar}\ and\ \citenamefont
  {Cerjan}(2022)}]{benalcazar2021}%
  \BibitemOpen
  \bibfield  {author} {\bibinfo {author} {\bibfnamefont {W.~A.}\ \bibnamefont
  {Benalcazar}}\ and\ \bibinfo {author} {\bibfnamefont {A.}~\bibnamefont
  {Cerjan}},\ }\bibfield  {title} {\bibinfo {title} {Chiral-symmetric
  higher-order topological phases of matter},\ }\href@noop {} {\bibfield
  {journal} {\bibinfo  {journal} {Physical Review Letters}\ }\textbf {\bibinfo
  {volume} {128}},\ \bibinfo {pages} {127601} (\bibinfo {year}
  {2022})}\BibitemShut {NoStop}%
\bibitem [{\citenamefont {Yariv}(1991)}]{yariv1991book}%
  \BibitemOpen
  \bibfield  {author} {\bibinfo {author} {\bibfnamefont {A.}~\bibnamefont
  {Yariv}},\ }\href@noop {} {\emph {\bibinfo {title} {Optical electronics}}}\
  (\bibinfo  {publisher} {Saunders College Publishing},\ \bibinfo {year}
  {1991})\BibitemShut {NoStop}%
\bibitem [{\citenamefont {Kruthoff}\ \emph {et~al.}(2017)\citenamefont
  {Kruthoff}, \citenamefont {de~Boer}, \citenamefont {van Wezel}, \citenamefont
  {Kane},\ and\ \citenamefont {Slager}}]{slager2017classification}%
  \BibitemOpen
  \bibfield  {author} {\bibinfo {author} {\bibfnamefont {J.}~\bibnamefont
  {Kruthoff}}, \bibinfo {author} {\bibfnamefont {J.}~\bibnamefont {de~Boer}},
  \bibinfo {author} {\bibfnamefont {J.}~\bibnamefont {van Wezel}}, \bibinfo
  {author} {\bibfnamefont {C.~L.}\ \bibnamefont {Kane}},\ and\ \bibinfo
  {author} {\bibfnamefont {R.-J.}\ \bibnamefont {Slager}},\ }\bibfield  {title}
  {\bibinfo {title} {Topological classification of crystalline insulators
  through band structure combinatorics},\ }\href
  {https://doi.org/10.1103/PhysRevX.7.041069} {\bibfield  {journal} {\bibinfo
  {journal} {Phys. Rev. X}\ }\textbf {\bibinfo {volume} {7}},\ \bibinfo {pages}
  {041069} (\bibinfo {year} {2017})}\BibitemShut {NoStop}%
\end{thebibliography}%

\clearpage

\appendix
\section{State counting mismatch}

\begin{figure}[h]
\includegraphics[width=0.8\columnwidth]{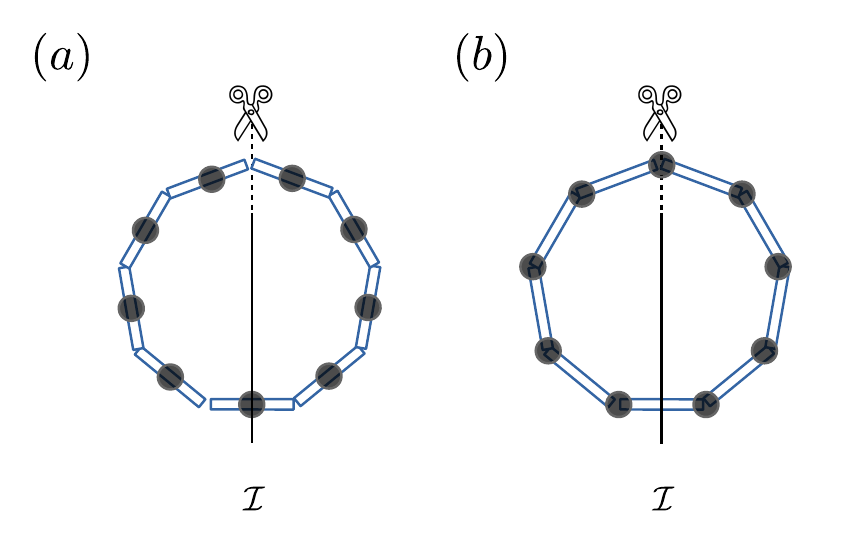}
\caption{Two Wannier center configurations in a finite 1D lattice with no boundaries. Blue rectangles represent unit cells, black circles represent Wannier centers. (a) Wannier centers in the $1a$ trivial phase. (b) Wannier centers in the $1b$ OAL phase. In both cases, the boundary is opened at the dotted line.}
\label{counting_mismatch_fig}
\end{figure}

Consider a 1D lattice under periodic boundary conditions. The lattice is gapped and the lowest band can exist in two phases, a trivial phase and an OAL phase, both protected by inversion symmetry ($\mathcal{I}$). In Fig.~\ref{counting_mismatch_fig}(a), the Wannier centers are at the middle of their unit cells (Wyckoff position $1a$). In this case, irrespective of the site symmetry representation, all Wannier states except the one that lies at the inversion center can be paired around the inversion center to span both representations, $\{+1,-1\}$, of $\mathcal{I}$. The Wannier state that lies at the inversion center simply transforms according to its site symmetry representation. If we open a boundary at the location indicated by the dotted line in Fig.~\ref{counting_mismatch_fig}(a), the Wannier states remain consistently paired about the inversion center.

Now consider the situation shown in Fig.~\ref{counting_mismatch_fig}(b), where the Wannier centers are located between unit cells (Wyckoff position $1b$). All but one Wannier states can be paired around the inversion center to span both representations, $\{+1,-1\}$, of $\mathcal{I}$. However, we are faced with a conundrum when we open a boundary at the location indicated by the dotted line in Fig.~\ref{counting_mismatch_fig}(b): since the boundary passes through a Wannier center, the corresponding state must be relocated to either the newly formed left edge or the right edge of the system. However, it cannot be moved to either edge since doing so would break inversion symmetry. The only possible resolution of this scenario comes about when the Wannier centers of a different band are in a similar (obstructed) situation. In this case, the two leftover Wannier states, one from each band, can reside at the two boundaries of the system and form a pair to span both representations of $\mathcal{I}$. 


The crucial observation here is that since $\mathcal{I}$ maps the boundary states to each other, any perturbation to the boundary that preserves $\mathcal{I}$ must affect both states similarly. This implies that counting states in the spectrum within the frequency range (or bandwidth) of a single band will always lead to either at least one missing state or one additional state. In general, for an OAL band with inversion symmetry, this counting mismatch is equal to $\overline{1}_2$ states (where $\overline{1}_2$ is any integer congruent to 1 mod 2). For $C_n$-symmetric systems in 2D, this counting mismatch is defined modulo $n$. If these $n$ states lie within a bandgap, they are localized to the $n$ corners of the finite system.

\section{Relation between Chern number and rotation invariants}
\label{app:RelationInvariants}
In this section, we derive relations between the Chern number and the rotation invariants at high-symmetry points of the BZ of $C_n$ symmetric crystals.
We provide the guidelines for such derivations; more detailed accounts of these calculations can be found in Ref.~\cite{fang_Chern_from_sym}. 

\begin{figure}[h!]
\includegraphics[width=\columnwidth]{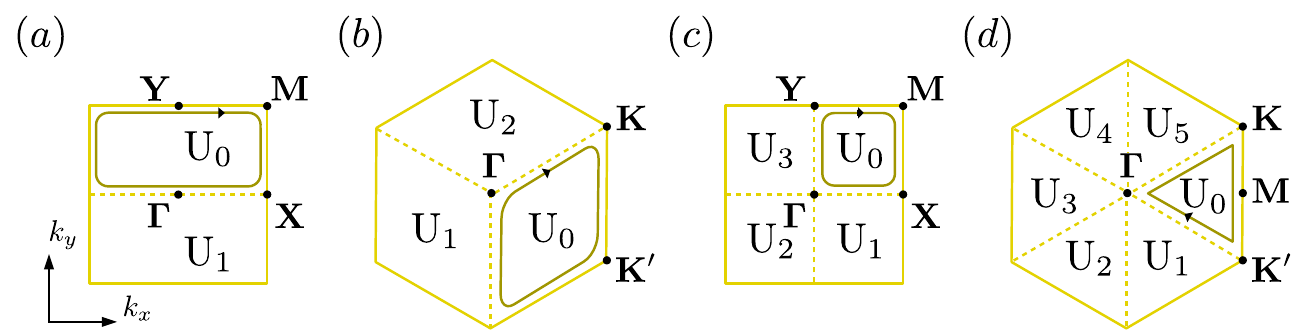}
\caption{The Brillouin zones of crystals with (a) $C_2$, (b) $C_3$, (c) $C_4$ and (d) $C_6$ symmetries. Each Brillouin zone is divided into copies of a fundamental domain over which an integral of the Berry connection is considered. The dark yellow loops indicate the reduced integral paths of Eq.~\ref{eq:ChernLineIntegrals}.}
\label{CnBZ}
\end{figure}

Consider the BZs of $C_n$ symmetric crystals shown in Fig.~\ref{CnBZ}. A nonzero Chern number represents an obstruction to choosing a smooth gauge for the electronic wave functions across the entire BZ. However, the BZs have a fundamental domain, $U_0$, over which a smooth gauge will be chosen; any discontinuities in the gauge are thus pushed to the boundaries \emph{between} symmetry-related fundamental domains. The entire Berry flux that gives rise to the Chern invariant can then be broken into $n$ identical contributions,
\begin{align}
    C &=\frac{\ii}{2\pi}\int_{\text{BZ}}\tr(\F)
    =\frac{\ii}{2\pi}\sum_i \int_{U_i} \tr(\F) \nonumber \\ &=\frac{\ii}{2\pi}\sum_i\int_{\partial U_i} \tr(\A_i).
\end{align}
That is, the Chern number can be calculated by computing the line integrals of the Berry connection along the boundaries of each domain $U_i$. Since the Chern number implies an obstruction to choosing a smooth gauge over the entire BZ, the line integrals along each domain do not cancel each other, instead, they are related by a gauge transformation. Using the fact that the contribution to the Chern number of each domain is equal, we have
\begin{widetext}
\begin{align}
C= 2 \frac{\ii}{2\pi} \int_{\overrightarrow{\bf X \Gamma} \cup \overrightarrow{\bf Y M}} \left(\tr (\A_0) - \tr(\A_1) \right) \quad \quad \text{$C_2$ symmetry},\nonumber\\
C= 3 \frac{\ii}{2\pi} \int_{\overrightarrow{\bf K \Gamma} \cup \overrightarrow{\bf K K'}} \left(\tr (\A_0) - \tr(\A_1) \right) \quad \quad \text{$C_3$ symmetry},\nonumber\\
C= 4 \frac{\ii}{2\pi} \int_{\overrightarrow{\bf X \Gamma} \cup \overrightarrow{\bf X' M}} \left(\tr (\A_0) - \tr(\A_1) \right) \quad \quad \text{$C_4$ symmetry},\nonumber\\
C= 6 \frac{\ii}{2\pi} \left[\int_{\overrightarrow{\bf K' \Gamma}} \left(\tr (\A_0) - \tr(\A_1)\right) +\int_{\overrightarrow{\bf KM}} \left(\tr (\A_0) - \tr(\A_3)\right) \right] \quad \quad \text{$C_6$ symmetry}.
    \label{eq:ChernLineIntegrals}
\end{align}
\end{widetext}
The line integral paths in Eq.~\ref{eq:ChernLineIntegrals} are shown in red in Fig.~\ref{CnBZ}. 

Notice that these line integrals contain the difference between two Berry connections at different domains, which results in the term
\begin{align}
\int_{\bf \Pi_0}^{\bf \Pi_1}\left(\tr(\A_i)-\tr(\A_j)\right)=\int_{\bf \Pi_0}^{\bf \Pi_1}\tr(g_{ij}^\dagger) dg_{ij}=\left.\ln \det g_{ij}\right|_{\bf \Pi_0}^{\bf \Pi_1},
\end{align}
where $g_{ij}$ is the gauge transformation matrix between Berry connections at domains $U_i$ and $U_j$. When evaluated at a high-symmetry point ${\bf \Pi}$, they are equal to the rotation operator projected into the subspace of bands of interest:
\begin{align}
    [g_{ij}({\bf \Pi})]_{\alpha \beta} = \matrixel{u^\alpha({\bf \Pi})}{r_n}{u^\beta({\bf \Pi})} \equiv [r_n({\bf \Pi})]_{\alpha \beta}.
\end{align}
At these HSPs, the projected rotation operator can be diagonalized into 
\begin{align}
    r_n({\bf \Pi})= \bigoplus_{p=1}^n \Pi_p I_{\# \Pi_p \times \# \Pi_p},
\end{align}
where $\#\Pi_p$ indicates the number of states at HSP ${\bf \Pi}$ with rotation eigenvalue $\Pi_p$. 

It will be useful to define the quantity
\begin{align}
    \delta_n({\bf \Pi})=\frac{n}{2\pi \ii} \ln \det r_n({\bf \Pi})=\sum_{p=1}^n (p-1) \#\Pi_p.
\end{align}
The integrals in Eq.~\ref{eq:ChernLineIntegrals} then imply the following relations 

\begin{align}
    C=-\delta_2({\bf \Gamma})+\delta_2({\bf X})-\delta_2({\bf M})+\delta_2({\bf Y})\quad \text{mod 2},\\
    C=-\delta_3({\bf \Gamma})+2\delta_3({\bf K})-\delta_3({\bf K'}) \quad \text{mod 3},\\
    C=-\delta_4({\bf \Gamma})-\delta_4({\bf M})+2\delta_2({\bf X}) \quad \text{mod 4},\\
    C=-\delta_6({\bf \Gamma})+4\delta_3({\bf K})-3\delta_2({\bf M}) \quad \text{mod 6},
\end{align}
or, in terms of the invariants in Eq~\ref{eq:ClassificationIndicesGeneral},

\begin{align}
    C+[X_1^{(2)}]+[Y_1^{(2)}]+[M_1^{(2)}]=0 \quad \text{mod 2}, \nonumber\\
    C+[K_1]+2[K_2]-2[K'_1]-[K'_2]=0 \quad \text{mod 3},\nonumber \\
    C-2[M_1^{(4)}]-[M_2^{(4)}]+[M_4^{(4)}]+2[X_1^{(2)}]=0 \quad \text{mod 4}, \nonumber\\
    C+8[K_1]+4[K_2]-3[M_1]=0 \quad \text{mod 6}.
\end{align}

\section{Invariants from induction of representations in 1D}
The maximal Wyckoff positions in a 1D $\mathcal{I}$-symmetric unit cell are $1a$ and $1b$, as shown in Fig.~\ref{fig:1DPhC_unitcell}(c) of the main text. The classification is given by the invariant $[X_1]$.

The values of this invariant can be enumerated exhaustively by working out the inverse problem, i.e., we start from the set of Wannier functions and derive the band representations at HSPs that such a set leads to. This inverse problem of band topology has been used to classify topological phases in insulators and is known variously as topological quantum chemistry~\cite{bradlyn2017} or symmetry indicators~\cite{symmetry_indicators_po, slager2017classification}.
To induce band representations, it is necessary to specify: (i) a Wannier center configuration and (ii) the symmetry representations of the Wannier functions, otherwise referred to as the ``site symmetry representations"~\cite{canoEBRs}. For a 1D system with inversion symmetry, the band representations at momentum $\mathbf{k}$ with Wannier centers at Wyckoff positions $1a$ and $1b$ are respectively
\begin{align}
    \rho_G^{\mathbf{k}} (\mathcal{I}) &= \rho(\mathcal{I}) \nonumber \hskip 36 pt (1a),\\
    \rho_G^{\mathbf{k}} (\mathcal{I}) &= e^{i|\mathbf{k}|a}\rho(\mathcal{I}), \hskip 11 pt (1b),
    \label{Band_rep_1D}
\end{align}
where $a$ is the lattice constant and $\rho(\mathcal{I})$ is the site symmetry representation, which under $\mathcal{I}$, admits the values $\pm 1$. The Wyckoff position $1b$ is invariant not under $\mathcal{I}$ but under $\mathcal{I}$ followed by a full lattice constant translation. This translation results in a phase factor of $e^{i|\mathbf{k}|a}$ for the band representation in momentum space in Eq.~\eqref{Band_rep_1D}.

For $\rho(\mathcal{I}) = +1$ and at the HSPs $\mathbf{\Gamma}$ and $\mathbf{X}$, the band representations are 
\begin{align}
    \rho_G^{\mathbf{\Gamma}} (\mathcal{I}) = +1 &, \quad \rho_G^{\mathbf{X}} (\mathcal{I}) = +1 \qquad (1a), \nonumber \\
    \rho_G^{\mathbf{\Gamma}} (\mathcal{I}) = +1 &, \quad \rho_G^{\mathbf{X}} (\mathcal{I}) = -1 \qquad (1b).
\end{align}
As a result of this the trivial phase $(1a)$ has $[X_1] = 0$ and the topological phase $(1b)$ has $[X_1] = -1$.

Similarly, for $\rho(\mathcal{I}) = -1$ and at the HSPs $\mathbf{\Gamma}$ and $\mathbf{X}$, the band representations are 
\begin{align}
    \rho_G^{\mathbf{\Gamma}} (\mathcal{I}) = -1 &, \quad \rho_G^{\mathbf{X}} (\mathcal{I}) = -1 \qquad (1a), \nonumber \\
    \rho_G^{\mathbf{\Gamma}} (\mathcal{I}) = -1 &, \quad \rho_G^{\mathbf{X}} (\mathcal{I}) = +1 \qquad (1b).
\end{align}
As a result of this the trivial phase $(1a)$ has $[X_1] = 0$ and the topological phase $(1b)$ has $[X_1] = +1$. 

The results are tabulated in Table \ref{tab:inversion_Invariants}. Since all possible combinations of $+1/-1$ are exhausted in table \ref{tab:inversion_Invariants}, all bands in 1D are atomic limits. 

\begin{table}[htb]
	\centering
		\centering
	\begin{tabular}{ccccc}
		\toprule 
		Wannier center & Site symm. & $\rho_G^{\mathbf{\Gamma}} (\mathcal{I})$ & $\rho_G^{\mathbf{X}} (\mathcal{I})$ & $[X_1]$\\
		\separatorrule
		$1a$ & $\rho(\mathcal{I})=+1$ & $+1$ & $+1$ & 0 \\
		$1a$ & $\rho(\mathcal{I})=-1$ & $-1$ & $-1$ & 0\\
		\separatorrule
		$1b$ & $\rho(\mathcal{I})=+1$ & $+1$ & $-1$ & $-1$\\
		$1b$ & $\rho(\mathcal{I})=-1$ & $-1$ & $+1$ & $+1$\\
		\bottomrule
	\end{tabular} 
	\caption{Inversion symmetry: Invariants induced from Wyckoff positions for different site symmetry representations}
	\label{tab:inversion_Invariants}
\end{table}

A single band in the trivial phase $(1a)$ has $[X_1] = 0$, while in the topological phase $(1b)$, has $[X_1] = +1$ or $-1$. A group of bands can also be characterized by $[X_1]$ since this invariant is linear  under the composition of bands, i.e., the indices $[X_1]_1$ and $[X_1]_2$ of bands 1 and 2 result in the index $[X_1]_1+[X_1]_2$ for the composed system of bands 1 and 2 taken together. This linear property of $[X_1]$  allows for the following possibility: Consider a set of two bands that are independently non-trivial with their individual Wannier centers at position $1b$ having $[X_1] = +1$ and $-1$ respectively. Taken together, these bands result in a trivial phase with $[X_1] = 0$. In such a case, the Wannier centers of the combined system of two bands are not each fixed to the maximal Wyckoff position, $1b$, but are generally separated away from $1b$, consistent with inversion symmetry. This separation can smoothly interpolate between the two maximal Wyckoff positions, $1a$ and $1b$, without closing a bandgap anywhere in the system or breaking the symmetry, and thus such a configuration of two bands is topologically identical to a trivial system. In contrast, consider two bands with Wannier centers located at $1b$ and $[X_1]_1=[X_1]_2=1$. The combined system has $[X_1]_1+[X_1]_2=2$, and the two Wannier centers remain pinned to $1b$. 

In general, we refer to bands that have Wannier centers fixed to positions away from the $1a$ position as obstructed atomic limits (OALs) and bands that have Wannier centers at the $1a$ position or movable Wannier centers that can be adiabatically brought to the $1a$ position as trivial atomic limits.

\section{Invariants from induction of representations in 2D}

In this section, we use the procedure developed in~\cite{canoEBRs} to determine the band representations induced from Wannier centers located at all possible Wyckoff positions and all site symmetry representations in 2D for class AI and class A. From them, we determine the classification indices $\chi_\mathcal{T}^{(n)}$ and $\chi^{(n)}$ specified in Tables \ref{tab:C2_Invariants_table}-\ref{tab:C6_Invariants} of the main text. Since all such bands, even in class A, are Wannierizable by definition, their Chern number is $0$.

When the Chern number vanishes, the topological class given by $\chi^{(n)}_\mathcal{T}$ or $\chi^{(n)}$ indicates both the Wyckoff position of the Wannier centers and the $C_n$ symmetry representation of the Wannier function itself (i.e., the site symmetry representation). The converse is also true. Therefore, the tables below and tables \ref{tab:C2_Invariants_table}-\ref{tab:C6_Invariants} in the main text show the correspondence between Wannier centers, site symmetry representations, and topological indices.

\subsection{\texorpdfstring{$C_2$}{Two-fold} symmetry}
The maximal Wyckoff positions in a $C_2$-symmetric unit cell are $1a$, $1b$, $1c$ and $1d$ as shown in Fig.~\ref{fig:2DWP} of the main text. The classification is given by $\chi_\mathcal{T}^{(2)} = \big( [X^{(2)}_1],[Y^{(2)}_1],[M^{(2)}_1]; N\big)$ (for TR-symmetric) and $\chi^{(2)} =\big( C\, \big| \, [X^{(2)}_1],[Y^{(2)}_1],[M^{(2)}_1]; N \big)$ (for TR-broken). 

The band representations induced from all maximal Wyckoff positions are given by
\begin{align}
\rho_G^{\bf k}(C_2) &= e^{\ii {\bf k.a_1}} \rho(C_2) \quad \text{(from $1c$)}\nonumber\\
\rho_G^{\bf k}(C_2) &= e^{\ii {\bf k.a_2}} \rho(C_2) \quad \text{(from $1d$)}\nonumber\\
\rho_G^{\bf k}(C_2) &= e^{\ii {\bf k.(a_1+a_2)}} \rho(C_2) \quad \text{(from $1b$)}
\label{eq:c2bandrepsfromall}
\end{align}
Using Eqs.~\eqref{eq:c2bandrepsfromall}, we calculate the rotation eigenvalues for all site symmetries when the Wannier centers are at $1b$, $1c$ and $1d$ in Table~\ref{tab:C2_Invariants}.

\subsection{\texorpdfstring{$C_3$}{Three-fold} symmetry}
The maximal Wyckoff positions in a $C_3$-symmetric unit cell are $1a$, $1b$, and $1c$ as shown in Fig.~\ref{fig:2DWP} of the main text. The classification is given by $\chi_\mathcal{T}^{(3)} = \big( [K^{(3)}_1], [K^{(3)}_2]; N \big)$ (for TR-symmetric) and $\chi^{(3)} =\big( C\, \big| \,[K^{(3)}_1], [K^{(3)}_2], [K'^{(3)}_1], [K'^{(3)}_2]; N\big)$ (for TR-broken). For both $C_6$ and $C_3$ symmetries, we use the following primitive vectors ${\bf a_1}=(1,0)$, ${\bf a_{2,3}}=(\pm \frac{1}{2}, \frac{\sqrt{3}}{2})$.

The band representations for Wannier centers at Wyckoff position $1b$ are given by
\begin{equation}
\rho_G^{\bf k}(C_3)= e^{\ii {\bf k}.a_2} \rho(C_3)
\label{eq:c3irrepsfrom1b}
\end{equation}

Using Eq.~\ref{eq:c3irrepsfrom1b}, we calculate the rotation eigenvalues for all site symmetries when the Wannier centers are at $1b$ in Table~\ref{tab:C3_inducedC3InvariantsFrom1b}.

The band representations for Wannier centers at Wyckoff position $1c$ are given by
\begin{align}
\rho_G^{\bf k}(C_3)= e^{\ii {\bf k}.a_1} \rho(C_3)
\label{eq:c3irrepsfrom1c}
\end{align}

Using Eq.~\ref{eq:c3irrepsfrom1c}, we calculate the rotation eigenvalues for all site symmetries when the Wannier centers are at $1c$ in Table~\ref{tab:C3_inducedC3InvariantsFrom1c}.

\subsection{\texorpdfstring{$C_4$}{Four-fold} symmetry}
The maximal Wyckoff positions in a $C_4$-symmetric unit cell are $1a$, $1b$, and $2c$, as shown in Fig.~\ref{fig:2DWP} of the main text. The classification is given by $\chi_\mathcal{T}^{(4)} = \big( [X^{(2)}_1],[M^{(4)}_1],[M^{(4)}_2]; N \big)$ (for TR-symmetric) and $\chi^{(4)} =\big( C\, \big| \, [X^{(2)}_1],[M^{(4)}_1],[M^{(4)}_2],[M^{(4)}_4]; N \big)$ (for TR-broken). 

The band representations for Wannier centers at Wyckoff position $2c$ are given by 
\begin{align}
\rho_G^{\bf k}(C_4) &=\left(\begin{array}{cc}
0 & e^{\ii {\bf k}\cdot{a}_1} \rho(C_2)\\
1 & 0
\end{array}\right), \nonumber\\
\rho_G^{\bf k}(C_2) &=\left(\begin{array}{cc}
e^{\ii {\bf k}\cdot{a}_1} & 0\\
0 & e^{\ii {\bf k}\cdot{a}_2}
\end{array}\right) \rho(C_2).
    \label{eq:c4BandRepFrom2c}
\end{align}
Using Eq.~\ref{eq:c4BandRepFrom2c}, we calculate the rotation eigenvalues for all site symmetries when the Wannier centers are at $2c$ in Table~\ref{tab:C4_InvariantsFrom2c}.

The band representations for Wannier centers at Wyckoff position $1b$ are given by
\begin{align}
\rho_G^{\bf k}(C_4) =
e^{\ii {\bf k}\cdot{a}_1}\rho(C_4), \quad
\rho_G^{\bf k}(C_2) =e^{\ii {\bf k}.(a_1+a_2)}\rho(C_2)
    \label{eq:c4BandRepFrom1b}
\end{align}
Using Eq.~\ref{eq:c4BandRepFrom1b}, we calculate the rotation eigenvalues for all site symmetries when the Wannier centers are at $1b$ in Table~\ref{tab:C4_InvariantsFrom1b}.

\subsection{\texorpdfstring{$C_6$}{Six-fold} symmetry}
The maximal Wyckoff positions in a $C_6$-symmetric unit cell are $1a$, $2b$, and $3c$ as shown in Fig.~\ref{fig:2DWP} of the main text. The classification is given by $\chi_\mathcal{T}^{(6)} = \big( [M^{(2)}_1],[K^{(3)}_1]; N \big)$ (for TR-symmetric) and $\chi^{(6)} =\big( C\, \big| \,[M^{(2)}_1],[K^{(3)}_1],[K^{(3)}_2]; N \big)$ (for TR-broken). For both $C_6$ and $C_3$ symmetries, we use the following primitive vectors ${\bf a_1}=(1,0)$, ${\bf a_{2,3}}=(\pm \frac{1}{2}, \frac{\sqrt{3}}{2})$.

The band representations for Wannier centers at Wyckoff position $2b$ are given by
\begin{align}
\rho_G^{\bf k}(C_3)&=\left(\begin{array}{cc}
e^{\ii {\bf k}\cdot{\bf a_1}} & 0\\
0 & e^{-\ii {\bf k}\cdot{\bf a_1}}
\end{array}\right) \rho(C_3),\nonumber\\
\rho_G^{\bf k}(C_2)&=\left(\begin{array}{cc}
0 & -1\\
1 & 0
\end{array}\right)
\label{eq:c6irrepsfrom2b}
\end{align}
Since the $C_2$ band representation, $\rho_G^{\bf k}(C_2)$, is independent of ${\bf k}$, the invariant $[M_1^{(2)}]$ vanishes. Using Eq.~\ref{eq:c6irrepsfrom2b}, we calculate the rotation eigenvalues for all site symmetries when the Wannier centers are at $2b$ in Table~\ref{tab:C6_inducedC3InvariantsFrom2b}.

The band representations for Wannier centers at Wyckoff position $3c$ are given by
\begin{align}
\rho_G^{\bf k}(C_3)&=\left(\begin{array}{ccc}
0 & 0 & 1\\
1 & 0 & 0\\
0 & 1 & 0
\end{array}\right), \nonumber\\
\rho_G^{\bf k}(C_2)&=\left(\begin{array}{ccc}
e^{\ii {\bf k}\cdot{a}_2}& 0\\
0 & e^{-\ii {\bf k}\cdot{a}_1} &0\\
0 & 0 & e^{-\ii {\bf k}\cdot{a}_3}
\end{array}\right) \rho(C_2),
\label{eq:c6irrepsfrom3c}
\end{align}
Since the $C_3$ band representation, $\rho_G^{\bf k}(C_3)$, is independent of ${\bf k}$, the invariants $[K_1^{(3)}]$ and $[K_2^{(3)}]$ vanish. Using Eq.~\ref{eq:c6irrepsfrom3c}, we calculate the rotation eigenvalues for all site symmetries when the Wannier centers are at $2b$ in Table~\ref{tab:C6_inducedC3InvariantsFrom3c}.

\begin{widetext}
    
\begin{table*}[htb]
	\centering
	\begin{tabular}{cccccccc}
		\toprule 
		Wyckoff pos. & 	Site symm. $\rho(C_2)$ & $\rho_G^{\bf \Gamma}(C_2)$ & $\rho_G^{\bf X}(C_2)$ & $\rho_G^{\bf Y}(C_2)$ & $\rho_G^{\bf M}(C_2)$ & $\chi_\mathcal{T}^{(2)}$ & $\chi^{(2)}$\\
		\midrule
		$1c$ &$+1$ & $\#\Gamma_1^{(2)}=1$ & $\#X_1^{(2)}=0$ & $\#Y_1^{(2)}=1$ & $\#M_1^{(2)}=0$ &  $(-1,0,-1; 1)$ & $(0 \, | \, -1,0,-1; 1)$\\
		& & $\#\Gamma_2^{(2)}=0$ & $\#X_2^{(2)}=1$ & $\#Y_2^{(2)}=0$ & $\#M_1^{(2)}=1$ & &\\
		\separatorrule
		& $-1$ & $\#\Gamma_1^{(2)}=0$ & $\#X_1^{(2)}=1$ & $\#Y_1^{(2)}=0$ & $\#M_1^{(2)}=1$ &  $(1,0,1; 1)$ & $(0 \, | \, 1, 0, 1; 1)$\\
		& & $\#\Gamma_2^{(2)}=1$ & $\#X_2^{(2)}=0$ & $\#Y_2^{(2)}=1$ & $\#M_1^{(2)}=0$ & &\\
		\separatorrule
		$1d$ &$+1$ & $\#\Gamma_1^{(2)}=1$ & $\#X_1^{(2)}=1$ & $\#Y_1^{(2)}=0$ & $\#M_1^{(2)}=0$ & $(0,-1,-1; 1)$ & $(0 \, | \, 0, -1, -1; 1)$\\
		& & $\#\Gamma_2^{(2)}=0$ & $\#X_2^{(2)}=0$ & $\#Y_2^{(2)}=1$ & $\#M_1^{(2)}=1$ & &\\
		\separatorrule
		& $-1$ & $\#\Gamma_1^{(2)}=0$ & $\#X_1^{(2)}=0$ & $\#Y_1^{(2)}=1$ & $\#M_1^{(2)}=1$ &  $(0,1,1; 1)$ & $(0 \, | \,0, 1, 1; 1)$\\
		& & $\#\Gamma_2^{(2)}=1$ & $\#X_2^{(2)}=1$ & $\#Y_2^{(2)}=0$ & $\#M_1^{(2)}=0$ & &\\
		\separatorrule
		$1b$ &$+1$ & $\#\Gamma_1^{(2)}=1$ & $\#X_1^{(2)}=0$ & $\#Y_1^{(2)}=0$ & $\#M_1^{(2)}=1$ & $(-1,-1,0; 1)$ & $(0 \, | \, -1, -1, 0; 1)$\\
		& & $\#\Gamma_2^{(2)}=0$ & $\#X_2^{(2)}=1$ & $\#Y_2^{(2)}=1$ & $\#M_1^{(2)}=0$ & &\\
		\separatorrule
		& $-1$ & $\#\Gamma_1^{(2)}=0$ & $\#X_1^{(2)}=1$ & $\#Y_1^{(2)}=1$ & $\#M_1^{(2)}=0$ &  $(1,1,0; 1)$ & $(0 \, | \, 1, 1, 0; 1)$\\
		& & $\#\Gamma_2^{(2)}=1$ & $\#X_2^{(2)}=0$ & $\#Y_2^{(2)}=0$ & $\#M_1^{(2)}=1$ & &\\
		\bottomrule
	\end{tabular} 
	\caption{$C_2$ symmetry: Invariants induced from the maximal Wyckoff positions for different site symmetry representations}
	\label{tab:C2_Invariants}
\end{table*}

\begin{table*}[htb]
	\centering
	\begin{tabular}{cccccc}
		\toprule 
		Site symm. $\rho(C_3)$ & $\rho_G^{\bf \Gamma}(C_3)$ & $\rho_G^{\bf K}(C_3)$ & $\rho_G^{\bf K'}(C_3)$ & $\chi_\mathcal{T}^{(3)}$ & $\chi^{(3)}$\\
		\midrule
		& $\#\Gamma_1^{(3)}=1$ & $\#K_1^{(3)}=0$ & $\#K'^{(3)}_1=0$ & & \\
		$1$ & $\#\Gamma_2^{(3)}=0$ & $\#K_2^{(3)}=1$ & $\#K'^{(3)}_2=0$ & $(-1,1; 1)$ & $(0\, | \, -1,1,-1,0; 1)$\\
		 & $\#\Gamma_3^{(3)}=0$ & $\#K_3^{(3)}=0$ & $\#K'^{(3)}_3=1$ & &\\
		\separatorrule
		& $\#\Gamma_1^{(3)}=0$ & $\#K_1^{(3)}=1$ & $\#K'^{(3)}_1=1$ & &\\
		$e^{\ii\frac{2\pi}{3}\sigma_z}$ & $\#\Gamma_2^{(3)}=1$ &  $\#K_2^{(3)}=0$ & $\#K'^{(3)}_2=1$ & $(1,-1; 2)$ &  $(0\, | \, 1,-1,1,0; 2)$\\
		& $\#\Gamma_3^{(3)}=1$ & $\#K_3^{(3)}=1$ & $\#K'^{(3)}_3=0$ & &\\
		\separatorrule
		& $\#\Gamma_1^{(3)}=0$ & $\#K_1^{(3)}=0$ & $\#K'^{(3)}_1=1$ & &\\
		$e^{\ii\frac{2\pi}{3}}$ & $\#\Gamma_2^{(3)}=1$ &  $\#K_2^{(3)}=0$ & $\#K'^{(3)}_2=0$ & $-$ & $(0\, | \, 0,-1,1,-1; 1)$\\
		& $\#\Gamma_3^{(3)}=0$ & $\#K_3^{(3)}=1$ & $\#K'^{(3)}_3=0$ & &\\
		\separatorrule
		& $\#\Gamma_1^{(3)}=0$ & $\#K_1^{(3)}=1$ & $\#K'^{(3)}_1=0$ & &\\
		$e^{\ii\frac{4\pi}{3}}$ & $\#\Gamma_2^{(3)}=0$ &  $\#K_2^{(3)}=0$ & $\#K'^{(3)}_2=1$ & $-$ & $(0\, | \, 1,0,0,1; 1)$\\
		& $\#\Gamma_3^{(3)}=1$ & $\#K_3^{(3)}=0$ & $\#K'^{(3)}_3=0$ & &\\
		\bottomrule		
	\end{tabular} 
	\caption{$C_3$ symmetry: Invariants induced from Wyckoff position $1b$ with different site symmetry representations}
	\label{tab:C3_inducedC3InvariantsFrom1b}
\end{table*}

\begin{table*}[htb]
	\centering
	\begin{tabular}{cccccc}
		\toprule 
		Site symm. $\rho(C_3)$ & $\rho_G^{\bf \Gamma}(C_3)$ & $\rho_G^{\bf K}(C_3)$ & $\rho_G^{\bf K'}(C_3)$ & $\chi_\mathcal{T}^{(3)}$ & $\chi^{(3)}$\\
		\midrule
		& $\#\Gamma_1^{(3)}=1$ & $\#K_1^{(3)}=0$ & $\#K'^{(3)}_1=0$ & & \\
		$1$ & $\#\Gamma_2^{(3)}=0$ & $\#K_2^{(3)}=0$ & $\#K'^{(3)}_2=1$ & $(-1,0; 1)$ & $(0\, | \, -1,0,-1,1; 1)$\\
		 & $\#\Gamma_3^{(3)}=0$ & $\#K_3^{(3)}=1$ & $\#K'^{(3)}_3=0$ & &\\
		\separatorrule
		& $\#\Gamma_1^{(3)}=0$ & $\#K_1^{(3)}=1$ & $\#K'^{(3)}_1=1$ & &\\
		$e^{\ii\frac{2\pi}{3}\sigma_z}$ & $\#\Gamma_2^{(3)}=1$ &  $\#K_2^{(3)}=1$ & $\#K'^{(3)}_2=0$ & $(1,0; 2)$ &  $(0\, | \, 1,0,1,-1; 2)$\\
		& $\#\Gamma_3^{(3)}=1$ & $\#K_3^{(3)}=0$ & $\#K'^{(3)}_3=1$ & &\\
		\separatorrule
		& $\#\Gamma_1^{(3)}=0$ & $\#K_1^{(3)}=1$ & $\#K'^{(3)}_1=0$ & &\\
		$e^{\ii\frac{2\pi}{3}}$ & $\#\Gamma_2^{(3)}=1$ &  $\#K_2^{(3)}=0$ & $\#K'^{(3)}_2=0$ & $-$ & $(0\, | \, 1,-1,0,-1; 1)$\\
		& $\#\Gamma_3^{(3)}=0$ & $\#K_3^{(3)}=0$ & $\#K'^{(3)}_3=1$ & &\\
		\separatorrule
		& $\#\Gamma_1^{(3)}=0$ & $\#K_1^{(3)}=0$ & $\#K'^{(3)}_1=1$ & &\\
		$e^{\ii\frac{4\pi}{3}}$ & $\#\Gamma_2^{(3)}=0$ &  $\#K_2^{(3)}=1$ & $\#K'^{(3)}_2=0$ & $-$ & $(0\, | \, 0,1,1,0; 1)$\\
		& $\#\Gamma_3^{(3)}=1$ & $\#K_3^{(3)}=0$ & $\#K'^{(3)}_3=0$ & &\\
		\bottomrule		
	\end{tabular} 
	\caption{$C_3$ symmetry: Invariants induced from Wyckoff position $1c$ with different site symmetry representations}
	\label{tab:C3_inducedC3InvariantsFrom1c}
\end{table*}

\begin{table*}[htb]
	\centering
	\begin{tabular}{cccccc}
		\toprule 
		Site symm. $\rho(C_2)$ & $\rho_G^{\bf \Gamma}(C_4)$ & $\rho_G^{\bf X}(C_2)$ & $\rho_G^{\bf M}(C_4)$ & $\chi_\mathcal{T}^{(4)}$ & $\chi^{(4)}$\\
		\midrule
		& $\#\Gamma_1^{(4)}=1$ & $\#X_1^{(2)}=1$ & $\#M_1^{(4)}=0$ & &\\
		$+1$ & $\#\Gamma_2^{(4)}=0$ & $\#X_2^{(2)}=1$ & $\#M_2^{(4)}=1$ & $(-1,-1,1; 2)$ & $(0 \, | \, -1,-1,1,1; 2)$\\
		 & $\#\Gamma_3^{(4)}=1$ &  & $\#M_3^{(4)}=0$ & &\\
 		 & $\#\Gamma_4^{(4)}=0$ & & $\#M_4^{(4)}=1$ & &\\
		\separatorrule
		& $\#\Gamma_1^{(4)}=0$ & $\#X_1^{(2)}=1$ & $\#M_1^{(4)}=1$ & &\\
		$-1$ & $\#\Gamma_2^{(4)}=1$ & $\#X_2^{(2)}=1$ & $\#M_2^{(4)}=0$ & $(1,1,-1; 2)$ & $(0 \, | \, 1,1,-1,-1; 2)$\\
		 & $\#\Gamma_3^{(4)}=0$ &  & $\#M_3^{(4)}=1$ & &\\
 		 & $\#\Gamma_4^{(4)}=1$ & & $\#M_4^{(4)}=0$ & &\\
		\bottomrule
	\end{tabular} 
	\caption{$C_4$ symmetry: Invariants induced from Wyckoff position $2c$ for different site symmetry representations}
	\label{tab:C4_InvariantsFrom2c}
\end{table*}

\begin{table*}[htb]
	\centering
	\begin{tabular}{cccccc}
		\toprule 
		Site symm. $\rho(C_2)$ & $\rho_G^{\bf \Gamma}(C_4)$ & $\rho_G^{\bf X}(C_2)$ & $\rho_G^{\bf M}(C_4)$ & $\chi_\mathcal{T}^{(4)}$ & $\chi^{(4)}$\\
		\midrule
		& $\#\Gamma_1^{(4)}=1$ & $\#X_1^{(2)}=0$ & $\#M_1^{(4)}=0$ & &\\
		$1$ & $\#\Gamma_2^{(4)}=0$ & $\#X_2^{(2)}=1$ & $\#M_2^{(4)}=0$ & $(-1,-1,0; 1)$ & $(0 \, | \, -1, -1, 0, 0; 1)$\\
		 & $\#\Gamma_3^{(4)}=0$ &  & $\#M_3^{(4)}=1$ & &\\
 		 & $\#\Gamma_4^{(4)}=0$ & & $\#M_4^{(4)}=0$ & &\\
		\separatorrule
		& $\#\Gamma_1^{(4)}=0$ & $\#X_1^{(2)}=0$ & $\#M_1^{(4)}=1$ & &\\
		$-1$ & $\#\Gamma_2^{(4)}=0$ & $\#X_2^{(2)}=1$ & $\#M_2^{(4)}=0$ & $(-1,1,0; 1)$ &$(0 \, | \, -1, 1, 0, 0; 1)$\\
		 & $\#\Gamma_3^{(4)}=1$ &  & $\#M_3^{(4)}=0$ & &\\
 		 & $\#\Gamma_4^{(4)}=0$ & & $\#M_4^{(4)}=0$ & &\\
		\separatorrule
		& $\#\Gamma_1^{(4)}=0$ & $\#X_1^{(2)}=2$ & $\#M_1^{(4)}=0$ & &\\
		$\ii\sigma_z$ & $\#\Gamma_2^{(4)}=1$ & $\#X_2^{(2)}=0$ & $\#M_2^{(4)}=1$ & $(2,0,0; 2)$ & $(0 \, | \, 2, 0, 0, 0; 2)$\\
		 & $\#\Gamma_3^{(4)}=0$ &  & $\#M_3^{(4)}=0$ & &\\
 		 & $\#\Gamma_4^{(4)}=1$ & & $\#M_4^{(4)}=1$ & &\\
		\separatorrule
		& $\#\Gamma_1^{(4)}=0$ & $\#X_1^{(2)}=1$ & $\#M_1^{(4)}=0$ & &\\
		$\ii$ & $\#\Gamma_2^{(4)}=1$ & $\#X_2^{(2)}=0$ & $\#M_2^{(4)}=0$ & $-$ & $(0 \, | \, 1, 0, -1, 1; 1)$\\
		 & $\#\Gamma_3^{(4)}=0$ &  & $\#M_3^{(4)}=0$ & &\\
 		 & $\#\Gamma_4^{(4)}=0$ & & $\#M_4^{(4)}=1$ & &\\
 		\separatorrule
		& $\#\Gamma_1^{(4)}=0$ & $\#X_1^{(2)}=1$ & $\#M_1^{(4)}=0$ & &\\
		$-\ii$ & $\#\Gamma_2^{(4)}=0$ & $\#X_2^{(2)}=0$ & $\#M_2^{(4)}=1$ & $-$ & $(0 \, | \, 1, 0, 1, -1; 1)$\\
		 & $\#\Gamma_3^{(4)}=0$ &  & $\#M_3^{(4)}=0$ & &\\
 		 & $\#\Gamma_4^{(4)}=1$ & & $\#M_4^{(4)}=0$ & &\\
 		\bottomrule
	\end{tabular} 
	\caption{$C_4$ symmetry: Invariants induced from Wyckoff position $1b$ for different site symmetry representations}
	\label{tab:C4_InvariantsFrom1b}
\end{table*}

\begin{table*}[htb]
	\centering
	\begin{tabular}{ccccc}
		\toprule 
		Site symm. $\rho(C_3)$ & $\rho_G^{\bf \Gamma}(C_3)$ & $\rho_G^{\bf K}(C_3)$ & $\chi_\mathcal{T}^{(6)}$ & $\chi^{(6)}$\\
		\midrule
		& $\#\Gamma_1^{(3)}=2$ & $\#K_1^{(3)}=0$ & &\\
		$1$ & $\#\Gamma_2^{(3)}=0$ & $\#K_2^{(3)}=1$ & $(0,-2; 2)$ & $(0 \, | \, 0, -2, 1; 2)$\\
		 & $\#\Gamma_3^{(3)}=0$ & $\#K_3^{(3)}=1$ & &\\
		 \separatorrule
		& $\#\Gamma_1^{(3)}=0$ & $\#K_1^{(3)}=2$ & &\\
		$e^{\ii\frac{2\pi}{3}\sigma_z}$ & $\#\Gamma_2^{(3)}=2$ & $\#K_2^{(3)}=1$ & $(0,2; 4)$ & $(0 \, | \, 0, 2, -1; 4)$\\
		 & $\#\Gamma_3^{(3)}=2$ & $\#K_3^{(3)}=1$ & &\\
		\separatorrule
		& $\#\Gamma_1^{(3)}=0$ & $\#K_1^{(3)}=1$ & &\\
		$e^{\ii\frac{2\pi}{3}}$ & $\#\Gamma_2^{(3)}=2$ & $\#K_2^{(3)}=0$ & $-$ & $(0 \, | \, 0, 1, -2; 2)$\\
		 & $\#\Gamma_3^{(3)}=0$ & $\#K_3^{(3)}=1$ & &\\
		\separatorrule
		& $\#\Gamma_1^{(3)}=0$ & $\#K_1^{(3)}=1$ & &\\
		$e^{\ii\frac{4\pi}{3}}$ & $\#\Gamma_2^{(3)}=0$ & $\#K_2^{(3)}=1$ & $-$ & $(0 \, | \, 0, 1, 1; 2)$\\
		 & $\#\Gamma_3^{(3)}=2$ & $\#K_3^{(3)}=0$ & &\\
		\bottomrule		
	\end{tabular} 
	\caption{$C_6$ symmetry: Invariants induced from Wyckoff position $2b$ with different site symmetry representations}
	\label{tab:C6_inducedC3InvariantsFrom2b}
\end{table*}

\clearpage

\begin{table*}[htb]
	\centering
	\begin{tabular}{ccccc}
		\toprule 
		Site symm. $\rho(C_2)$ & $\rho_G^{\bf \Gamma}(C_2)$ & $\rho_G^{\bf M}(C_2)$ & $\chi_\mathcal{T}^{(6)}$ & $\chi^{(6)}$\\
		\midrule
		$1$ & $\#\Gamma_1^{(2)}=3$ & $\#M_1^{(2)}=1$ & $(-2,0; 3)$ & $(0 \, | \, -2, 0, 0; 3)$\\
		& $\#\Gamma_2^{(2)}=0$ & $\#M_2^{(2)}=2$ & &\\
		 \separatorrule
		$-1$ & $\#\Gamma_1^{(2)}=0$ & $\#M_1^{(2)}=2$ & $(2,0; 3)$ & $(0 \, | \, 2, 0, 0; 3)$\\
		& $\#\Gamma_2^{(2)}=3$ & $\#M_2^{(2)}=1$ & &\\
		\bottomrule
	\end{tabular} 
	\caption{$C_6$ symmetry: Invariants induced from Wyckoff position $3c$ with different site symmetry representations}
	\label{tab:C6_inducedC3InvariantsFrom3c}
\end{table*}

\end{widetext}

\section{Calculation of corner charges}
\begin{figure*}[htb]
\includegraphics[width=\textwidth]{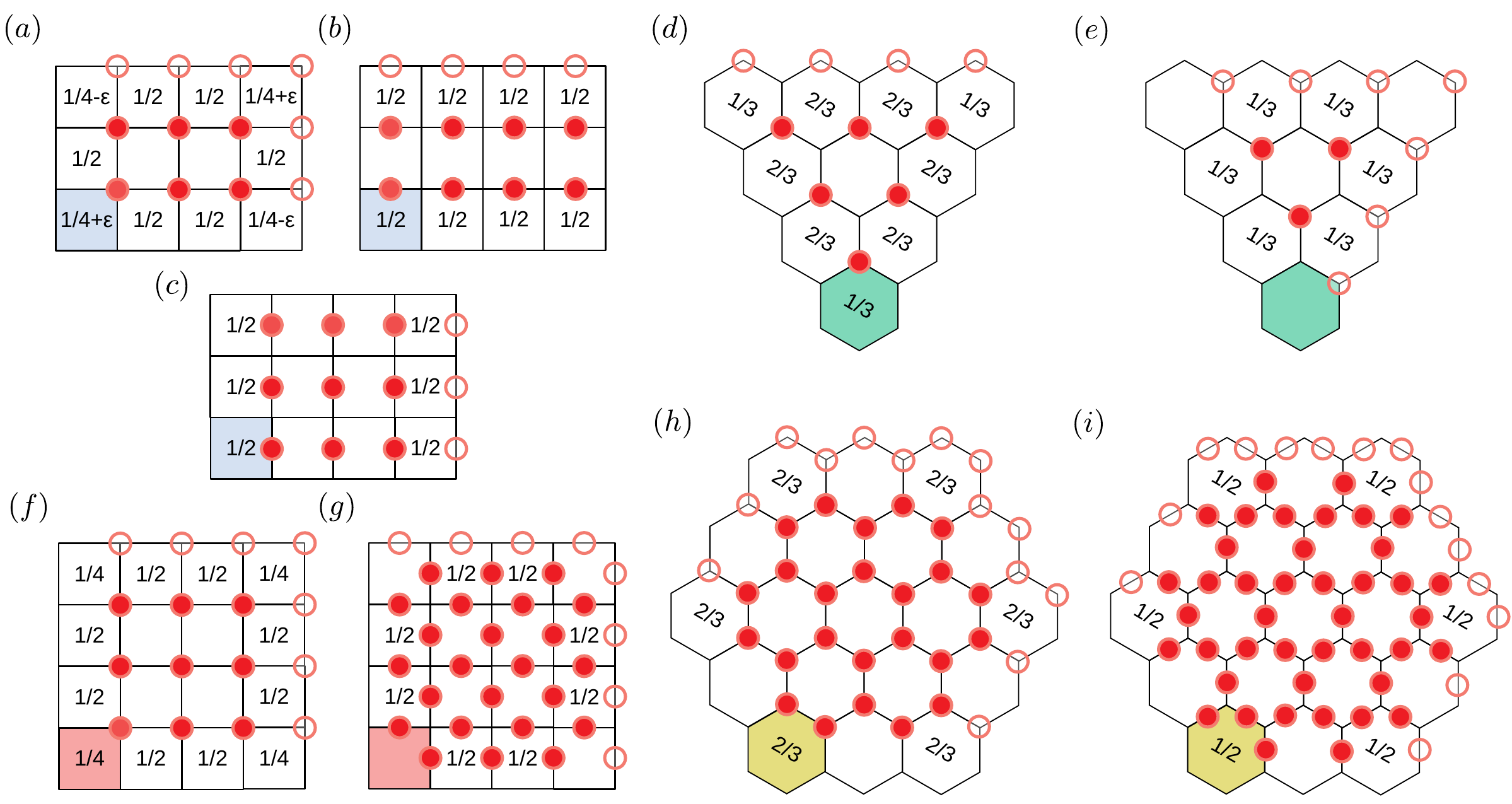}
\caption{$C_2$-symmetric tilings of one Wannier center placed at the Wyckoff position (a) $1b$ (b) $1d$ and (c) $1c$. $C_3$-symmetric tilings of one Wannier center placed at Wyckoff position (d) $1b$ and (e) $1c$. $C_4$-symmetric tilings of (f) one Wannier center placed at Wyckoff position $1b$ and (g) two Wannier centers placed at the Wyckoff position $2c$. $C_6$-symmetric tilings of (h) two Wannier centers placed at Wyckoff position $2b$ and (i) three Wannier centers placed at Wyckoff position $3c$. In all figures, some Wannier centers (hollow circles) have been removed to maintain $C_n$ symmetry. The resultant fractional occupation along the boundaries is labeled, which helps identify the quantization of fractional corner charges in each case.}
\label{fig:P_Q_calc}
\end{figure*}

The corner charges $(Q_{\text{corner}}^{(n)})$ can be determined by considering finite tilings of unit cells with $C_n$ symmetry for each Wannier center configuration. The formula for $Q_{\text{corner}}^{(n)}$ in terms of the symmetry-indicator invariants can then be determined by the procedure described below. In this procedure, a choice of a set of linearly independent $\chi$ indices is made and it is important to note that the corner charge formulae are not unique and depend on this choice. However, the corner charge itself is a physical quantity and is independent of this choice. 

The particular finite tilings considered here and the corner charges that they host are shown in Fig.~\ref{fig:P_Q_calc} for all $C_n$ symmetries. It is also important to note that in the case of $C_3$ symmetry, the corner charges depend on the exact type of finite tiling considered. The system formed by placing one Wannier center at $1b$ hosts a corner charge of $1/3$ only in an inverted triangle tiling, as shown in Fig.~\ref{fig:P_Q_calc}(d). If instead an upright triangle tiling is considered, the corner charge would be $0$ (and vice versa for $1c$). This subtlety does not arise in $C_3$-symmetric systems when bands with vanishing $\mathbf{P}^{(3)}$ are being considered.

With these two caveats in mind, we now derive the corner charge formulas given in the main text for all $C_n$ symmetries, with and without TRS.

\subsection{$C_2$ symmetry}
For both TR-symmetric and TR-broken cases, $C_2$-symmetric lattices have three symmetry-indicator invariants: $[X^{(2)}_1]$, $[Y^{(2)}_1]$ and $[M^{(2)}_1]$. The corner charge is given by a linear combination of these invariants
\begin{equation}
Q_{\text{corner}}^{(2)} = \lambda_1[X^{(2)}_1] + \lambda_2[Y^{(2)}_1] + \lambda_3[M^{(2)}_1].
\end{equation}
To determine $\lambda_1, \lambda_2, \lambda_3$, we solve for $Q_i = \chi_{ij}^{(2)}\lambda_j$, where $Q_i$ is the corner charge that corresponds to the $\chi^{(2)}$ formed by the $i$-th row of $\chi_{ij}^{(2)}$. Since the $\chi$-indices for different site symmetry representations are linearly dependent, we choose the following linearly independent set of $\chi$-indices that form a basis:
$1b: (-1,-1,0; 1)$, $1c: (-1,0,-1; 1)$, $1d: (0,-1,-1; 1)$. By examining the finite systems formed by the tilings of the $C_2$-symmetric unit cell in Fig.~\ref{fig:P_Q_calc}(a), (b), and (c), we see that the corner charges for $1b$, $1c$ and $1d$ are $1/2$, $0$ and $0$ respectively. This implies that
\begin{equation}
    \left( \begin{array}{c}
       \frac{1}{2} \\
        0 \\
        0
   \end{array} \right) = \left( \begin{array}{ccc}
            -1 & -1 & 0\\
            -1 & 0 & -1\\
             0 & -1 & -1
        \end{array} \right) \left( \begin{array}{c}
       \lambda_1 \\
        \lambda_2 \\
        \lambda_3
   \end{array} \right),
\end{equation}
which gives $\lambda_{1,2,3} = -\frac{1}{4}, -\frac{1}{4}, \frac{1}{4}$. Therefore,
\begin{align} 
Q_{\mathcal{T},\text{corner}}^{(2)} = Q_{\text{corner}}^{(2)} = \frac{1}{4}\left( -[X^{(2)}_1]-[Y^{(2)}_1]+[M^{(2)}_1] \right).
\end{align}

For $C_2$-symmetry, it is possible to find a situation where the charges at each corner are not quantized. The system shown in Fig.~\ref{fig:P_Q_calc}(a) is a valid $C_2$-symmetric configuration where the corner charges need not be quantized but the corner charge per $C_2$-symmetric sector is quantized to $1/2$.

\subsection{$C_3$ symmetry}
For the TR-symmetric case, $C_3$-symmetric lattices have two symmetry-indicator invariants: $[K^{(3)}_1]$ and $[K^{(3)}_2]$.  The corner charge is given by a linear combination of these invariants
\begin{equation}
Q_{\mathcal{T},\text{corner}}^{(3)} = \lambda_1[K^{(3)}_1] + \lambda_2[K^{(3)}_2].
\end{equation}
We choose the following linearly independent set of $\chi$-indices that form a basis: $1b: (-1,1; 1)$, $1c: (-1,0; 1)$. By examining the finite systems formed by the tilings of the $C_3$-symmetric unit cell in Fig.~\ref{fig:P_Q_calc}(d) and (e), we see that the corner charges for $1b$ and $1c$ are $1/3$ and $0$ respectively. This implies that
\begin{equation}
    \left( \begin{array}{c}
        \frac{1}{3} \\
        0
   \end{array} \right) = \left( \begin{array}{cc}
            -1 & 1 \\
            -1 & 0
        \end{array} \right) \left( \begin{array}{c}
       \lambda_1 \\
        \lambda_2
   \end{array} \right),
\end{equation}
which gives $\lambda_{1,2} = 0, \frac{1}{3}$. Therefore,
\begin{align}
 Q_{\mathcal{T},\text{corner}}^{(3)} = \frac{1}{3}[K^{(3)}_2].
\end{align}

For the TR-broken case, $C_3$-symmetric lattices have four symmetry-indicator invariants: $[K^{(3)}_1]$, $[K^{(3)}_2]$, $[K'^{(3)}_1]$, $[K'^{(3)}_2]$.  The corner charge is given by a linear combination of these invariants
\begin{equation}
Q_{\text{corner}}^{(3)} = \lambda_1[K^{(3)}_1] + \lambda_2[K^{(3)}_2] + \lambda_3[K'^{(3)}_1] + \lambda_4[K'^{(3)}_2].
\end{equation}
We choose the following linearly independent set of $\chi$-indices that form a basis: $1b_1: (0 \, | \, -1,1,-1,0; 1)$, $1b_2: (0 \, | \, 1,0,0,1; 1)$, $1c_1: (0 \, | \, -1,0,-1,1; 1)$, $1c_2: (0 \, | \, 0,1,1,0; 1)$. By examining the finite systems formed by the tilings of the $C_3$-symmetric unit cell in Fig.~\ref{fig:P_Q_calc}(d) and (e), we see that the corner charges for $1b_1$, $1b_2$, $1c_1$ and $1c_2$ are $1/3$, $1/3$, $0$ and $0$ respectively. This implies that
\begin{equation}
    \left( \begin{array}{c}
        \frac{1}{3} \\
        \frac{1}{3} \\
        0 \\
        0
   \end{array} \right) = \left( \begin{array}{cccc}
            -1 & 1 & -1 & 0 \\
            1 & 0 & 0 & 1 \\
            -1 & 0 & -1 & 1 \\
            0 & 1 & 1 & 0 
        \end{array} \right) \left( \begin{array}{c}
       \lambda_1 \\
        \lambda_2 \\
        \lambda_3 \\
        \lambda_4
   \end{array} \right),
\end{equation}
which gives $\lambda_{1,2,3,4} = \frac{1}{3}, \frac{1}{3}, -\frac{1}{3}, 0$. Therefore,
\begin{align}
 Q_{\text{corner}}^{(3)} = \frac{1}{3}\left([K^{(3)}_1] + [K^{(3)}_2] - [K'^{(3)}_1]  \right).
\end{align}

\subsection{$C_4$ symmetry}
For the TR-symmetric case, $C_4$-symmetric lattices have three symmetry-indicator invariants: $[X^{(4)}_1]$, $[M^{(4)}_1]$ and $[M^{(4)}_2]$.  The corner charge is given by a linear combination of these invariants
\begin{equation}
Q_{\mathcal{T},\text{corner}}^{(4)} = \lambda_1[X^{(4)}_1] + \lambda_2[M^{(4)}_1] + \lambda_3[M^{(4)}_2].
\end{equation}
We choose the following linearly independent set of $\chi$-indices that form a basis: $2c: (-1,-1,1; 2)$, $1b_1: (-1,1,0; 1)$, $1b_2: (2,0,0; 2)$. By examining the finite systems formed by the tilings of the $C_4$-symmetric unit cell in Fig.~\ref{fig:P_Q_calc}(f) and (g), we see that the corner charges for $2c$, $1b_1$ and $1b_2$ are $0$, $1/4$ and $1/2$ respectively (note that $1b_2: (2,0,0; 2)$ is induced by two bands, each with Wannier centers at $1b$. Therefore, the net corner charge is $(2\times 1/4) \text{ mod } 1 = 1/2$). This implies that
\begin{equation}
    \left( \begin{array}{c}
        0 \\
        \frac{1}{4} \\
        \frac{1}{2}
   \end{array} \right) = \left( \begin{array}{ccc}
            -1 & -1 & 1\\
            -1 & 1 & 0\\
             2 & 0 & 0
        \end{array} \right) \left( \begin{array}{c}
       \lambda_1 \\
        \lambda_2 \\
        \lambda_3
   \end{array} \right),
\end{equation}
which gives $\lambda_{1,2,3} = \frac{1}{4}, \frac{1}{2}, \frac{3}{4}$. Therefore,
\begin{align}
 Q_{\mathcal{T},\text{corner}}^{(4)} = \frac{1}{4}\left( [X^{(2)}_1]+2[M^{(4)}_1]+3[M^{(4)}_2] \right)   .
\end{align}

For the TR-broken case, $C_4$-symmetric lattices have four symmetry-indicator invariants: $[X^{(4)}_1]$, $[M^{(4)}_1]$, $[M^{(4)}_2]$ and $[M^{(4)}_4]$. The corner charge is given by a linear combination of these invariants
\begin{equation}
Q_{\text{corner}}^{(4)} = \lambda_1[X^{(4)}_1] + \lambda_2[M^{(4)}_1] + \lambda_3[M^{(4)}_2] + \lambda_4[M^{(4)}_4].
\end{equation}
We choose the following linearly independent set of $\chi$-indices that form a basis: $2c: (0 \, | \, -1,-1,1,1; 2)$, $1b_1: (0 \, | \, -1,1,0,0; 1)$, $1b_2: (0 \, | \, 1,0,-1,1; 1)$, $1b_3: (0 \, | \, 1,0,1,-1; 1)$. By examining the finite systems formed by the tilings of the $C_4$-symmetric unit cell in Fig.~\ref{fig:P_Q_calc}(f) and (g), we see that the corner charges for $2c$, $1b_1$, $1b_2$ and $1b_3$ are $0$, $1/4$, $1/4$ and $1/4$ respectively. This implies that
\begin{equation}
    \left( \begin{array}{c}
        0 \\
        \frac{1}{4} \\
        \frac{1}{4} \\
        \frac{1}{4}
   \end{array} \right) = \left( \begin{array}{cccc}
            -1 & -1 & 1 & 1\\
            -1 & 1 & 0 & 0\\
             1 & 0 & -1 & 1\\
             1 & 0 & 1 & -1
        \end{array} \right) \left( \begin{array}{c}
       \lambda_1 \\
        \lambda_2 \\
        \lambda_3 \\
        \lambda_4
   \end{array} \right),
\end{equation}
which gives $\lambda_{1,2,3,4} = \frac{1}{4}, \frac{1}{2}, \frac{3}{8}$, $\frac{3}{8}$. Therefore,
\begin{align}
 Q_{\text{corner}}^{(4)} = \frac{1}{4}\left( [X^{(2)}_1]+2[M^{(4)}_1]+\frac{3}{2}[M^{(4)}_2] + \frac{3}{2}[M^{(4)}_4] \right)   .
\end{align}

\subsection{$C_6$ symmetry}
For the TR-symmetric case, $C_6$-symmetric lattices have two symmetry-indicator invariants: $[M^{(2)}_1]$ and $[K^{(3)}_1]$.  The corner charge is given by a linear combination of these invariants
\begin{equation}
Q_{\mathcal{T},\text{corner}}^{(6)} = \lambda_1[M^{(2)}_1] + \lambda_2[K^{(3)}_1].
\end{equation}
We choose the following linearly independent set of $\chi$-indices that form a basis: $2b: (0,2; 4)$, $3c: (2,0; 3)$. By examining the finite systems formed by the tilings of the $C_6$-symmetric unit cells in Fig.~\ref{fig:P_Q_calc}(h) and (i), we see that the corner charges for $2b$ and $3c$ are $1/3$ and $1/2$ respectively (note that $2b: (0,2; 4)$ is induced by four bands, with Wannier centers of each pair at $2b$. The net corner charge is therefore ($2\times 2/3) \text{ mod } 1 = 1/3$). This implies that
\begin{equation}
    \left( \begin{array}{c}
        \frac{1}{3} \\
        \frac{1}{2}
   \end{array} \right) = \left( \begin{array}{cc}
            0 & 2 \\
            2 & 0
        \end{array} \right) \left( \begin{array}{c}
       \lambda_1 \\
        \lambda_2
   \end{array} \right),
\end{equation}
which gives $\lambda_{1,2} = \frac{1}{4}, \frac{1}{6}$. Therefore,
\begin{equation}
 Q_{\mathcal{T},\text{corner}}^{(6)} = \frac{1}{4}[M^{(2)}_1] + \frac{1}{6}[K^{(3)}_1].
\end{equation}

For the TR-broken case, $C_6$-symmetric lattices have three symmetry-indicator invariants: $[M^{(2)}_1]$,  $[K^{(3)}_1]$ and $[K^{(3)}_2]$.  The corner charge is given by a linear combination of these invariants
\begin{equation}
Q_{\text{corner}}^{(6)} = \lambda_1[M^{(2)}_1] + \lambda_2[K^{(3)}_1] + \lambda_3[K^{(3)}_2].
\end{equation}
We choose the following linearly independent set of $\chi$-indices that form a basis: $2b_1: (0 \, | \, 0,1,-2; 2)$, $2b_2: (0 \, | \, 0, 1, 1; 2)$, $3c: (0 \, | \, 2, 0, 0; 3)$. By examining the finite systems formed by the tilings of the $C_6$-symmetric unit cells in Fig.~\ref{fig:P_Q_calc}(h) and (i), we see that the corner charges for $2b_1$, $2b_2$ and $3c$ are $2/3$, $2/3$ and $1/2$ respectively. This implies that
\begin{equation}
    \left( \begin{array}{c}
        \frac{2}{3} \\
        \frac{2}{3} \\
        \frac{1}{2}\\
   \end{array} \right) = \left( \begin{array}{ccc}
            0 & 1 & -2 \\
            0 & 1 & 1 \\
            2 & 0 & 0
        \end{array} \right) \left( \begin{array}{c}
       \lambda_1 \\
        \lambda_2 \\
        \lambda_3
   \end{array} \right),
\end{equation}
which gives $\lambda_{1,2,3} = \frac{1}{4}, \frac{2}{3}, 0$. Therefore,
\begin{equation}
 Q_{\text{corner}}^{(6)} = \frac{1}{4}[M^{(2)}_1] + \frac{2}{3}[K^{(3)}_1].
\end{equation}

\begin{figure*}[htb]
\includegraphics[width=0.9\textwidth]{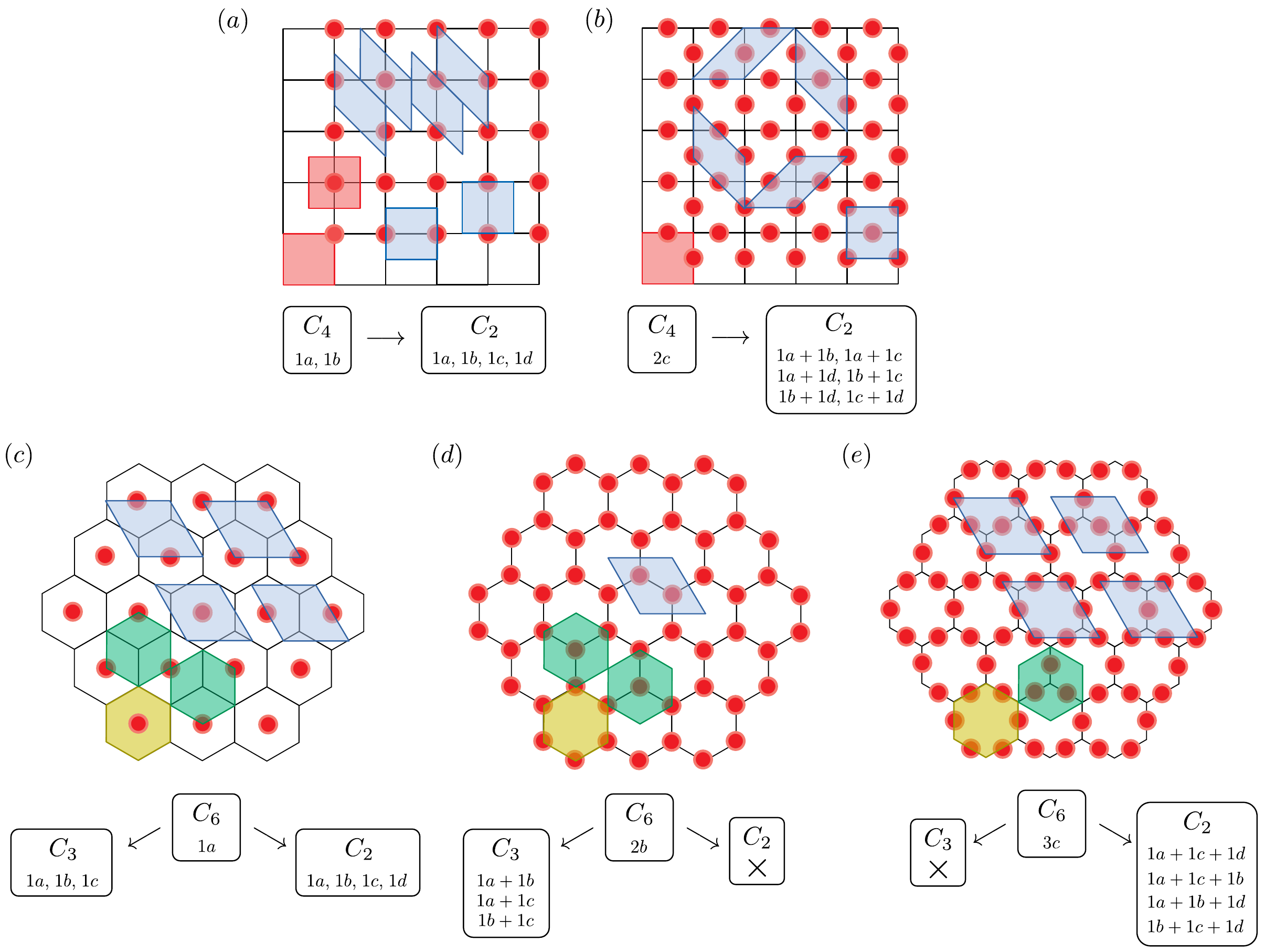}
\caption{All possible atomic limits that have (a) one and (b) two Wannier centers with $C_4$- (red) and $C_2$- (blue) symmetric unit cells that correspond to the same infinite lattice. All possible atomic limits that have (d) one, (e) two, and (f) three Wannier centers with $C_6$- (yellow), $C_3$- (green), and $C_2$- (blue) symmetric unit cells that correspond to the same infinite lattice. Note that some atomic limits with reduced symmetry do not have Wannier centers fixed to maximal Wyckoff positions but are instead movable (marked with a $\times$).}
\label{fig:UC_C4_C6}
\end{figure*}

\section{Effect of unit cell choices on boundary states}

When considering 2D PhCs with OAL bands, a particular choice of unit cell can affect the relevant symmetries for the topological  classification of the bulk and the presence of boundary states in a finite tiling of that unit cell. For example, consider all possible unit cell choices shown in Fig.~\ref{fig:UC_C4_C6}(a). The unit cells marked in red are $C_4$ symmetric and correspond to Wannier centers located at $1a$ or $1b$ positions. The same infinite structure is also consistent with unit cells that have reduced symmetry, in this case, $C_2$ symmetry, marked in blue. These correspond to Wannier centers located at the $1a$, $1b$, $1c$, or $1d$ positions. A finite tiling of any of these unit cells will result in edge or corner states depending on the dipole moment and corner charge of their respective  Wannier center configurations. This analysis is performed diagrammatically for all possible Wannier center configurations in Fig.~\ref{fig:UC_C4_C6}. 

Furthermore, when a choice of the unit cell reduces the symmetry of the system, the new symmetry-reduced invariants may be found using the following relations:
Under TRS, the $C_2$ invariants of a $C_4$-symmetric PhC obey $[X_1^{(2)}]=[Y_1^{(2)}]$ and $[M_1^{(2)}]=-2[M_2^{(4)}]$, and the $C_3$ invariants of a $C_6$-symmetric PhC obey $[K^{(3)}_1]=[K^{(3)}_2]$.
Under broken TRS, the $C_2$ invariants of a $C_4$-symmetric PhC obey $[X_1^{(2)}]=[Y_1^{(2)}]$ and $[M_1^{(2)}]=-2[M_2^{(4)}]$, and the $C_3$ invariants of a $C_6$-symmetric PhC obey $[K^{(3)}_1]=[K^{\prime (3)}_1]$ and $[K^{(3)}_2]=[K^{\prime (3)}_2]$.

These considerations are important for PhC design since a second ``cladding" material is often required to confine the boundary states of the topological ``core". With a different choice of unit cell made for the cladding material, both the core and cladding can have identical band structures and therefore conveniently overlapping bandgaps while having different topological invariants.

\end{document}